\documentclass[%
preprint,
 amsmath,amssymb,
 aps,
]{revtex4-2}

\usepackage{graphicx}
\usepackage{dcolumn}
\usepackage{bm}

\usepackage{subfigure}         
\usepackage{graphicx}          
\usepackage{dcolumn}           
\usepackage{bm}                
\usepackage[mathlines]{lineno} 

\usepackage{bookmark}  
\usepackage{graphicx, multirow,soul}
\usepackage{graphicx} 
\usepackage{booktabs}
\usepackage{comment}
\usepackage{amssymb}
\usepackage{amsmath}

\DeclareGraphicsExtensions{.eps,.pdf,.png,.jpg,.jpeg}

\usepackage[normalem]{ulem}

\usepackage{float}

\newcommand{\cteq}{\texttt{CT18 }}
\newcommand{\cteqpunctuation}{\texttt{CT18}}
\newcommand{\resbos}{{\sc ResBos }}
\newcommand{\chris}{\cite{PhysRevD.99.054004} }
\newcommand{\chrispunctuation}{\cite{PhysRevD.99.054004}}
\newcommand{\hera}{\texttt{CT14HERA2} }
\newcommand{\herapunctuation}{\texttt{CT14HERA2}}
\newcommand{\fb}{fb$^{-1}$}

\begin{document}

\preprint{MSUHEP-24-024}

\title{Further Reduction of the PDF Uncertainty in the High-Mass Drell-Yan Spectrum Utilizing Neutral and Charged Current Inputs} 

\author{Yao Fu}
\email{fuyao3@msu.edu}
\author{Raymond Brock}
\email{brock@pa.msu.edu}
\author{Daniel Hayden}
\email{haydend5@msu.edu}
\author{Chien-Peng Yuan}
\email{yuanch@msu.edu}

\affiliation{Department of Physics and Astronomy, Michigan State University, East Lansing, Michigan, 48823, USA}

\date{\today}

\begin{abstract}
 Uncertainties in the parametrization of Parton Distribution Functions are  a serious limiting systematic uncertainty in Large Hadron Collider searches for Beyond the Standard Model physics. This is especially true for measurements at high scales induced by quark and anti-quark collisions, where Drell-Yan continuum backgrounds are dominant. In Phys. Rev. D\textbf{99}, 054004 (2019) we presented a unique strategy for improving uncertainties using neutral current Drell-Yan backgrounds and here we update that strategy and include charged current Drell-Yan final states in the program and demonstrate significant improvements. Through a judicious selection of measurable kinematical quantities can reduce the assigned systematic PDF uncertainties by significant factors in limit-setting or discovery for neutral and charged, high mass Intermediate Vector Bosons. This approach will be take advantage of the huge statistical precision of future High Luminosity, Large Hadron Collider Standard Model datasets and could also improve uncertainties in the high statistics results from LHC Run 3.
 
 \vspace{0.5em}

 \vspace{0.5em}
\end{abstract}



\maketitle
\newpage 
\tableofcontents
\newpage

\section{Introduction}
Searches for physics beyond the standard model (BSM) are limited by systematic uncertainties of statistical, experimental, modeling, and theoretical origins. Signals and limits are all predicated on precise and accurate simulations of what standard model (SM) predictions would be for rates and especially kinematical distributions of measured objects and combinations of objects. Among the theoretical uncertainties, in many reactions precision knowledge of Parton Distributions Functions (PDFs) is becoming a limiting precision limit for BSM searches. 

Despite recent progress in lattice Quantum Chromodynamics (QCD), PDFs cannot be calculated analytically over a wide range of parton momentum fraction $x$ within
the framework of QCD, their shapes must be modeled by fitting measured distributions from many combinations of varied experimental data. These data come from combining  results of many experiments. Generally, these inputs include data from Deep Inelastic Scattering (DIS) experiments, various fixed target hadron experiments, the Fermilab Tevatron, and now the Large Hadron Collider (LHC) experiments (for example, the \texttt{CT18} ensembles of the CTEQ-TEA~\cite{ref:CT18} group combine 39 separate experimental inputs, including 13 from low energy LHC experiments). The experiments publish data templates of their measured cross sections plus correlated error matrices and these are fit by the three major PDF fitting groups: CTEQ-TEA~\cite{ref:CT18}, MSHT~\cite{ref:MSHT20}, and NNPDF~\cite{ref:NNPDF40},  into different ``brands'' of parameterized PDF functions for the individual parton species. Plus, in order to be used in modeling the SM background for BSM searches, the PDF groups characterize their theoretical and fitting uncertainties through published error matrices --- for CTEQ, the product is tables of nominal best fits for gluon, up, anti-up, down, anti-down, strange, charm, and bottom partons and 56 Hessian sets and two extreme sets for gluon at the small $x$ region. (Two of us are CTEQ members and so our reference to PDF fitting is through the lens of the CT18 PDF sets \cite{ref:CT18}.)
\begin{figure}[htp]
\centering
\includegraphics[width= 0.99\linewidth]{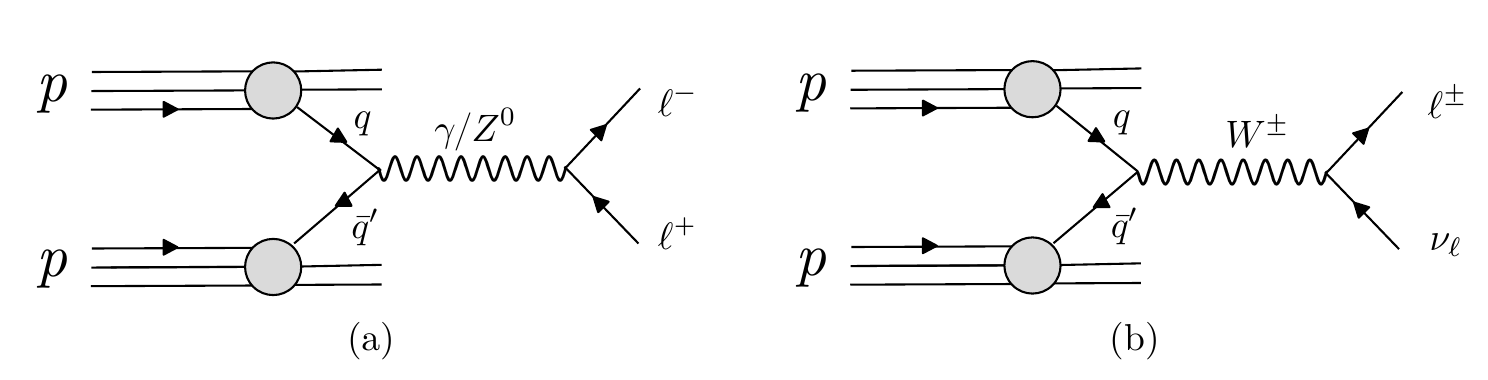}
\caption{Lowest order diagrams for (a) the production of neutral IVBs, $\gamma/Z^0$ bosons in proton--proton collisions and (b)the production of charged IVBs, $W^\pm$.}
\label{ZW_LO} 
\end{figure}

Our interest is in reducing the uncertainties inherent in BSM high mass Drell-Yan (DY) searches for charged dilepton final states and lepton plus missing energy final states. The former comes in lowest-order (LO) as shown in Figure~\ref{ZW_LO} from the reaction  $pp \to \ell^{+}\ell^{-} + X$, where $\ell = e \text{ or } \mu$ and the latter from reaction $pp \to \ell^{\pm} \nu + X$, where $\ell = e \text{ or } \mu$. Forming the invariant mass of all $ \ell^{+}\ell^{-}$ pairs enables a search for resonant production of new intermediate vector boson (IVB) states such as, ``$Z'\,$'' through a neutral current (NC) process, and analyzing the transverse mass of $ \ell^{\pm}\nu$ is a search for new $W$ boson states, ``$W'\,$'' through a charged current (CC) process.  The limits on such searches are dominated by systematic uncertainties and among them, are those due to theoretical contributions dominated by PDFs.

In  \chris we suggested a new strategy for selecting experimental input to new, specialized PDF sets (``boutique'')  as worth exploring and in this work we revisit that effort in the light of more recent global fits and we expand it to other reactions. Others have also explored this topic as well. \cite{Fiaschi:2022wgl, Fiaschi:2021sin, Fiaschi:2021okg, Amoroso:2020fjw, Accomando:2019vqt, Accomando:2018nig, Accomando:2017scx, Bodek:2015ljm,hammou2023hideseekpdfsconceal, costantini2024simunetopensourcetoolsimultaneous, Iranipour_2022, Greljo_2021, hammou2024unravellingnewphysicssignals}

\section{Review of IVB Searches}
\begin{figure}[!th]
  \centering
  \subfigure[]{
    \label{ATLAS_Zee_36fb}
    \includegraphics[width=0.45\textwidth]{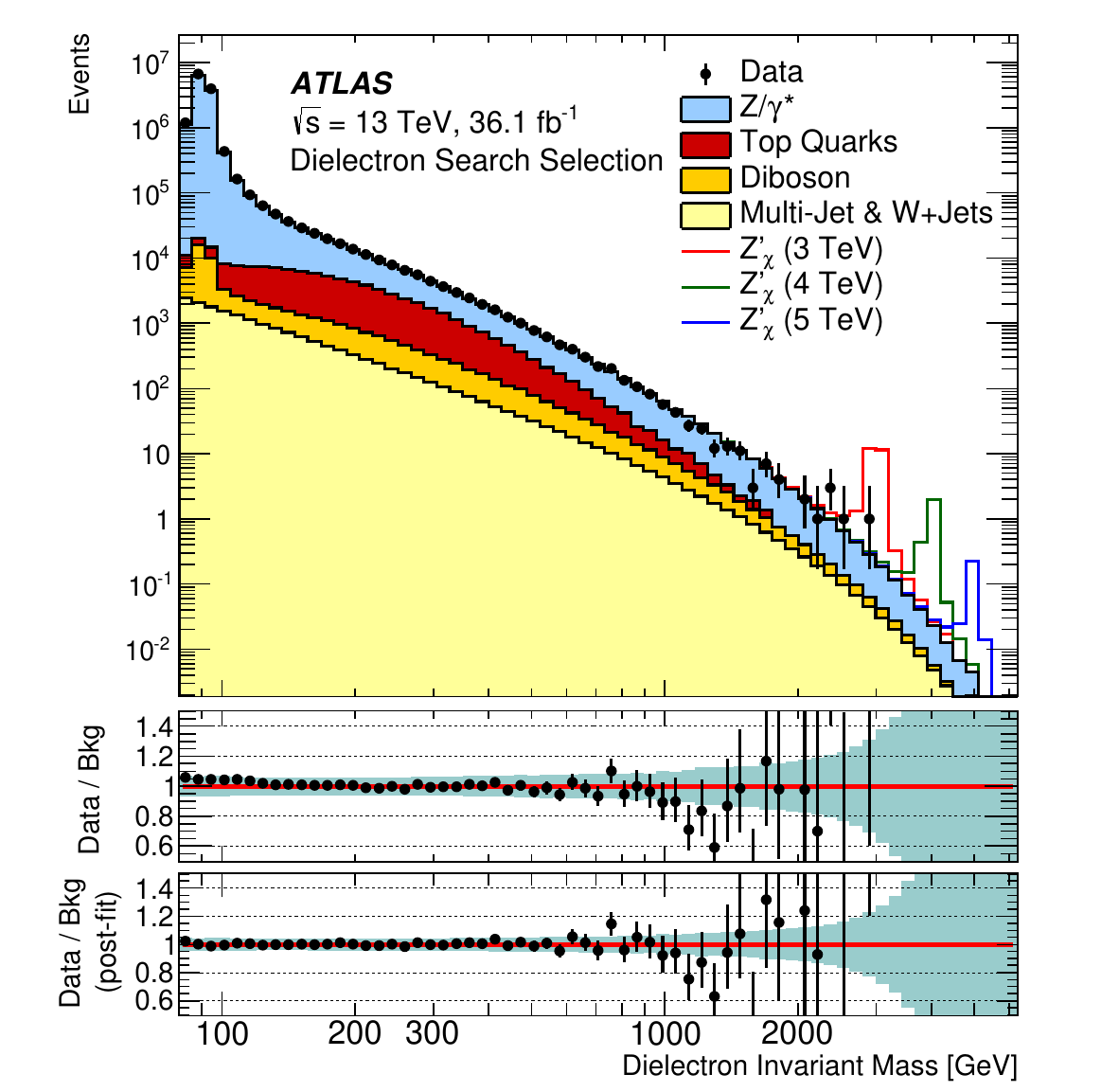}
  }
  \subfigure[]{
    \label{ATLAS_Zee_139fb}
    \includegraphics[width=0.45\textwidth]{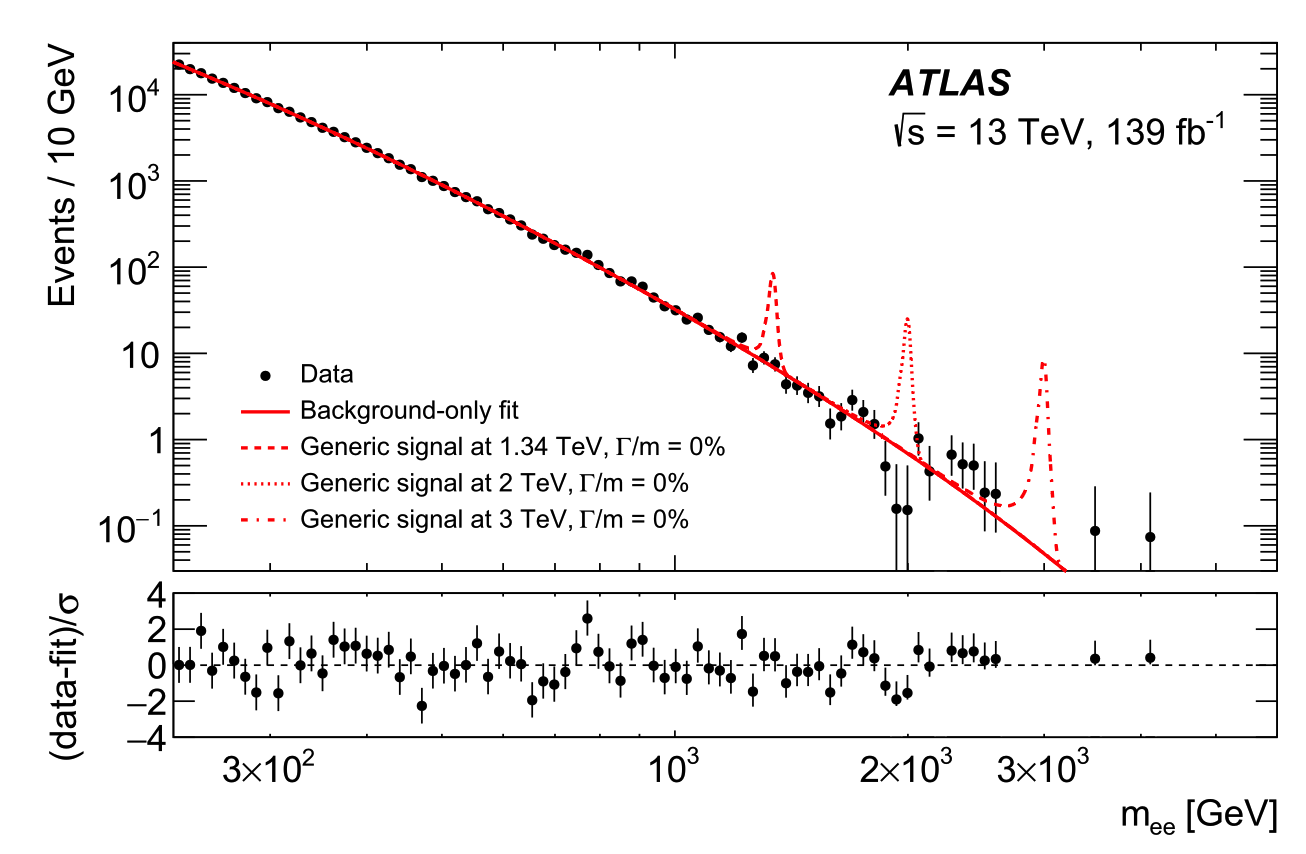}
  }
    \subfigure[]{
    \label{CMS_Zee_36fb}
    \includegraphics[width=0.45\textwidth]{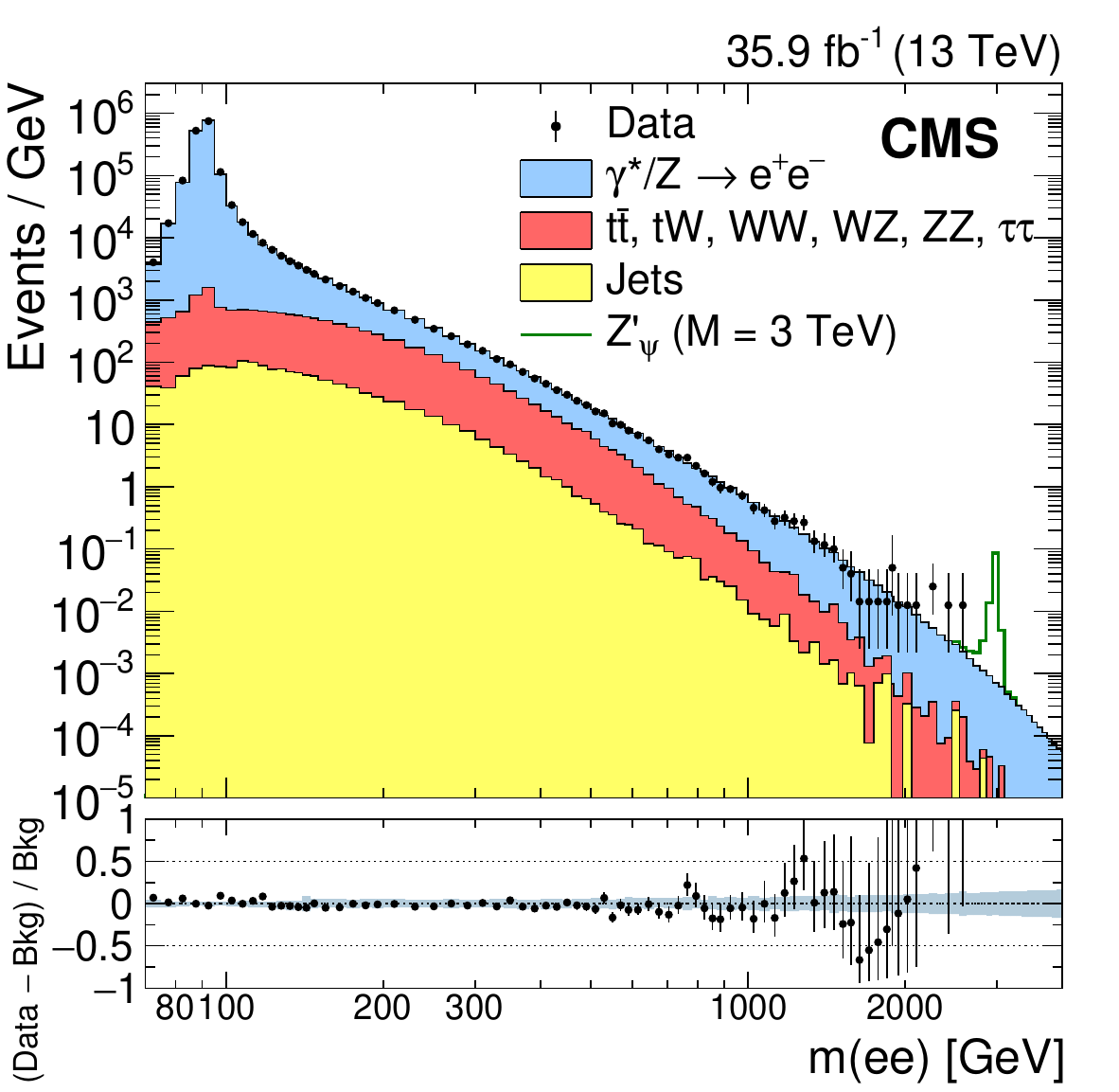}
  }
    \subfigure[]{
    \label{CMS_Zee_137fb}
    \includegraphics[width=0.45\textwidth]{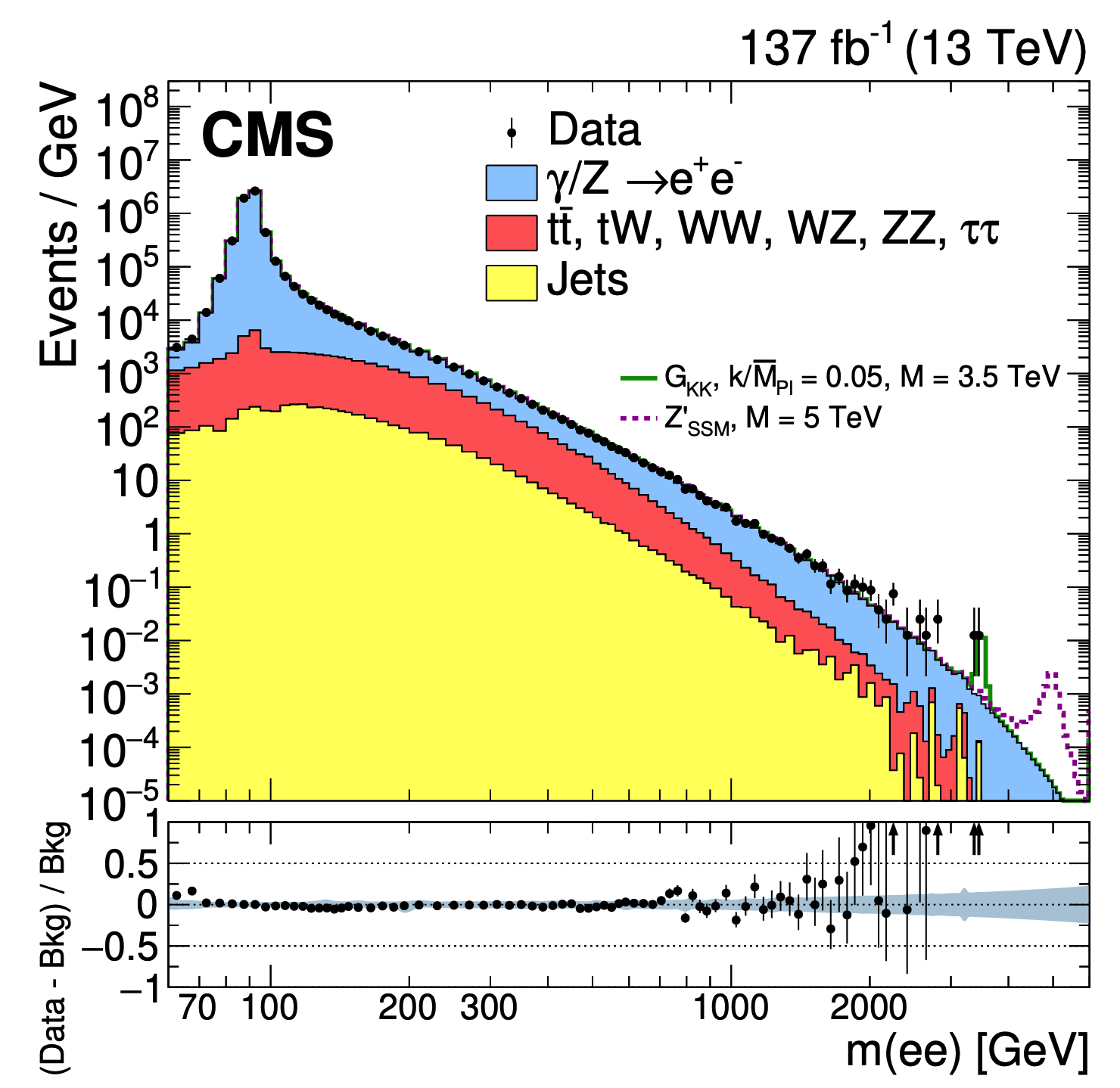}
  }
  \caption{Historical distributions of the dilepton invariant mass in a hypothetical $ q\overline{q}  \to \gamma^{*}/Z'  \to \ell^{+}\ell^{-}$ reaction in of both ATLAS (a)~\cite{ref:ATLAS13TeV:ZPrime2017} and (b)~\cite{ref:ATLAS13TeV:ZPrime2019} and CMS (c)~\cite{ref:CMS13TeV:ZPrime2018} and (d)~\cite{ref:CMS13TeV:ZPrime2021}. ATLAS (b) has chosen to use a data-driven approach, essentially extrapolating the low mass DY continuum into the search region.}
  \label{ATLAS_CMS_Zee}
\end{figure}

\subsection{Resonant $Z'\,$ Search Status}

Searches for IVB of high mass have been undertaken by the ATLAS and CMS collaborations using LHC Run 2 proton--proton collisions ($pp$) at a center-of-mass energy of 13~TeV and lower~\cite{ref:ATLAS13TeV:ZPrime2017,ref:CMS13TeV:ZPrime2018,ref:ATLAS13TeV:ZPrime2019, ref:CMS13TeV:ZPrime2021}. Figure~\ref{ATLAS_CMS_Zee} shows the DY spectra for $pp \to e^{+}e^{-} + X$ from the two LHC searches for evidence of resonant $Z'\,$ states.

The bands in Figures~\ref{ATLAS_Zee_36fb}, \ref{CMS_Zee_36fb}, and \ref{CMS_Zee_137fb} indicate the direct uncertainty for experimental and DY SM backgrounds, dominated by PDF uncertainties. In Run 2, ATLAS took a data-driven approach to calculating the backgrounds and the uncertainties from PDFs contribute less of the overall uncertainty in the background-modeling approach of the other three measurements. However, PDF uncertainties also dominate below the maximum search region in this measurement as well. 
\begin{table}[h!]
\begin{ruledtabular}
\begin{tabular}{|l|cc|cc|}
\hline
\multirow{2}{*}{Systematic Uncertainty} &
\multicolumn{2}{c|}{CMS} &
\multicolumn{2}{c|}{ATLAS } \\
 & $ee$ [\%] & $\mu\mu$ [\%] & $ee$ [\%] & $\mu\mu$ [\%] \\
\hline
$\int L\,dt =$                & \multicolumn{2}{c|}{35.9 fb$^{-1}$~\cite{ref:CMS13TeV:ZPrime2018} }& \multicolumn{2}{c|}{36.1 fb$^{-1}$ ~\cite{ref:ATLAS13TeV:ZPrime2017}} \\
\hline
PDF Uncertainty                 & 7   & 7   & 20.8 & 13.1 \\
Combined Experimental         & 12  & 15  & 12.8 & 18.9 \\
\hline
$\int L\,dt =$                & \multicolumn{2}{c|}{137 fb$^{-1}$~\cite{ref:CMS13TeV:ZPrime2021}} & \multicolumn{2}{c|}{139 fb$^{-1}$~\cite{ref:ATLAS13TeV:ZPrime2019}} \\
\hline
PDF Uncertainty                 & 20   & 20   & NA & NA \\
Combined Experimental         & 7  & 10  & 12.5 &47.2 \\
\hline
\end{tabular}
\end{ruledtabular}
  \caption{Published uncertainties for all LHC NCDY dilepton $Z'\,$ searches for CMS and ATLAS for the 13~TeV LHC running. The uncertainties have a scale dependence and can differ according to the di-lepton channel. For the 35.9/36.1~fb$^{-1}$ the results quoted  are evaluated at a mass of approximately 4~TeV for ATLAS and 5~TeV  for CMS. The experimental systematic uncertainties for the 137/139~fb$^{-1}$ running are for 6~TeV for CMS and what's listed is an average of ranged quoted. For the ATLAS 139~fb$^{-1}$ result the collaboration used a ``data-driven'' approach that extrapolates the continuum into the search region, so no monte carlo predictions for SM backgrounds and their errors are included. The large uncertainty quoted for ATLAS muons is due to the ``Good muon requirement'' which is worth about 40\%.}
  \label{tab:PDFuncertaintiesLHCZ}
\end{table}
\subsection{Resonant $W'\,$ Search Status}
\begin{figure}[!th]
  \centering
  \subfigure[]{
    \label{LABEL}
    \includegraphics[width=0.48\textwidth]{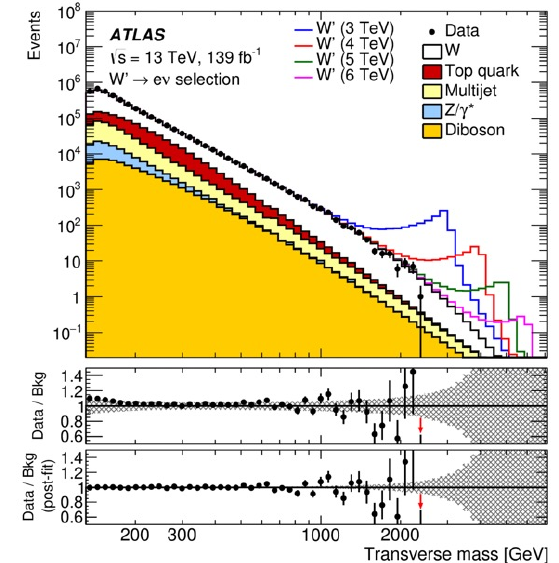}
  }
  \subfigure[]{
    \label{LABEL}
    \includegraphics[width=0.48\textwidth]{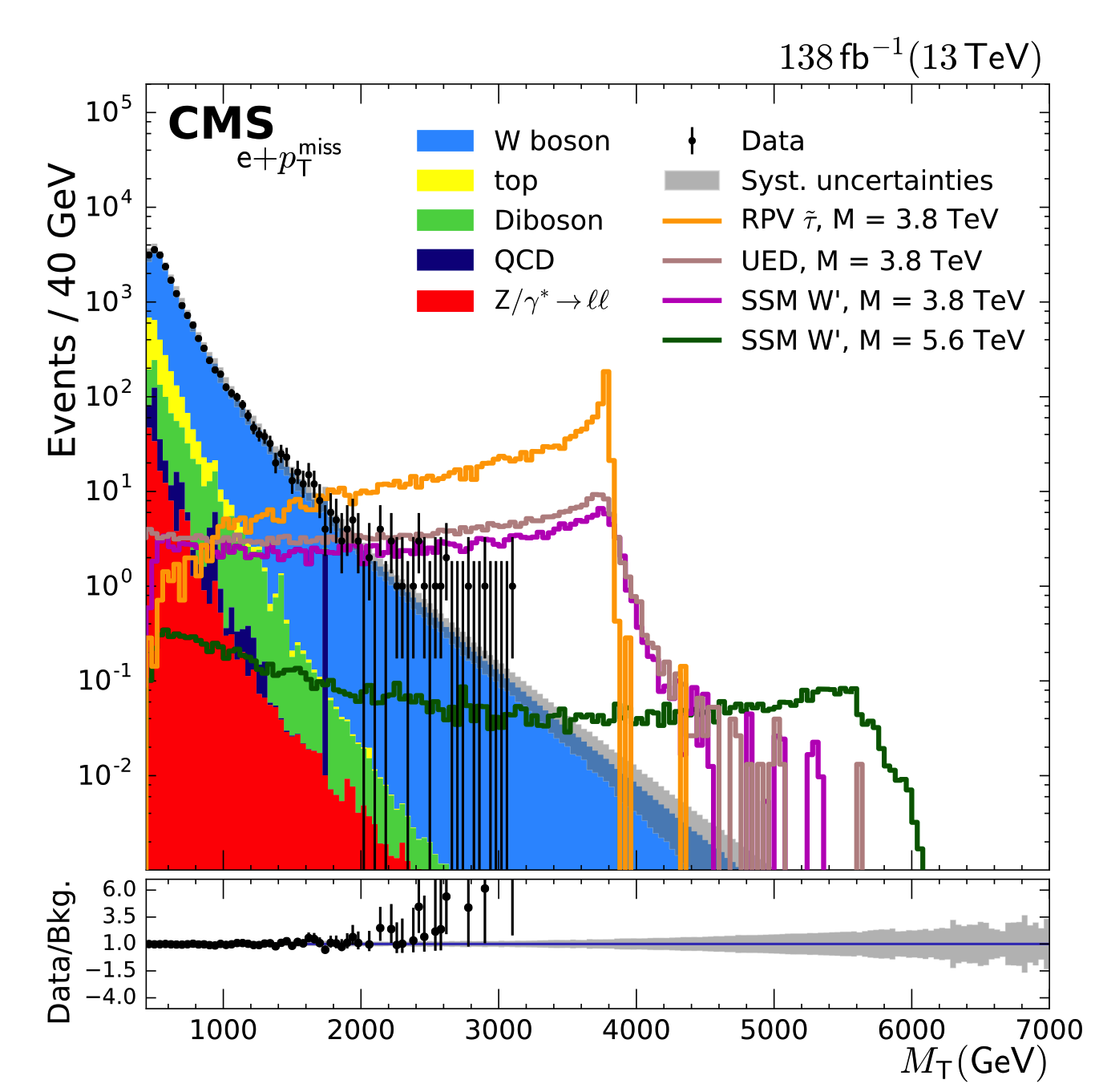}
  }
  \caption{Most recent distributions of the transverse mass in a hypothetical $q\overline{q}  \to W^\pm \to \ell^{\pm}\nu_{\ell}$ reaction in of both (a) ATLAS and (b) CMS.}
  \label{CMS_Wenu_CC_Mt_138}
\end{figure}

Searching for BSM CC IVBs is also a regular project for LHC collaborations. For $W'\,$  decays, the lack of information about the invisible neutrino compromises precision. But as discussed below, the transverse mass distribution which relies on determination of an event's charged lepton momentum and its missing transverse energy still leads to measurable distributions which, in terms of a discovery, would create deviations from the steeply falling $m_T$ distribution. Figure~\ref{CMS_Wenu_CC_Mt_138} shows the DY spectra for $pp \to \ell^{\pm} \nu + X$ from the two LHC searches for evidence of resonant $W'\,$ states.

\begin{table}[ht]
\begin{ruledtabular}
\begin{tabular}{|l|cc|cc|}
\hline
\multirow{2}{*}{Systematic Uncertainty} &
\multicolumn{2}{c|}{CMS } &
\multicolumn{2}{c|}{ATLAS } \\
 & $M_T(e)$ [\%] & $M_T(\mu)$ [\%] & $M_T(e)$ [\%] & $M_T(\mu)$ [\%] \\
\hline
$\int L\,dt =$                & \multicolumn{2}{c|}{138 fb$^{-1}$~\cite{2022CMSW} }& \multicolumn{2}{c|}{139 fb$^{-1}$ ~\cite{2025ATLASW}} \\
\hline
PDF Uncertainty                 & $\sim$ 40   & $\sim$ 40   & $\sim$ 8.2 & $\sim$ 7.4 \\ Combined Experimental         & $\sim$ 5  & $\sim$ 5  & $\sim$ 12 (16.4) & $\sim$ 17 (14.0) \\
\hline
\end{tabular}
\end{ruledtabular}
  \caption{Most recent published uncertainties for CCDY transverse mass $W'$ searches for CMS and ATLAS for the 13~TeV LHC running. The systematic uncertainties for the 137/139~fb$^{-1}$ running are at 4~TeV for CMS. The total systematic uncertainties for the ATLAS measurement are determined at 2~TeV while the parenthetical results are their PDF uncertainties evaluated at 6~TeV.}
  \label{tab:PDFuncertaintiesLHCW}
\end{table}

\subsection{Current Lower Limits for the Production of High-Mass, BSM IVB States}

The results for lower bound limits on a Sequential Standard Model $Z'$  are $M_{Z'}<5.1$~TeV at 95\% confidence level for ATLAS and $M_{Z'}<5.15$~TeV at 95\% confidence level for CMS. Table~\ref{tab:PDFuncertaintiesLHCZ} show approximate amounts attributed to PDF and total uncertainties in the four historical LHC measurements.

The results for lower bound limits on a Standard Model-like $W'$  are $M_{W'}(e\nu)<6.0$~TeV and $M_{W'}(\mu\nu)<5.1$~TeV at 95\% confidence level for ATLAS and $M_{W'}(e\nu)<5.4$~TeV and $M_{W'}(\mu\nu)<5.6$~TeV at 95\% confidence level for CMS. Table~\ref{tab:PDFuncertaintiesLHCW} show a summary of PDF and total uncertainties for the two most recent measurements of CMS  and ATLAS. 

The two experiments are obviously in agreement that any potential BSM IVB production is for a state with mass beyond $M>5-6$~TeV. Discovery signals now at the LHC would be tiny and highly compromised by PDF uncertainties.

\section{The Drell-Yan Process}

The general NC Drell-Yan (NCDY) process~\cite{PhysRevLett.25.316} of $pp  \to \ell^{+}\ell^{-} + X$ at leading order originates from an $s$-channel exchange of a virtual photon or a $Z^0$ electroweak boson at the parton level:
\begin{equation}\label{eq:s}
  q\overline{q}  \to \gamma^{*}/Z  \to \ell^{+}\ell^{-}.
\end{equation}
Here, $X$ denotes any additional final-state particles (radiated partons, the underlying event, multi-parton interactions, etc.). 

The general CC Drell-Yan (CCDY) process of $pp  \to \ell^{\pm}\nu_{\ell} + X$ involves an intermediate $W$ boson that decays into a charged lepton and its neutrino,

\begin{equation}\label{eq:s}
  q\overline{q}  \to W^\pm \to \ell^{\pm}\nu_{\ell}.
\end{equation}

In next-to-leading order (NLO), the real corrections introduce additional $t$ and $u$-channel contributions  for each subprocess:
\begin{align}
  qg &\rightarrow \gamma^{*}/Z/(W) \rightarrow \ell^{+}\ell^{-}(\nu_{\ell}) + q \label{eq:t1}  \\ 
  \overline{q}g &\rightarrow \gamma^{*}/Z/(W) \rightarrow \ell^{+}\ell^{-}(\nu_{\ell}) + q \label{eq:t2} \\
  q\overline{q} &\rightarrow \gamma^{*}/Z/(W) \rightarrow \ell^{+}\ell^{-}(\nu_{\ell}) + g. \label{eq:t3} 
\end{align}
The leading order process for NC and CC  is depicted in Fig.~\ref{ZW_LO}.
In each case, the vector boson decays into a pair of same-flavor, lepton-anti-lepton pairs. For simplicity, our discussion will focus on the leading order process, but all of our results are based on the {\sc ResBos2} package~\cite{Isaacson:2022rts,Isaacson:2023iui}, which provides predictions at next-to-next-to-leading logarithmic plus next-to-next-to-leading order (N$^3$LL+NNLO)
accuracy in QCD interactions, including the correct treatment of angular functions.
The {\sc ResBos2} package is the updated version of the original {\sc ResBos} package~\cite{PhysRevD.50.R4239,PhysRevD.56.5558,PhysRevD.67.073016}.
Throughout this work, the resummation calculations are performed using the canonical choices for the resummation scales and the BLNY nonperturbative functional form~\cite{Landry:2002ix}. In addition, the  renormalization and factorization scales are set to the transverse mass of the di-lepton pair in the calculation of the fixed-order contribution  for both the NC and CC channels.

Notice that the PDFs of interest in each reaction are quark and anti-quark species and not significantly gluons. So the premium in each case is knowledge of the $u(x), \bar {u}(x), d(x), \text{ and } \bar{d}(x)$ parton densities at highscales. A reminder of the kinematics for searches at the scale of interest is worth describing.

%

\section{NC PDF Reduction}
The most granular NC DY triple-differential cross section can be represented as a function of the dilepton invariant mass $m_{\ell\ell}$, the dilepton rapidity $y_{\ell\ell}$, and the cosine of the lepton polar angle in the Collins-Soper rest frame $\cos\theta^{*}$ \cite{PhysRevD.16.2219}.

For the LO $s$-channel process, the DY triple-differential cross section can be written as
\begin{equation} \label{Zcross}
  \frac{d^{3}\sigma}{dm_{\ell\ell}dy_{\ell\ell}d\cos\theta^{*}} = \frac{\pi\alpha^{2}}{3m_{\ell\ell}s} %
  \sum_{q} C_{q} \left[ f_{q/P_{1}}(x_{1},Q^{2}) f_{\bar{q}/P_{2}}(x_{2},Q^{2}) %
    + \left( q\leftrightarrow \overline{q} \right) \right].
\end{equation}
Here $\sqrt{s}$ is the center of mass energy of the LHC, and $P_1$ and $P_2$ are the 4-momenta of protons 1 and 2. In the standard fashion, $x_{1}$ and $x_{2}$ are the incoming parton momentum fractions such that  $p_{1}=x_{1}P_{1}$ and  $p_{2}=x_{2}P_{2}$. We take our notation from~\cite{Aaboud2017}.

The functions $f_{q/P_{1}}(x_{1},Q^{2})$ and $f_{\bar{q}/P_{2}}(x_{2},Q^{2})$ are the PDFs for quark flavors $q$ and $\bar{q}$, respectively. The term $(q\leftrightarrow \overline{q})$ accounts for the fact that either proton can carry a sea quark, as the LHC is a proton-proton collider. \par
Finally, the quantity $C_{q}$ accounts for the parton-level dynamics in terms of important electroweak parameters, and exhibits dependencies on both dilepton mass and $\cos\theta^\star$.
The transferred four-momentum, $Q$, sets the energy scale of the collision, which can be identified with the square of the dilepton invariant mass, $m^{2}_{\ell\ell}$. Well-known kinematic definitions include
\begin{align}
  Q^{2} = \left(p_{1}+p_{2}\right)^{2} = x_{1}x_{2}s
\end{align}
and
\begin{align}
  y_{\ell\ell} = \frac{1}{2} \ln \left( \frac{x_{1}}{x_{2}} \right),
\end{align}
which parametrizes the dilepton rapidity in terms of the $x$ fractions of the initial-state partons at LO. From these, the variables are related, also at LO, by
\begin{equation} \label{eq:xqxqb}
  x_{1}=\frac{m_{\ell\ell}}{\sqrt{s}}e^{+y_{\ell\ell}}, \quad x_{2}=\frac{m_{\ell\ell}}{\sqrt{s}}e^{-y_{\ell\ell}}.
\end{equation}

\begin{figure}[!h]
  \centering
  \subfigure[]{
    \label{fig:uvl3000R}
    \includegraphics[width=0.43\textwidth]{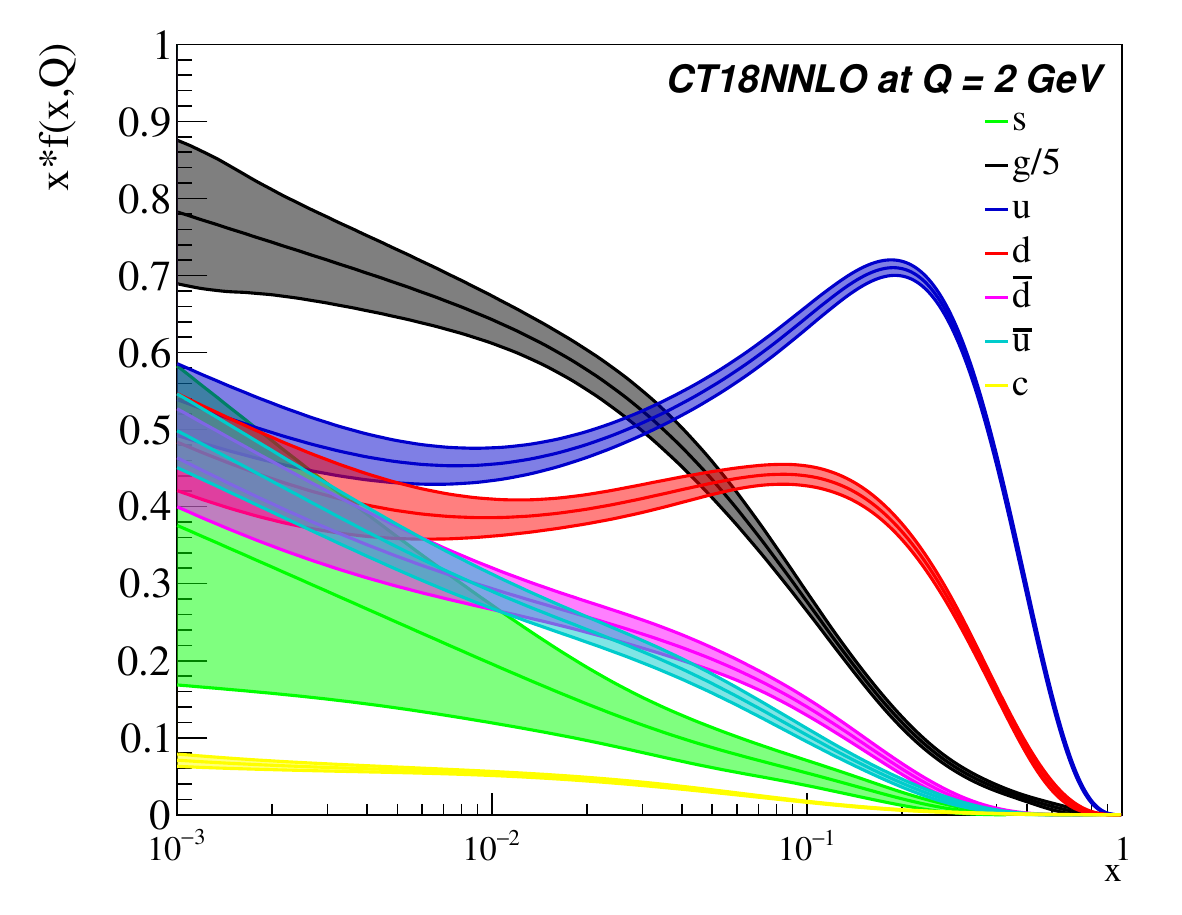}
  }
  \subfigure[]{
    \label{fig:dvl3000R}
    \includegraphics[width=0.43\textwidth]{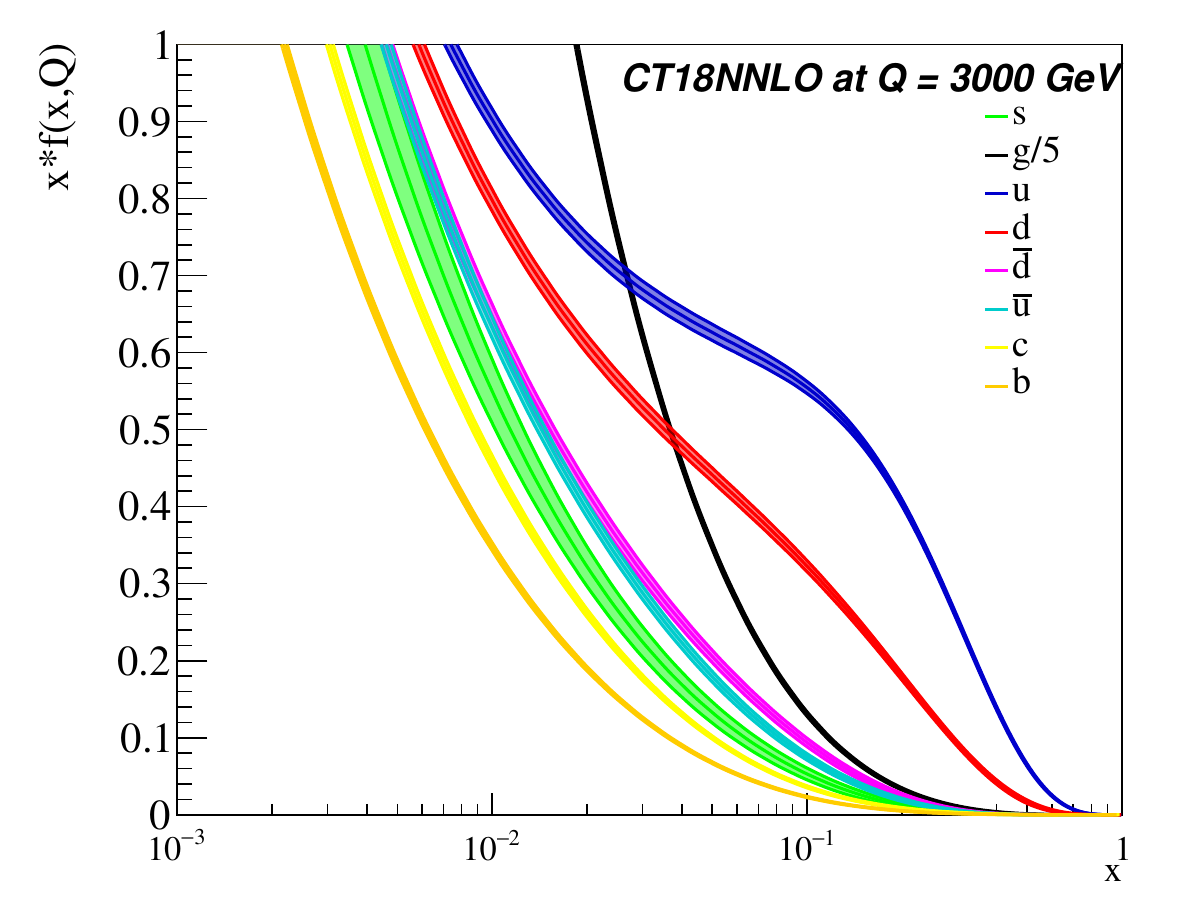}
  }\\
  \subfigure[]{
    \label{fig:ubar3000R}
    \includegraphics[width=0.43\textwidth]{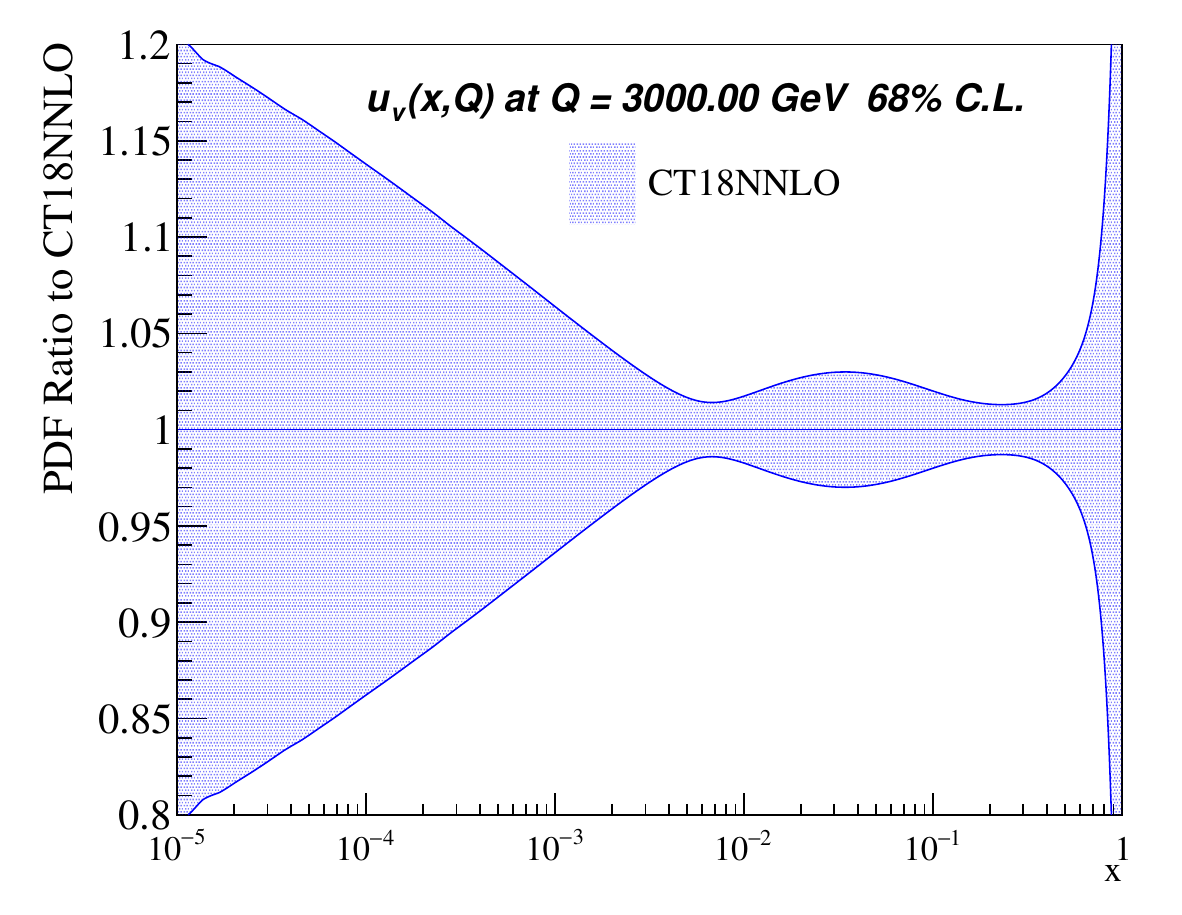}
  }
  \subfigure[]{
    \label{fig:dbar3000R}
    \includegraphics[width=0.43\textwidth]{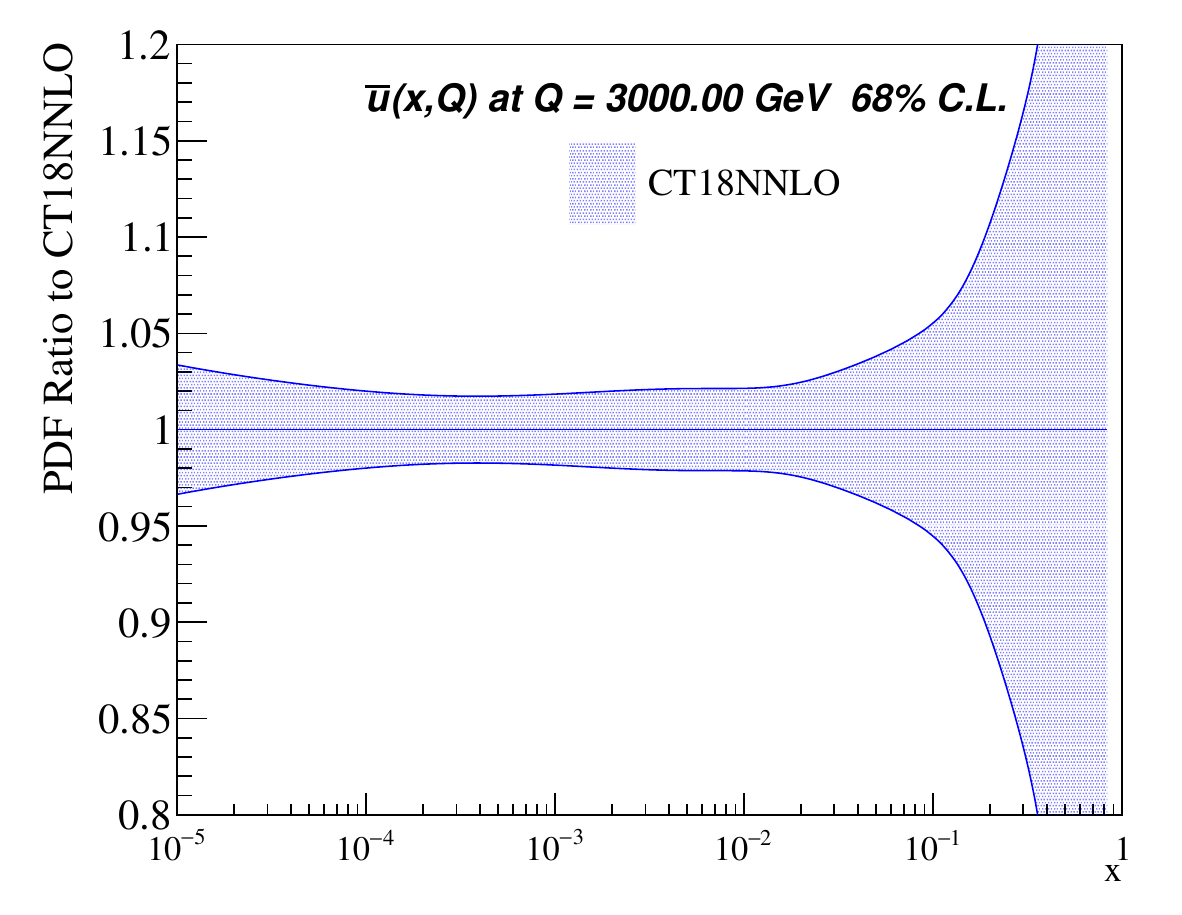}
  }
  \caption{The CT18 quark PDFs evaluated at a scales of (a) $Q=2$~GeV and (b) $Q=3$~TeV, while their uncertainties are shown in (c) and (d).}
  \label{fig:PDFUncertainties}
\end{figure}
\clearpage
Eq.~(\ref{eq:xqxqb}) provides the first hint to the source of the large PDF uncertainty in high-mass DY production. The $\sqrt{s}=13$ TeV LHC is now probing extremely large values of $m_{\ell\ell}$, beyond a few TeV. As such, a central dilepton event with an invariant mass of $m_{\ell\ell}=4$~TeV and rapidity of $y_{\ell\ell}=0$ requires $x$ fractions beyond $x\simeq 0.2$.
\begin{figure}[H]
  \centering
  \subfigure[]{
    \label{L2Sensitivity_uv}
    \includegraphics[width=0.55\textwidth]{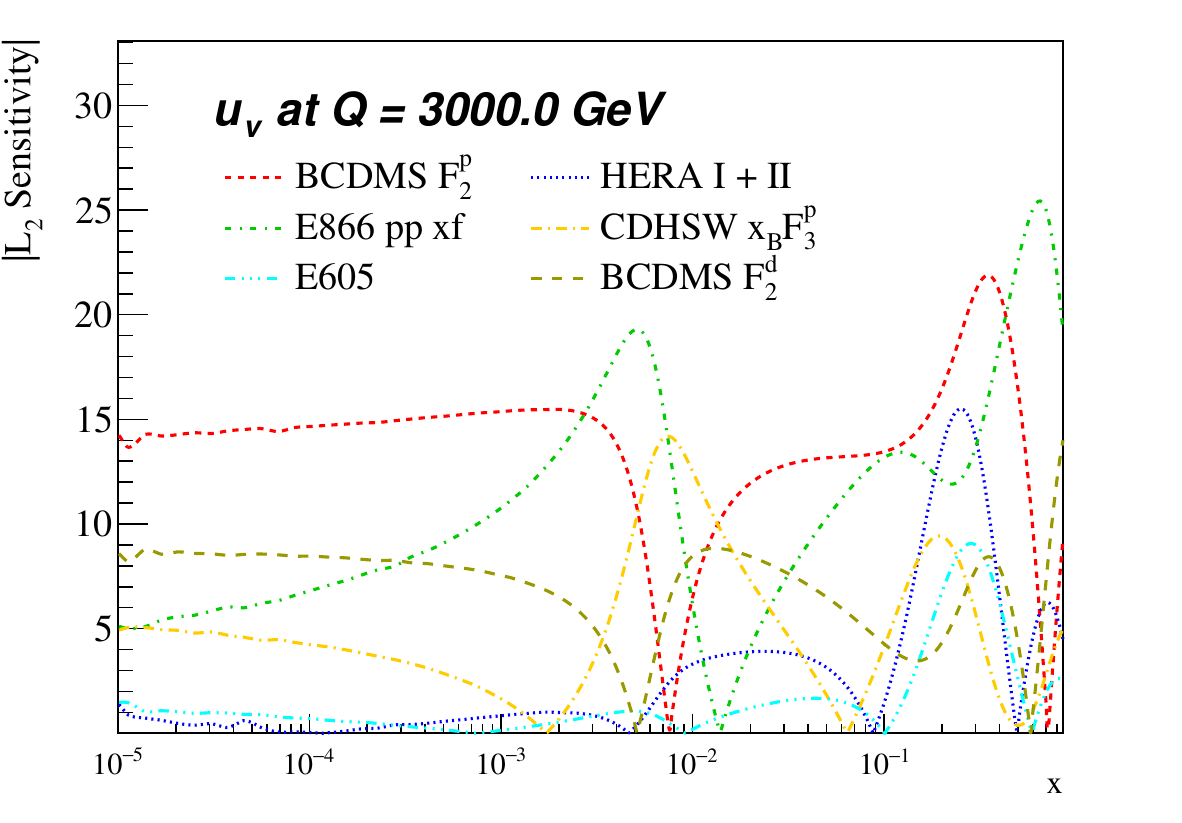}
  }
  \subfigure[]{
    \label{L2Sensitivity_ubar}
    \includegraphics[width=0.55\textwidth]{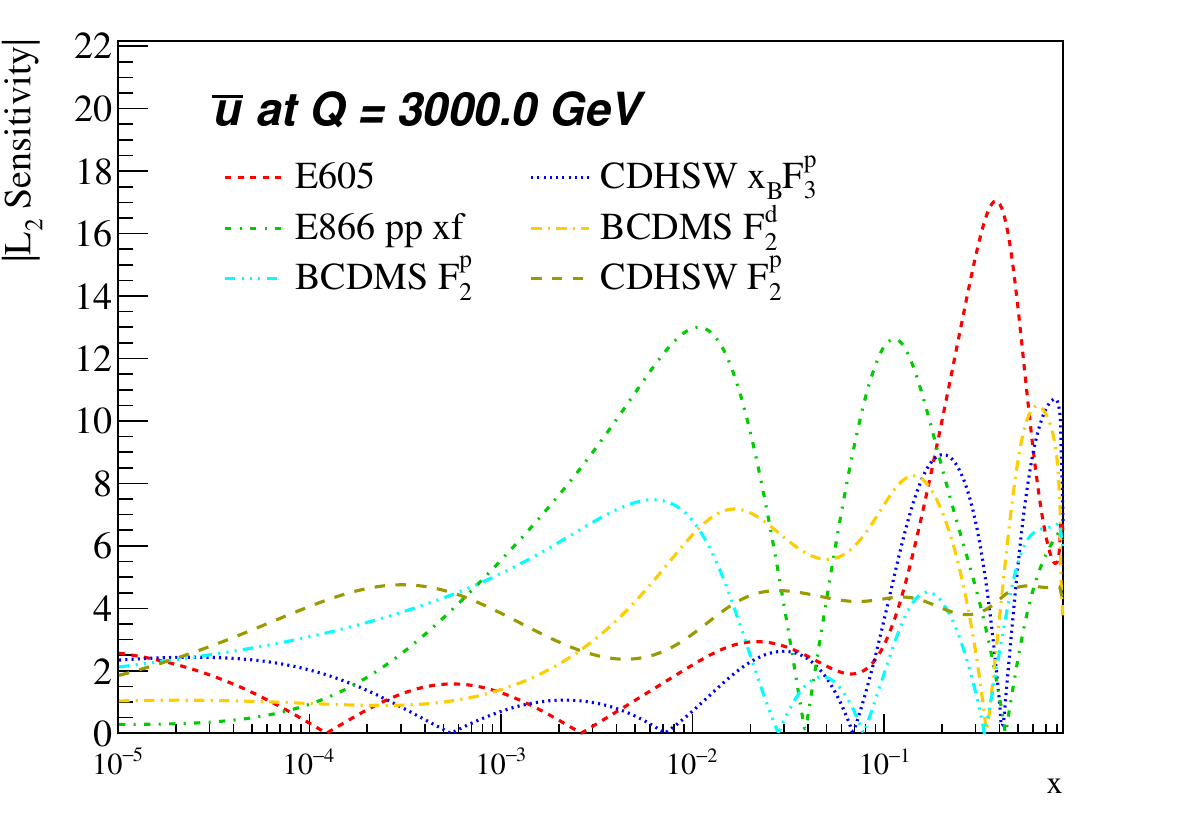}
  }
  \caption{The various experimental inputs and their relative importance in the overall fits for (a) $u_v$  and (b) $\bar{u}$ are shown in terms of the sensitivity quantity $|L_2|$. See the text for explanation.}
\end{figure}
%

%
%
\begin{figure}[!th]
  \centering
  \subfigure[]{
    \label{region_xQ}
    \includegraphics[width=0.48\textwidth]{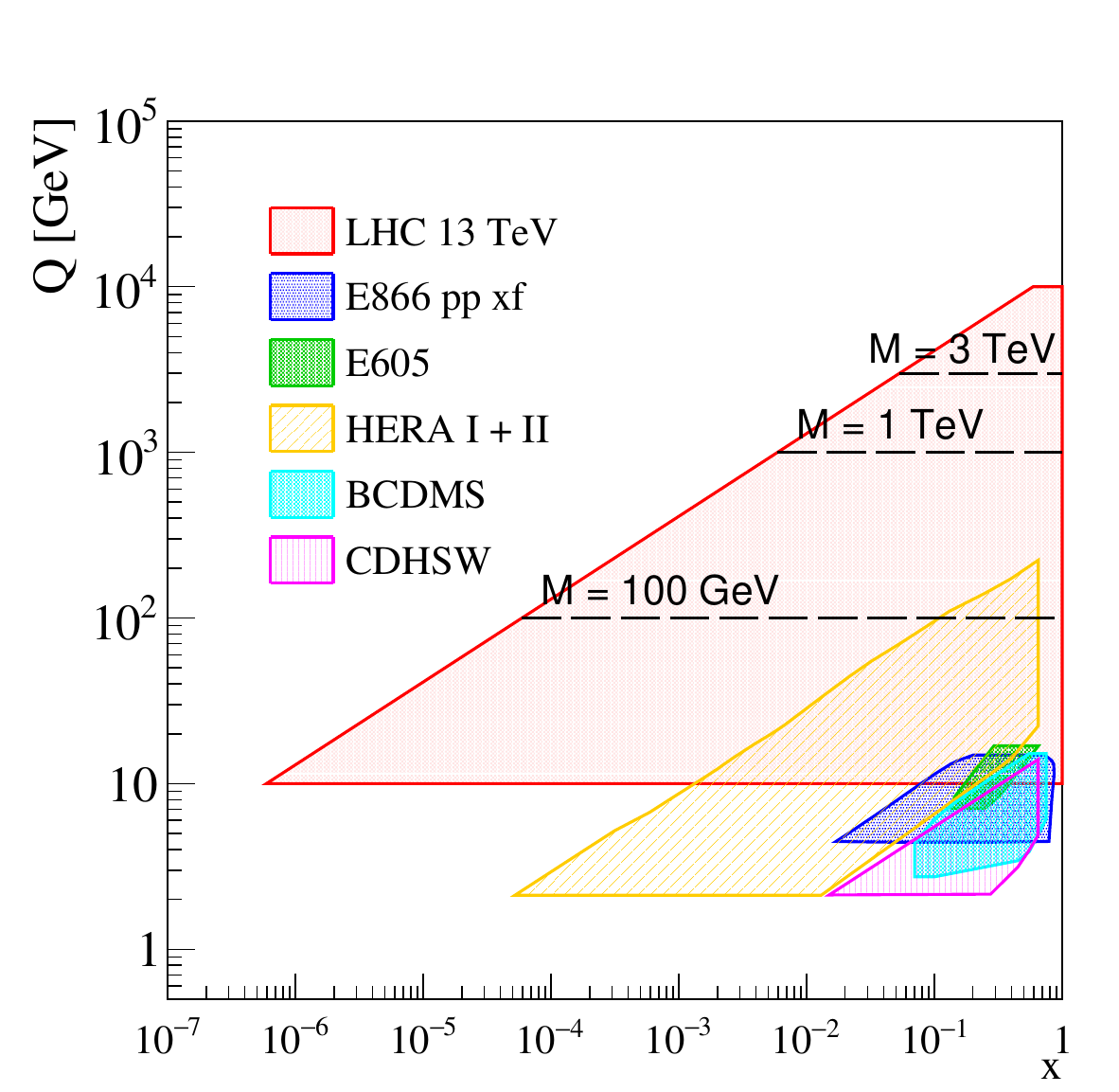}
  }
  \subfigure[]{
    \label{dileptonmass18}
    \includegraphics[width=0.48\textwidth]{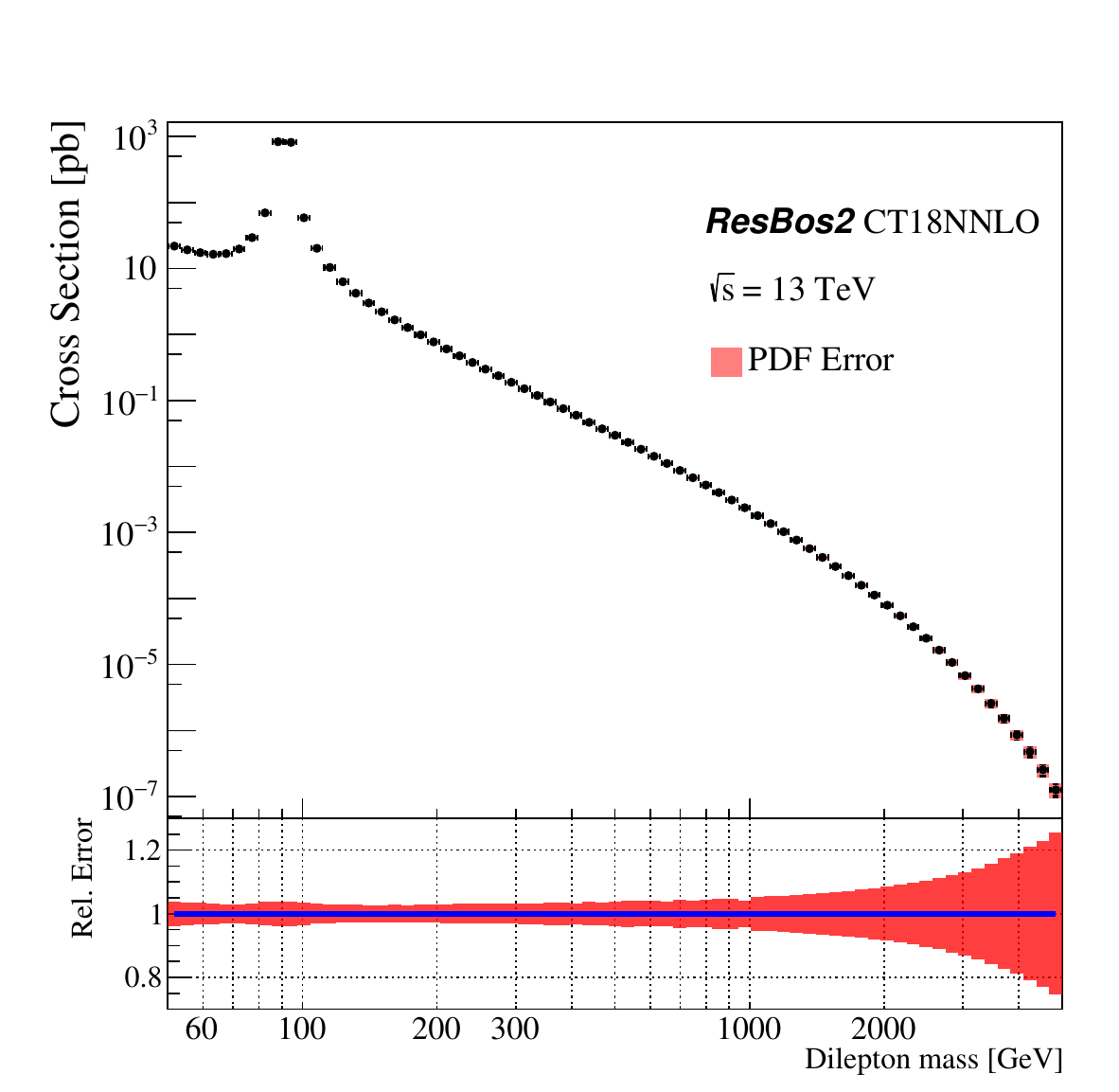}
  }
  \caption{(a) The regions of $Q$ and $x$ covered by the many of the inputs to \texttt{CT18} PDF fitting. The LHC 13 TeV region is not an input, but traces out the kinematical range of LHC collisions. (b) From \chrispunctuation, the $m_{\ell\ell}$  distribution for NCDY events as simulated in {\sc ResBos}, highlighting the current uncertainty due to PDFs using the \texttt{CT18} global fits.}
  \label{LABEL}
\end{figure}

This region of $x$ is hard to characterize precisely as it begins to probe regions of sea and even valence quark momentum fractions which are not well constrained by mostly DIS inputs. Figure~\ref{fig:uvl3000R} shows (a) the $u$ and (b) $\bar{u}$ momentum fractions from the \texttt{CT18} PDF set evaluated at two scales (a) $Q = 2$~GeV and (b) $Q = 3$~TeV. The parton distribution shapes are dramatically altered by evolving from the region of inputs to the fits to the scales of BSM IVB searches. The uncertainties on those parton distributions from the \texttt{CT18} fits are shown in (c) for $u_V$ and (d) $\bar{u}$.  The CTEQ-TEA group \cite{Hou_2021} has quantified the relative contributions of data inputs to a measure of ``$L_2$ sensitivity'' \cite{Wang:2018lut,Wang:2018lut,Hobbs:2019gob,Jing:2023isu} and Figures~\ref{L2Sensitivity_uv} and \ref{L2Sensitivity_ubar} show this quantity for the most relevant 15 data inputs. For this parameter, the higher the value of $L_2$, the more sensitive these PDFs are to each experiment. ($L_2$ is actually a signed quantity with positive and negative values, which encode additional features of the data contributions. But for our purposes, we've folded them into absolute value to make a general point about individual relative contributions.) Notice that in the region of $x\sim 0.2-0.3$ that the most important experiments for the $u$ and $\bar{u}$ fits are:
\begin{itemize}
\item $u$: $F_2$ from BCDMS, E605, and HERA I and II.
\item $\bar{u}$: E605 and $F_3$ from CDHSW.
\end{itemize}
LHC experimental data have not contributed meaningfully to the $x$ region of interest so far, and the determination of NC and CC backgrounds required of BSM searches currently depends on classic DIS and DY experiments from the 1980s and 1990s. Figure~\ref{region_xQ} shows the region of $Q$ and $x$ covered by the data inputs to the \texttt{CT18} fits, which are relevant to DY kinematics at the LHC. Extrapolation from the scale of global fit inputs to the region of interest is more than three orders of magnitude in $Q$.

Figure~\ref{dileptonmass18} shows the iconic invariant mass distribution of dilepton pairs calculated using the {\sc ResBos}~\cite{PhysRevD.50.R4239,PhysRevD.56.5558,PhysRevD.67.073016} Monte Carlo (MC) generator and the \texttt{CT18} PDFs.  The ratio band is the quoted CTEQ-TEA~\cite{ref:CT18}  PDF uncertainties of about 20\% at $m_{\ell\ell} = 4$~TeV. 

The questions for us are: can this lack of precision in the PDFs and, consequently, the uncertainties in BSM IVB searches be improved?

\section{CC PDF Reduction}
The differential cross section for the production of $W$ bosons in $p$-$p$ collisions is limited to rapidity of the outgoing charged lepton, $y_\ell$ and the transverse mass $$m_T^W \;=\; \sqrt{\,2\,p_T^{\ell}\,p_T^{\nu}\,\bigl(1 - \cos\Delta\phi(\ell,\nu))}.$$ Here $p_T^{\ell}$ is the transverse charged lepton momentum; $\,p_T^{\nu}$ is that of the outgoing neutrino, or in practice, the missing transverse momentum; and  $\Delta\phi(\ell,\nu)$ is the azimuthal angle between the charged lepton and the neutrino. 
%
%
\begin{equation}
\label{Wcross}
  \frac{d^{2}\sigma^{W^\pm}}{dm_T\,dy_\ell}
  \; =\;
  \frac{\pi\,G_F\,M_W^{2}}{3\sqrt{2}\,s}\;
  \frac{m_T}{M_W^{2}\sqrt{1-\tfrac{m_T^{2}}{M_W^{2}}}}\;
  | V_{q\bar{q}}|^2 \nonumber\\
  \Big[
    f_{q/P_1}\!\big(x_1,Q^{2}\big)\,
    f_{\bar q'/P_2}\!\big(x_2,Q^{2}\big)
    + (q\leftrightarrow\bar q',\,P_1\leftrightarrow P_2)
  \Big]
\end{equation}
The quantities inside the brackets have the same interpretation as that of Eq.~\ref{Zcross} and $V_{q\bar{q}}$ is the quark mixing matrix parameter set.
For $W$ boson production, only a double differential cross section in $dm_T\,dy_\ell$  is measurable due to the lack of full momentum information for the outgoing neutrino. 

\begin{figure}
    \centering
    \includegraphics[width=0.5\linewidth]{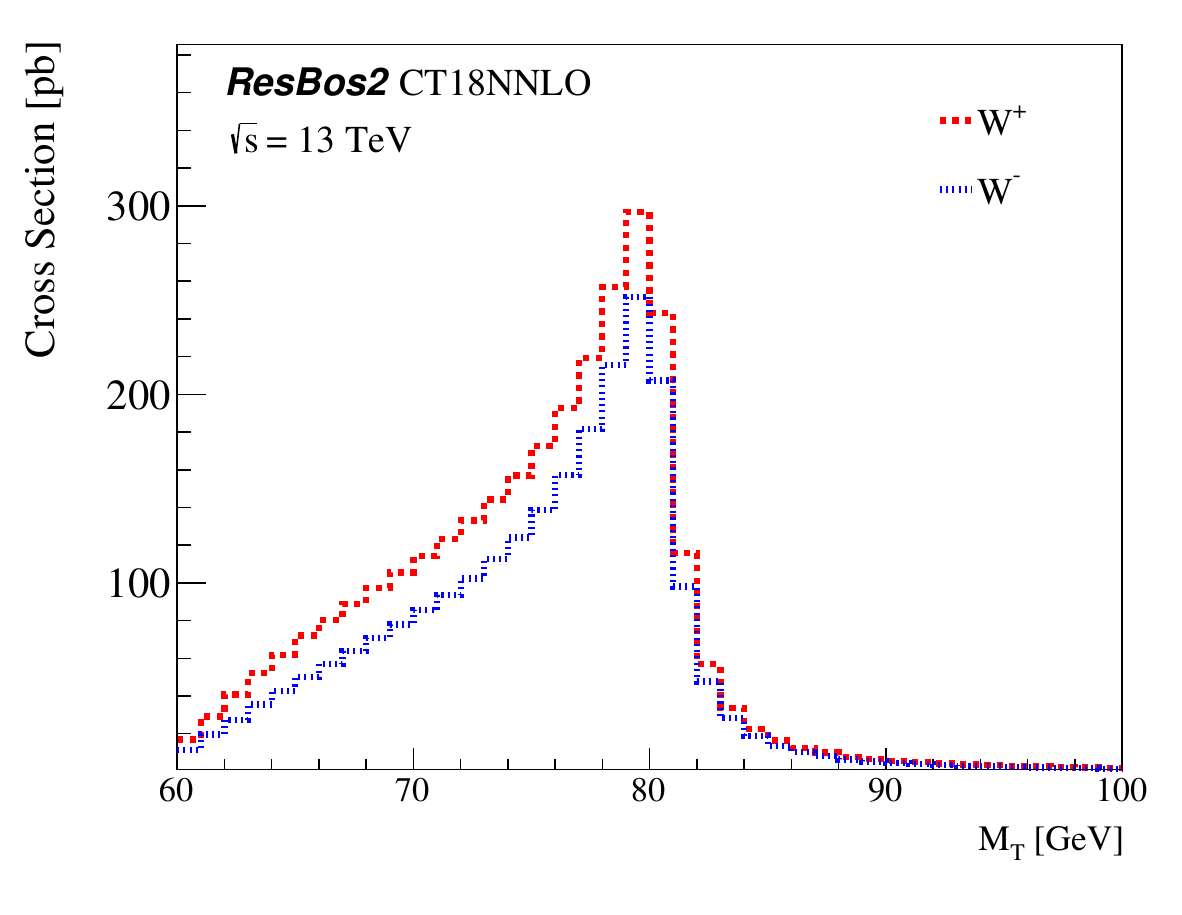}
    \caption{The transverse mass of the two signs of $W$ bosons as generated from \texttt{CT18} PDFs and {\sc ResBos}.}
    \label{fig:transversmass}
\end{figure}

The production process is still the annihilation of a valence $u$($d$) quark with, now a $\bar{d}$($\bar{u}$) quark. And the same lack of precision experienced by the NC BSM IVB searches plagues the CC limits. Figure~\ref{fig:transversmass} shows the distribution of transverse mass from CCDY events calculated using the {\sc ResBos2} program and the \texttt{CT18} PDFs. 
The uncertainty due to PDFs for this quantity is approximately 20\% at $m_T = 4$~TeV. This measurement is more sensitive to experimental uncertainties than for NCDY measurements.

We ask the same question as above: is there a way to reduce the PDF uncertainty in BSM CCDY searches? And, in addition, is there a global strategy to improve both BSM NCDY and CCDY background uncertainties?


\section{A  Strategy For PDF Uncertainty Reduction}
In \chris we found a significant reduction in the PDF uncertainties relevant for high-mass, BSM NCDY  searches. The hint was intriguing power in the Collins-Soper angle, $\cos{\theta^*}$, and the result was a suggested plan to reduce the PDF uncertainties specifically in backgrounds for high mass BSM IVB searches using a new set of inputs to PDF fitting, namely the triple differential cross section of Eq.~\ref{Zcross}. We addressed two questions in that paper:

\begin{enumerate}
\item If  $\cos{\theta^*}$ data were incorporated in PDF fitting, how significantly might the reduction in PDF uncertainties be?
\item Would these decreased errors be a signficant reduction in the overall theoretical uncertainties in future BSM, high-mass DY searches?
\end{enumerate}
We concluded that the answers were significant reductions in both. In this paper we extend this strategy to include new features.

\subsection{Review and Update of Neutral Current DY Strategy from 2019}
\begin{figure}[!t]
  \centering
  \includegraphics[width=\textwidth]{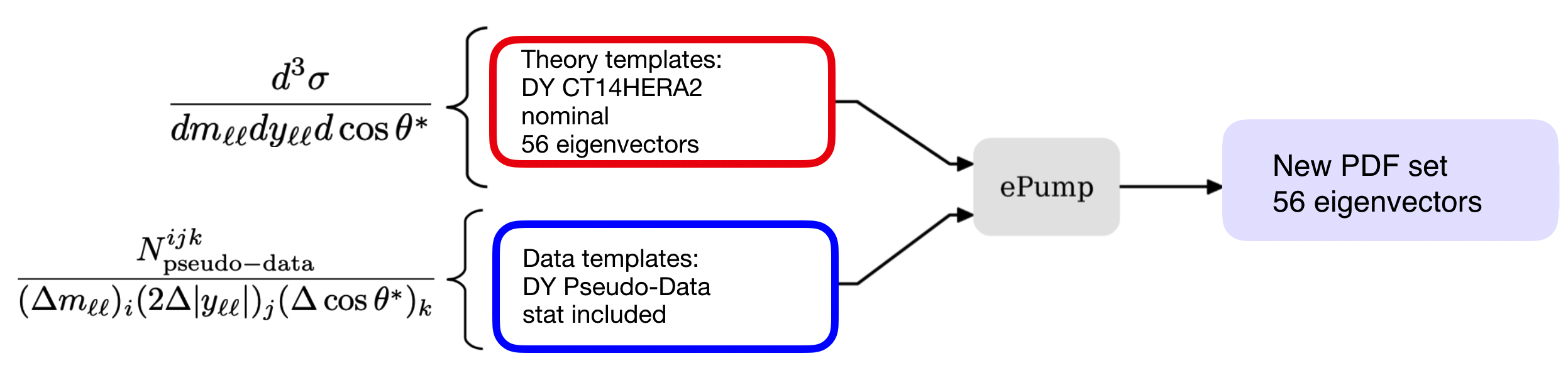}
  \caption{The \texttt{ePump} package requires two inputs to generate an updated PDF set: an existing Theory template of a PDF set %
    (parameters + uncertainties) and binned Data template of (pseudo-) data, including statistical uncertainties from integrated luminosity assumptions.}
  \label{fig:ePumpFlow}
\end{figure}

Here we repeat our results of \chris, but updating them to CT18, use \resbos  and, significantly, we add CC inputs to our strategy as outlined in Fig.~\ref{fig:ePumpFlow}. Overall, the strategy is the same as  before: we create pseudodata inputs for a particular integrated luminosity as templates in kinematical quantities in one region of the DY spectrum below a boundary in mass, incorporate them in a fit for new PDF sets using \texttt{ePump}~\cite{Schmidt:2018hvu,Hou:2019gfw} along with the \texttt{CT18} templates. The luminosities we use assume contributions from both ATLAS and CMS, so 6000~fb$^{-1}$ for High Luminosity LHC (HL-LHC) running and 600~fb$^{-1}$ for LHC Run 2 plus partial Run 3 data. This results in a new,  ``boutique'' PDF sets characterizing background predictions tailored for the BSM IVB high mass searches. We choose 1~TeV as that boundary for the NC channel, as  sufficiently below the rejected mass range for NCDY as to not be contaminated  by new physics. The CC channel we describe below. The NCDY fiducial region considered for our analysis is designed explicitly to probe the PDFs at high $x$, and is defined by
\begin{equation} \label{eq:fiducial}  
40\, \mathrm{GeV} < m_{\ell\ell} < 1000 \, \mathrm{GeV}, \quad |y_{\ell\ell}|<3.6, \quad -1<\cos\theta^{*}<1.
\end{equation}
In \chris DY samples were generated using the {\sc ResBos} MC generator with the \texttt{CT14HERA2} PDF set for the $\sqrt{s}=13$~TeV LHC. Events are further required to pass a loose event selection in order to construct the finalized data templates. Dilepton events with an invariant mass of $m_{\ell\ell}>80$~GeV were required to satisfy $p_{T}^\ell>30$~GeV, while low-mass events in the interval of $40<m_{\ell\ell}<80$ must satisfy $p_{T}^\ell>15$~GeV.
In addition, events must consist of leptons which are distributed as central-central or central-forward. 

Events passing these selections were binned in \texttt{ePump} template histograms, which parametrize the triple-differential cross section of Eq.~(\ref{Zcross}), according to 
\begin{equation} 
\label{eq:sigmaBins}
  \mathcal{L}_{\mathrm{int}} \left( \frac{d^{3}\sigma}{dm_{\ell\ell}d|y_{\ell\ell}|d\cos\theta^{*}} \right)_{ijk} %
  = 
  \frac{N^{ijk}_{\mathrm{pseudo-data}}}{(\Delta m_{\ell\ell})_{i} (2\Delta |y_{\ell\ell}|)_{j} (\Delta\cos\theta^{*})_{k}},
\end{equation}

\noindent where $i$, $j$, and $k$ correspond to the bin indices of each distribution of interest.  Note that in a realistic measurement, the numerator of Eq.~(\ref{eq:sigmaBins}) would be replaced by $N^{ijk}_{\mathrm{data}}-N^{ijk}_{\mathrm{bkg}}$, where the background component arises from the standard dilepton background processes.

The total number of pseudo-data events are given by $N^{ijk}_{\mathrm{pseudo-data}}$, the integrated luminosity of the pseudo-dataset is $\mathcal{L}_{\mathrm{int}}$, and $(\Delta m_{\ell\ell})_{i}$, $(2\Delta |y_{\ell\ell}|)_{j}$, and $(\Delta \cos\theta^{*})_{k}$ are the corresponding bin widths. The factor of two in the denominator accounts for the modulus in the rapidity bin width. The bins used to parametrize Eq.~(\ref{eq:sigmaBins}) include 12 bins for $40<m_{\ell\ell}<1000$~GeV, 18  bins for $0<|y_{\ell\ell}|<2.4$, and 6 bins for $-1<\cos\theta^{*}<1$.

\begin{figure}[!t]
  \centering
  \subfigure[]{
    \label{fig:cos_40_66}
    \includegraphics[width=0.43\textwidth]{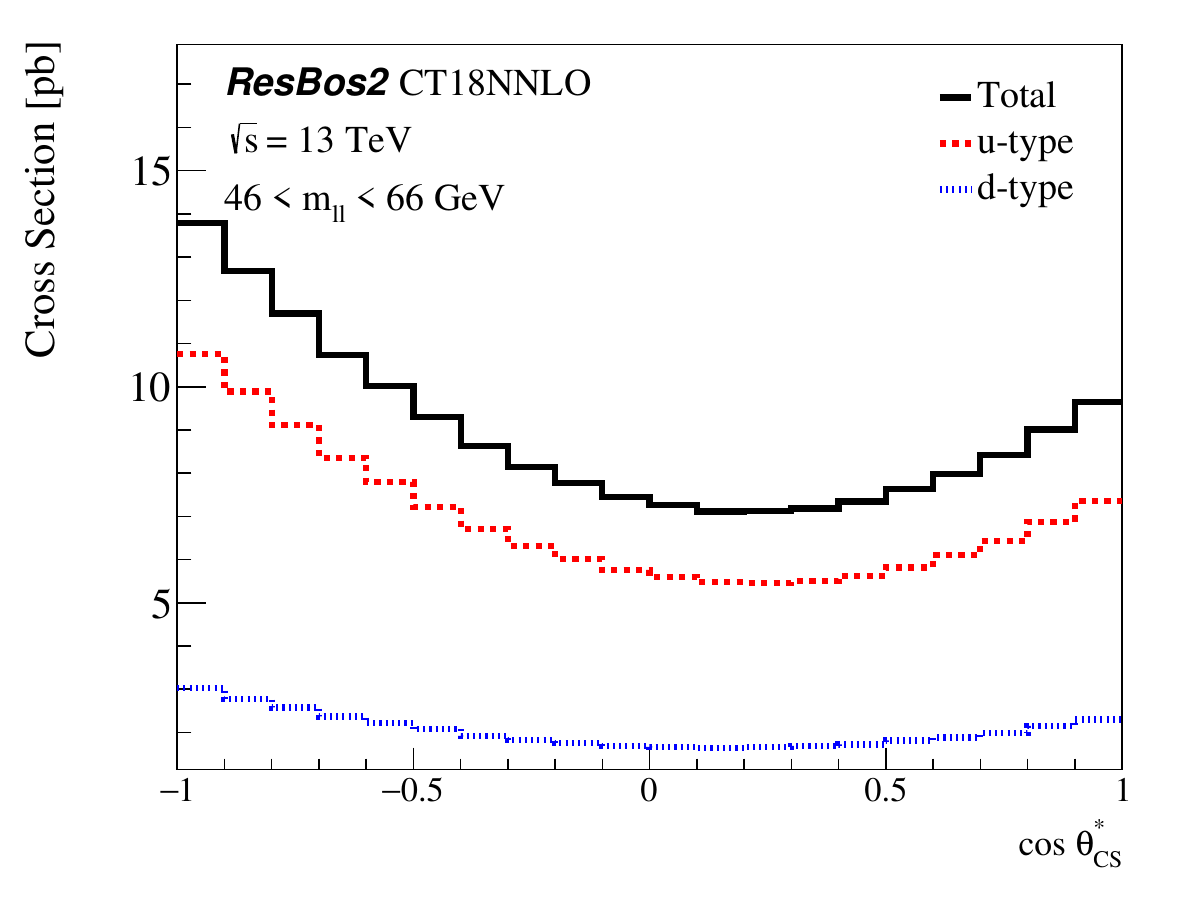}
  }
  \subfigure[]{
    \label{fig:cos_66_116}
    \includegraphics[width=0.43\textwidth]{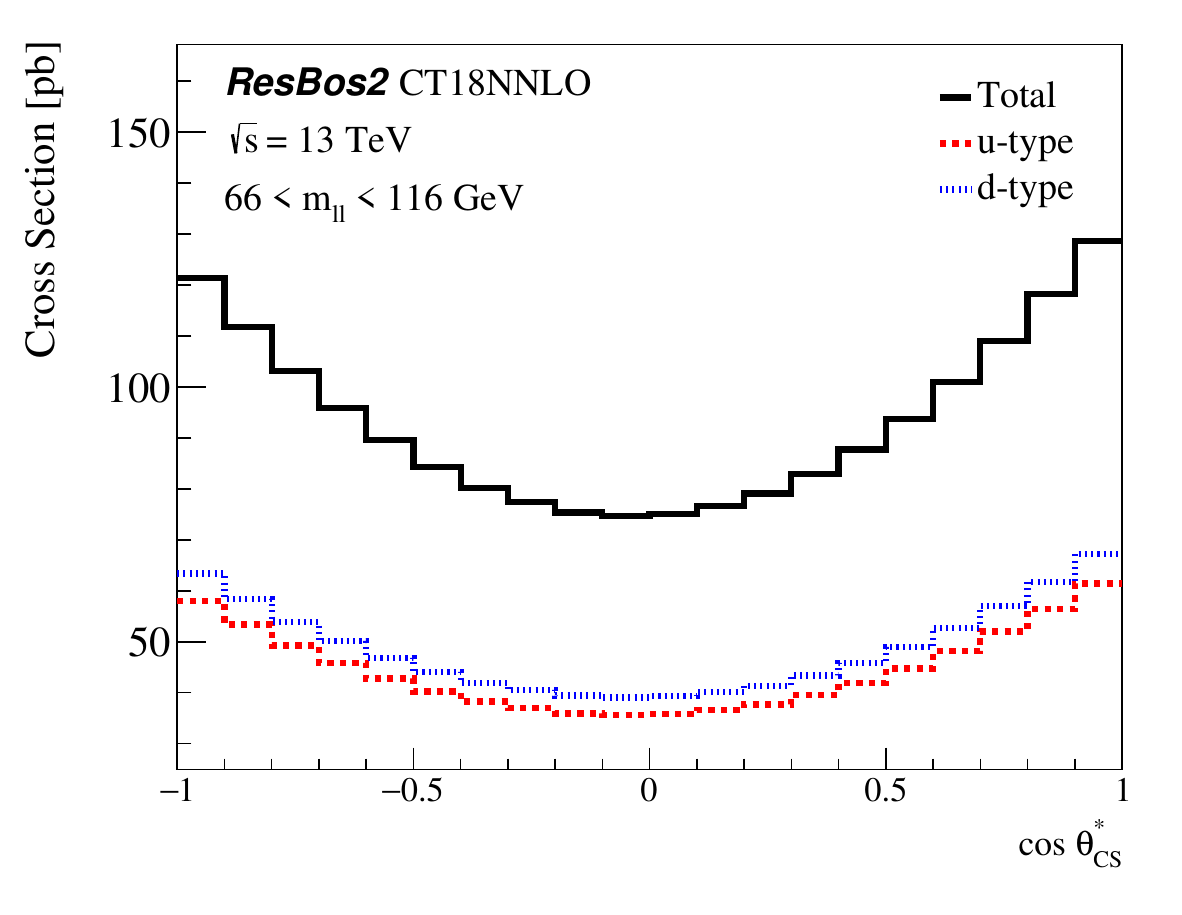}
  }\\
  \subfigure[]{
    \label{fig:cos_116_250}
    \includegraphics[width=0.43\textwidth]{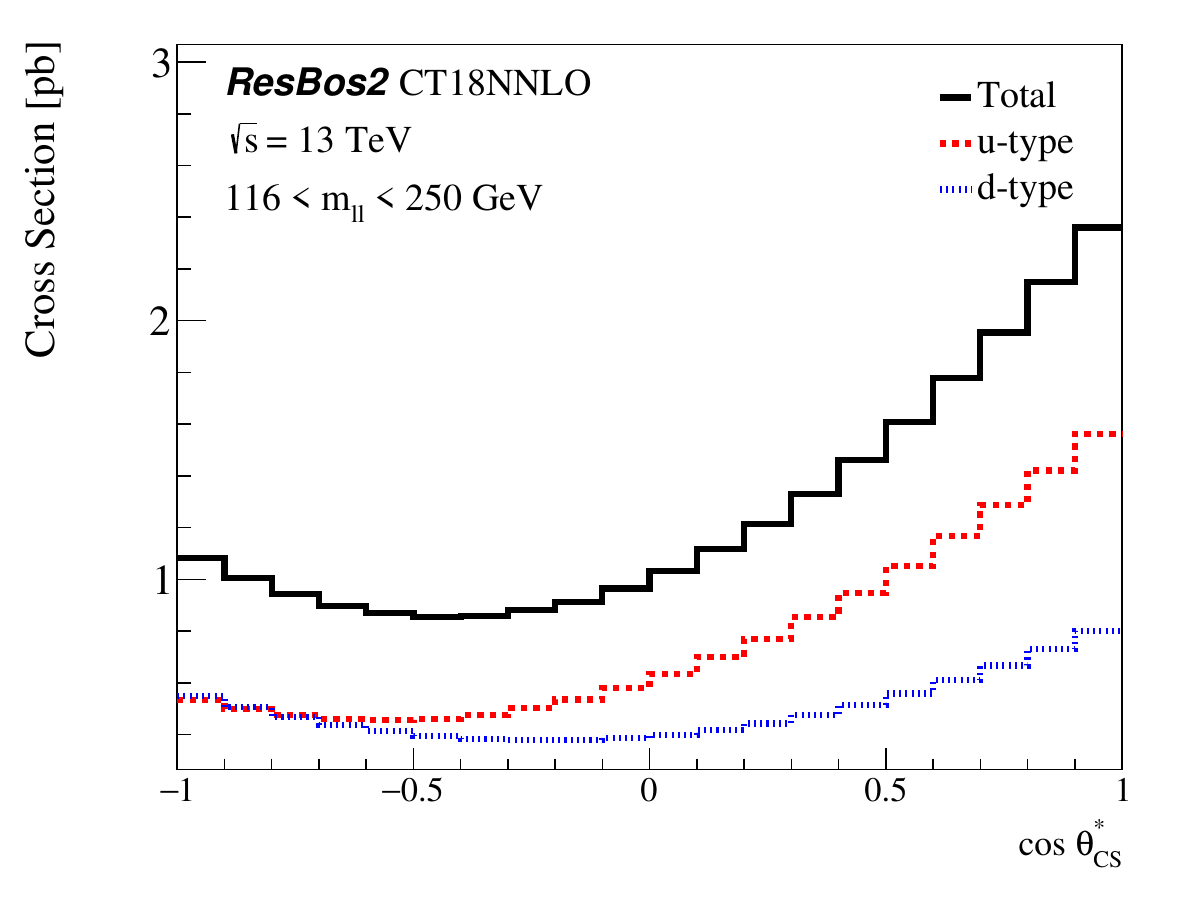}
  }
  \subfigure[]{
    \label{fig:cos_250_400}
    \includegraphics[width=0.43\textwidth]{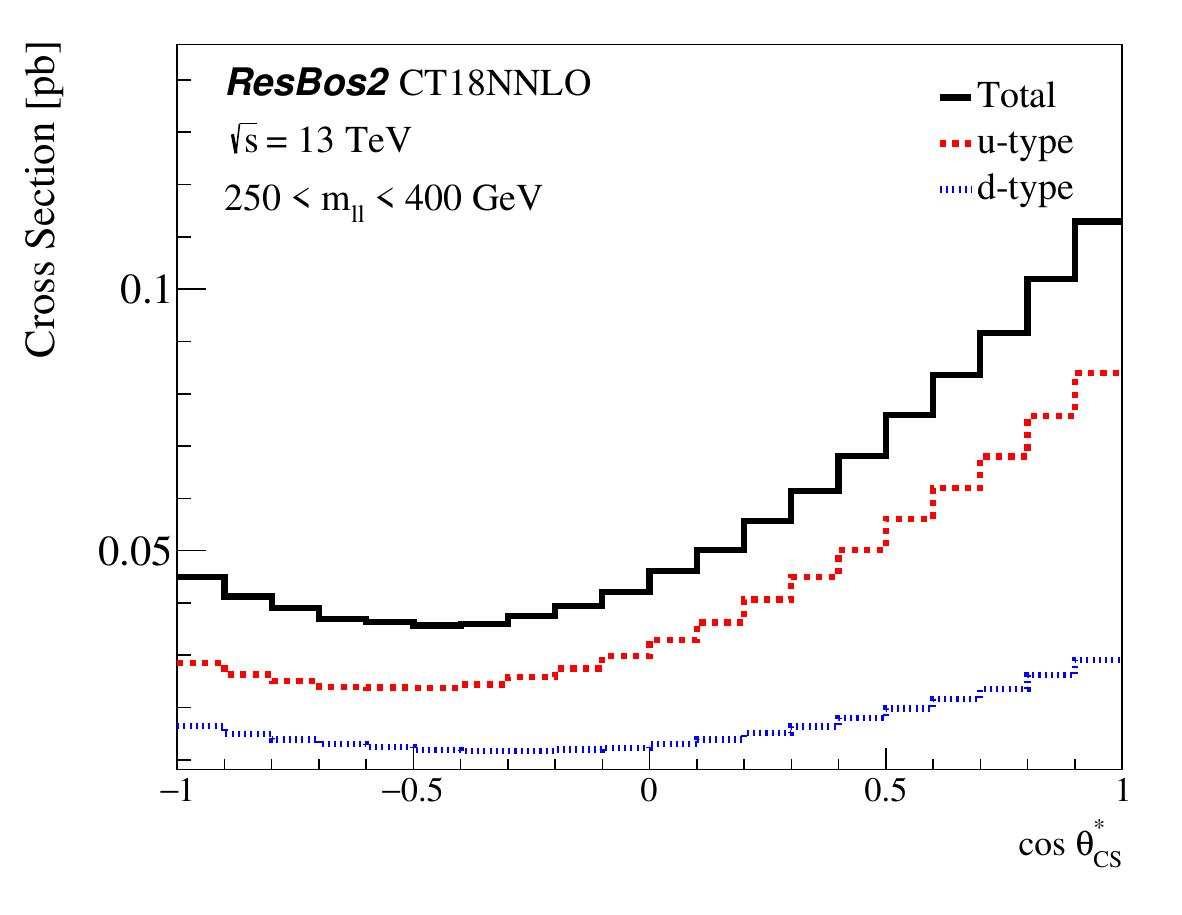}
  }\\
  \subfigure[]{
    \label{fig:cos_400_600}
    \includegraphics[width=0.43\textwidth]{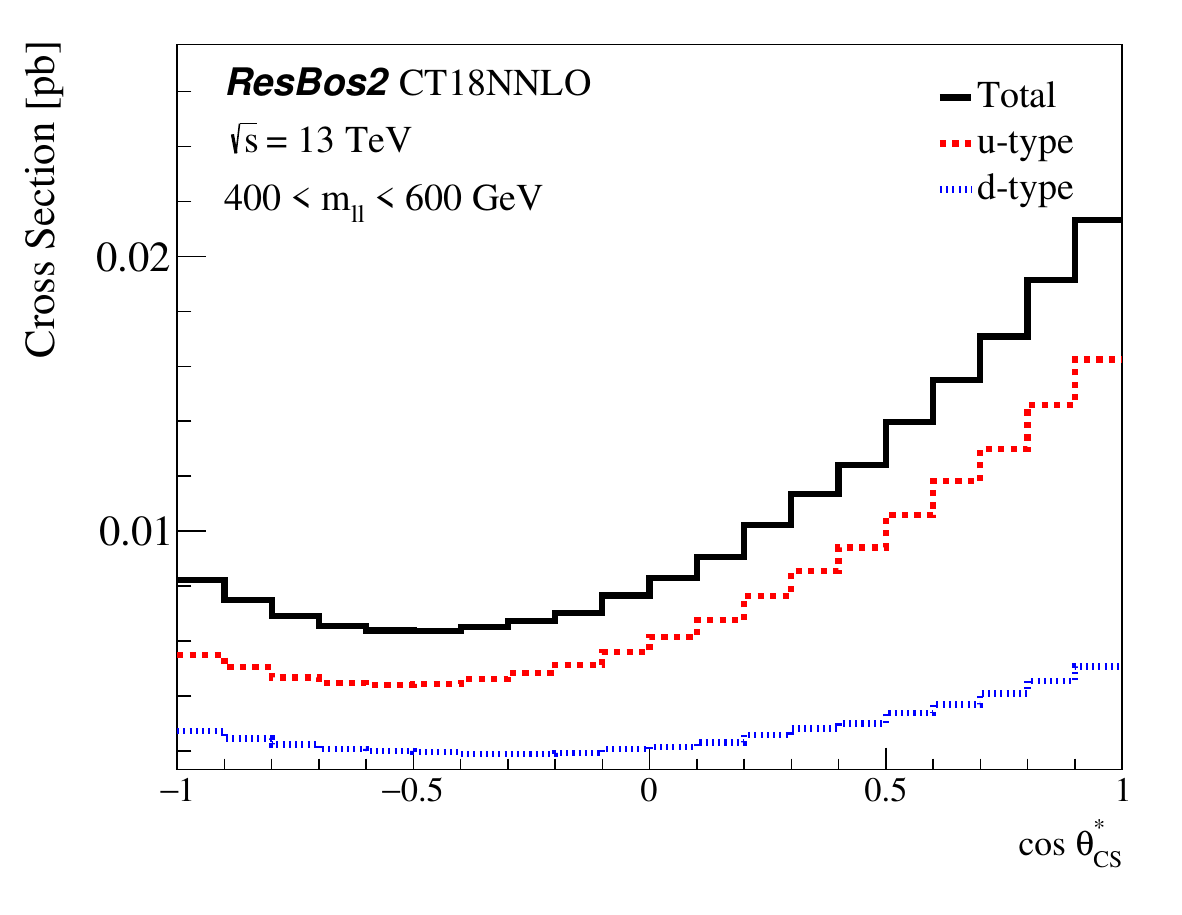}
  }
  \subfigure[]{
    \label{fig:cos_600_1000}
    \includegraphics[width=0.43\textwidth]{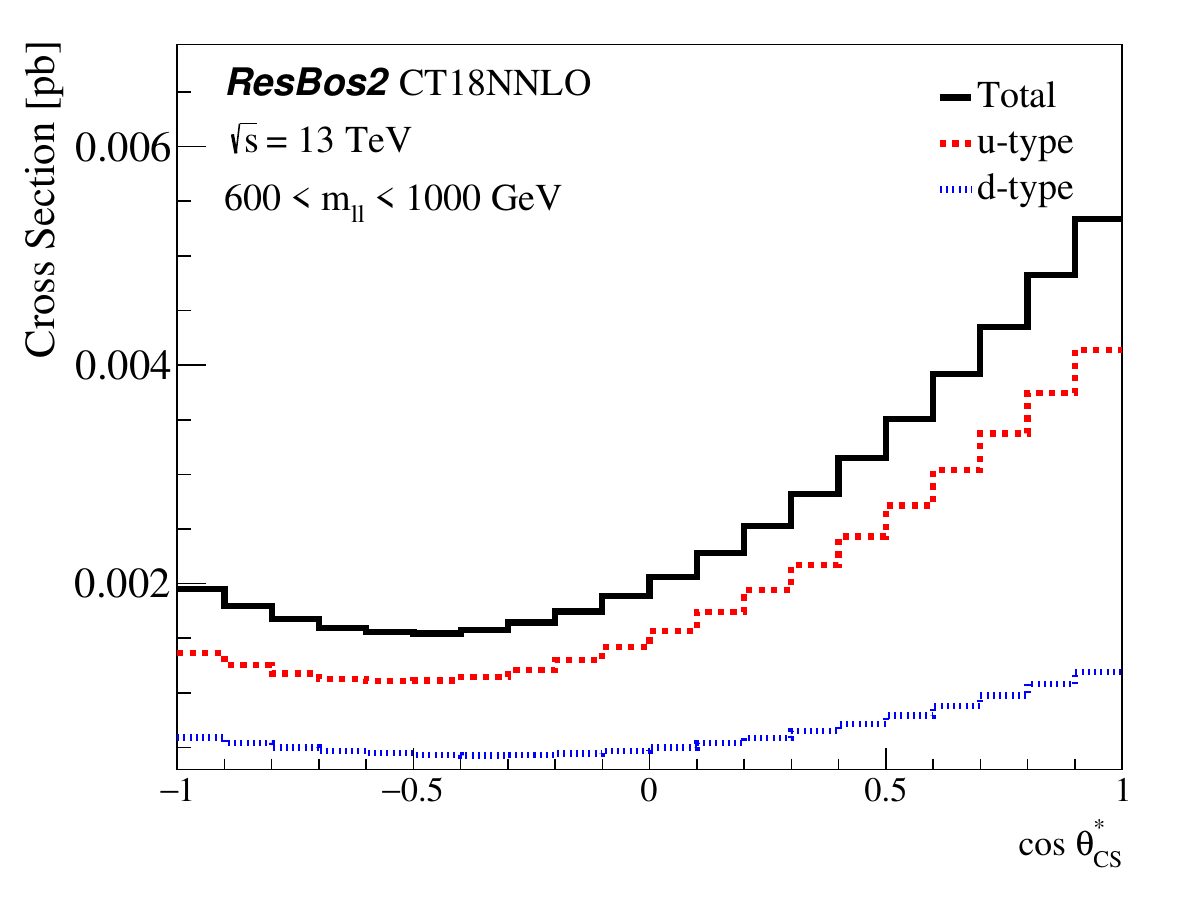}
  }
  \caption{The lepton angular distribution $\cos\theta^{*}$ in slices of dilepton invariant mass, $m_{\ell\ell}$, ranging %
    from 40~GeV to 1~TeV~\subref{fig:cos_40_66}-\subref{fig:cos_600_1000}. The up-type and down-type DY sub-processes %
    are shown as well, which exhibit a strong angular dependence, especially at high mass. The \texttt{CT18} PDF set is used.}
  \label{costheta18}
\end{figure}

The total number of measurement bins is $N_{\mathrm{bins}} = 12 \times 18 \times 6 = 1296$ for the fiducial region considered and they define the $N_{\mathrm{new}}$ data points. We saw significant improvement in a predicted background uncertainty due to PDFs and showed that improvements were due in part to the separation of the up and down quark contributions to the cross section. As the mass of the input DY data approached that 1~TeV threshold for as masses of DY leptons passed 250~GeV, the $\cos{\theta^*}$ contributions were increasingly from up quark (and hence $\bar{u}$ quarks). We noted at the time that. ``At high mass and high polar angle, the LHC DY process proceeds almost entirely through the $u\bar{u}$ sub-process, effectively making the LHC a $u\bar{u}$ collider.'' This $u$ to $d$ dominance is about a factor of four. This progression is shown in Figure~\ref{costheta18} for \cteqpunctuation.

In this paper we update the 2019 NCDY results using modern \texttt{CT18} global fits and expand on this strategy to include (1) differential CCDY data inputs to the new PDF fitting and (2) to use these new PDFs to model SM backgrounds in the search region for both CCDY as well as NCDY BSM searches.

We have redone the previous results with the triple differential cross section templates, replacing the \hera  global set with those from \texttt{CT18} in the \texttt{ePump} fits. Using our strategy for creating tailored PDF fits using the above strategy resulted in a reduction of the uncertainty in background due to PDFs from 31\% to 8.9\%. By using the \texttt{CT18} fits as a baseline, our strategy results in a reduction of PDF uncertainty in the background from 25\% to 8.3\%. The \texttt{CT18} base results were more precise (31\% versus 25\%) than the \herapunctuation, but our strategy still leads to a significant reduction of more than a factor of 3 over the \texttt{CT18} global fits alone. 
\begin{figure}[!th]
  \centering
  \subfigure[]{
    \label{LABEL1}
    \includegraphics[width=0.48\textwidth]{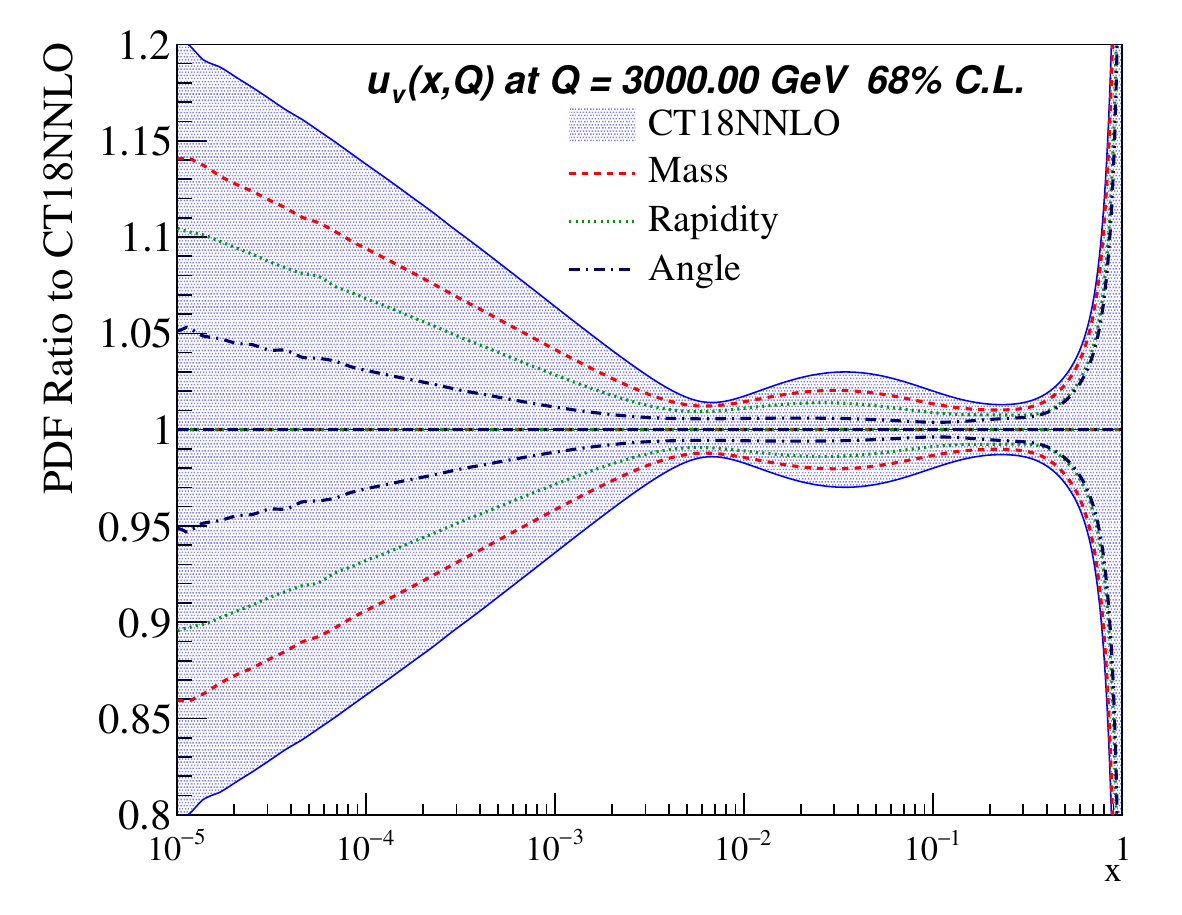}
  }
  \subfigure[]{
    \label{LABEL2}
    \includegraphics[width=0.48\textwidth]{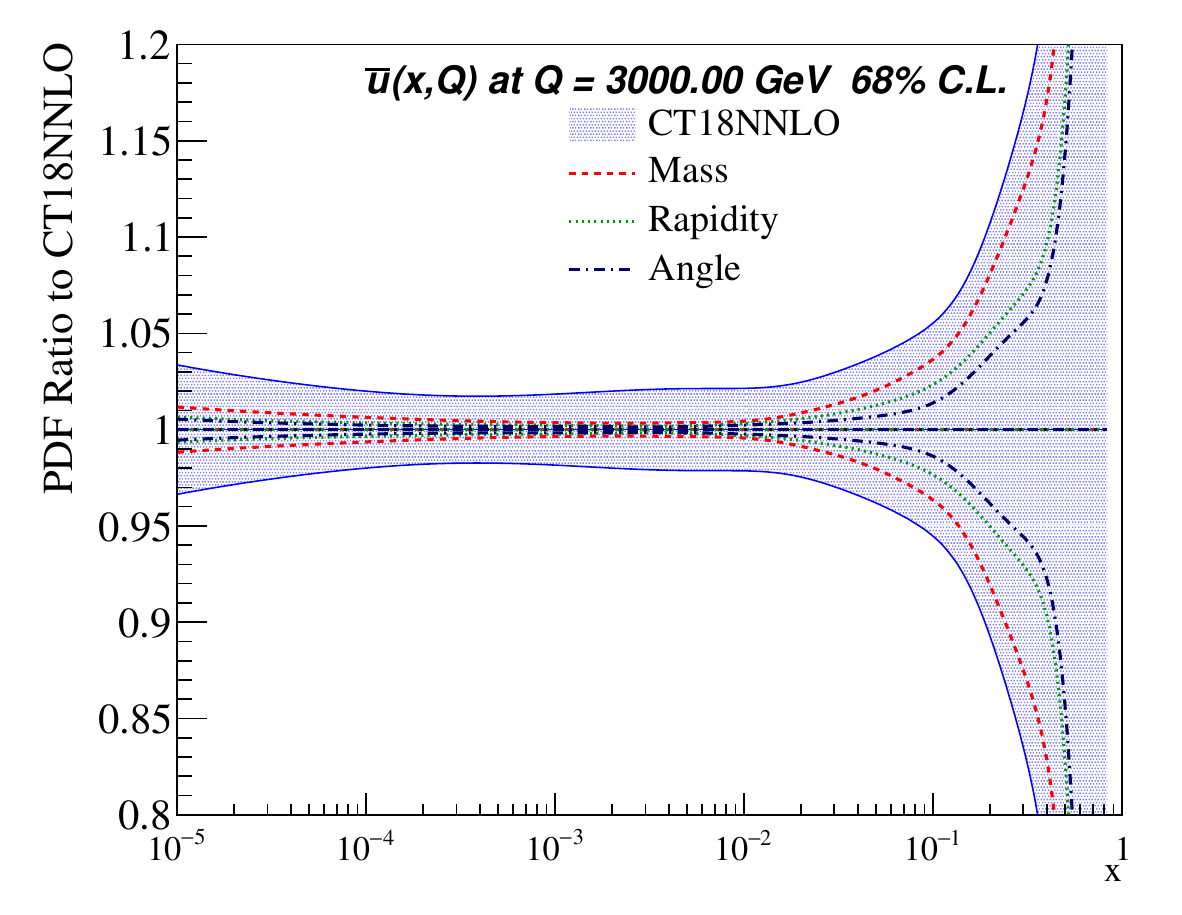}
  }
    \subfigure[]{
    \label{LABEL2}
    \includegraphics[width=0.48\textwidth]{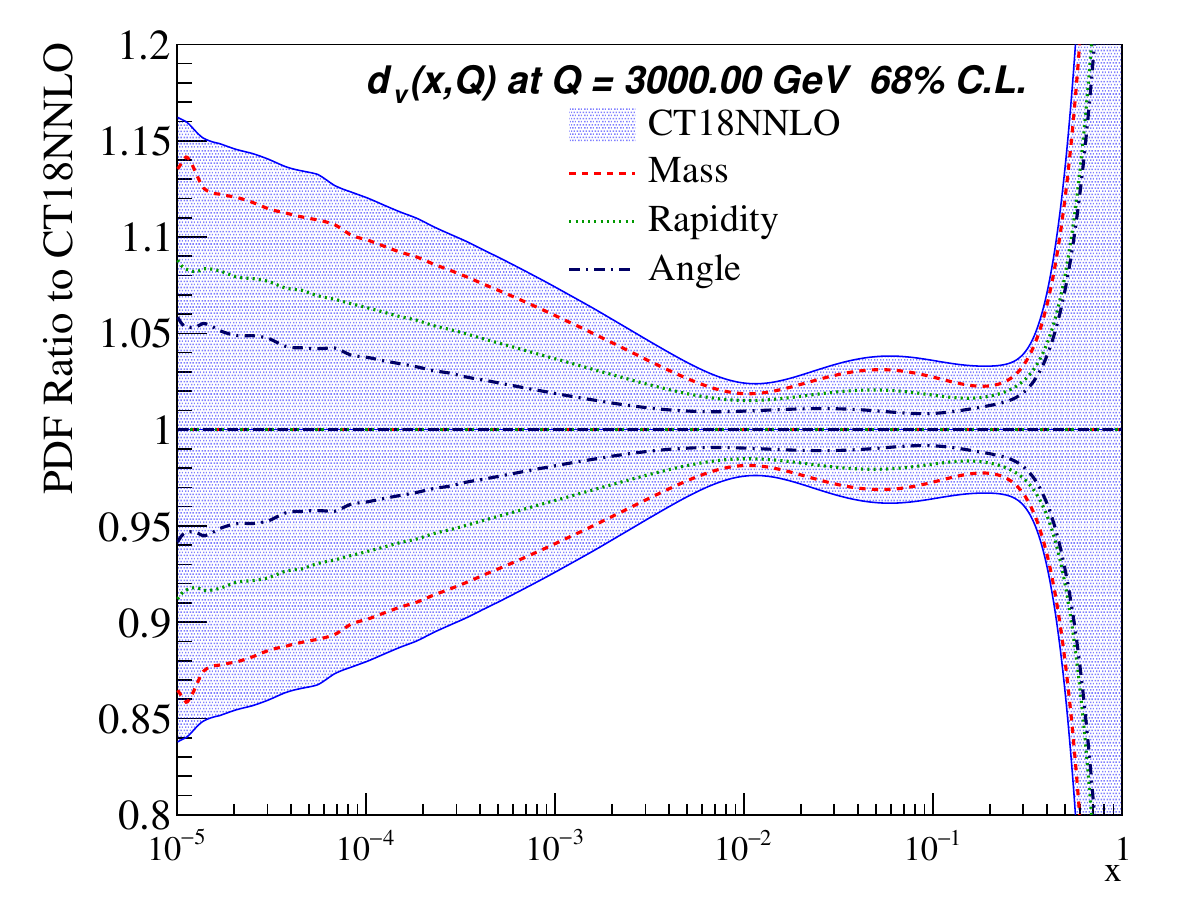}
  }
    \subfigure[]{
    \label{LABEL2}
    \includegraphics[width=0.48\textwidth]{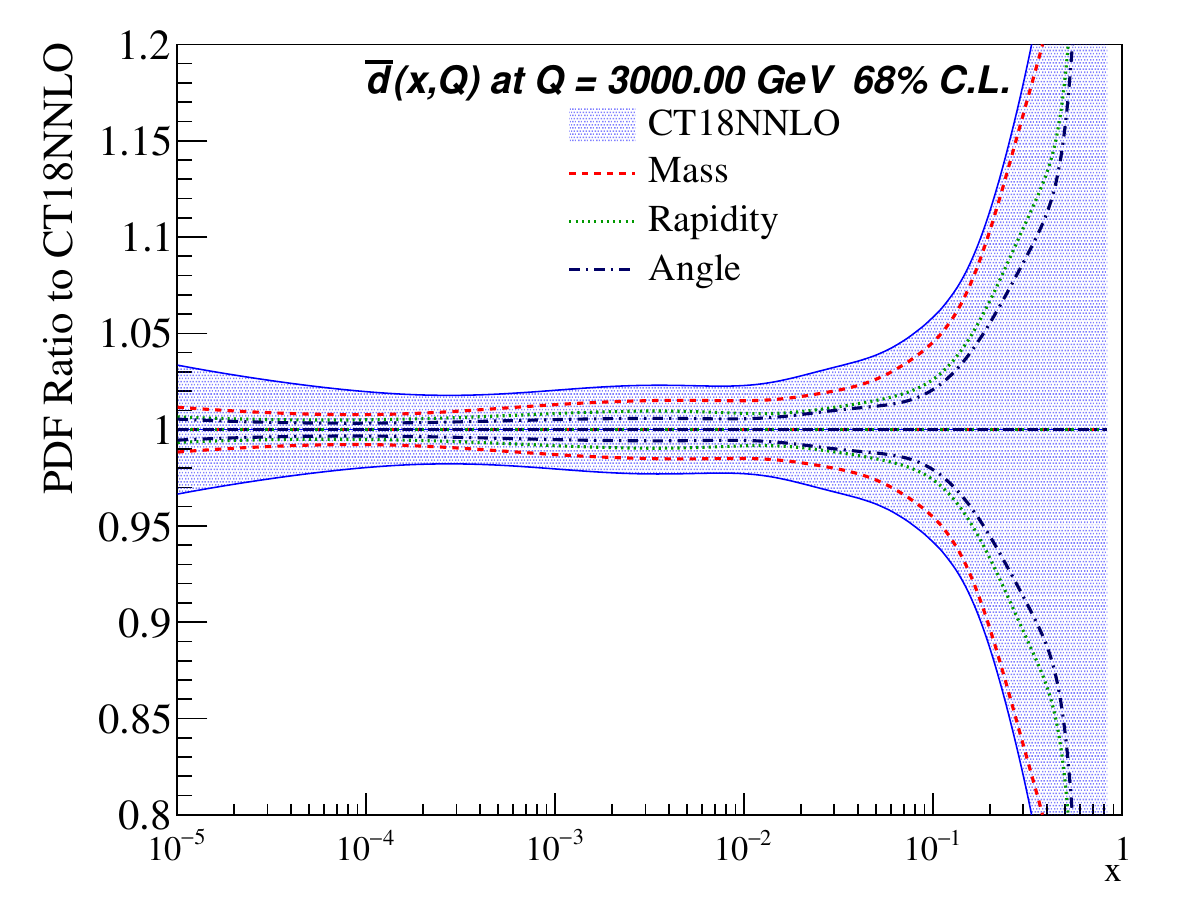}
  }
  \caption{Relevant parton distribution improvements for the NCDY project, successively adding each of the three selections of variables sent to \texttt{ePump}.}
  \label{partonsNC18}
\end{figure}

\begin{figure}[!th]
  \centering
  \subfigure[]{
    \label{M_NC_3000}
    \includegraphics[width=0.48\textwidth]{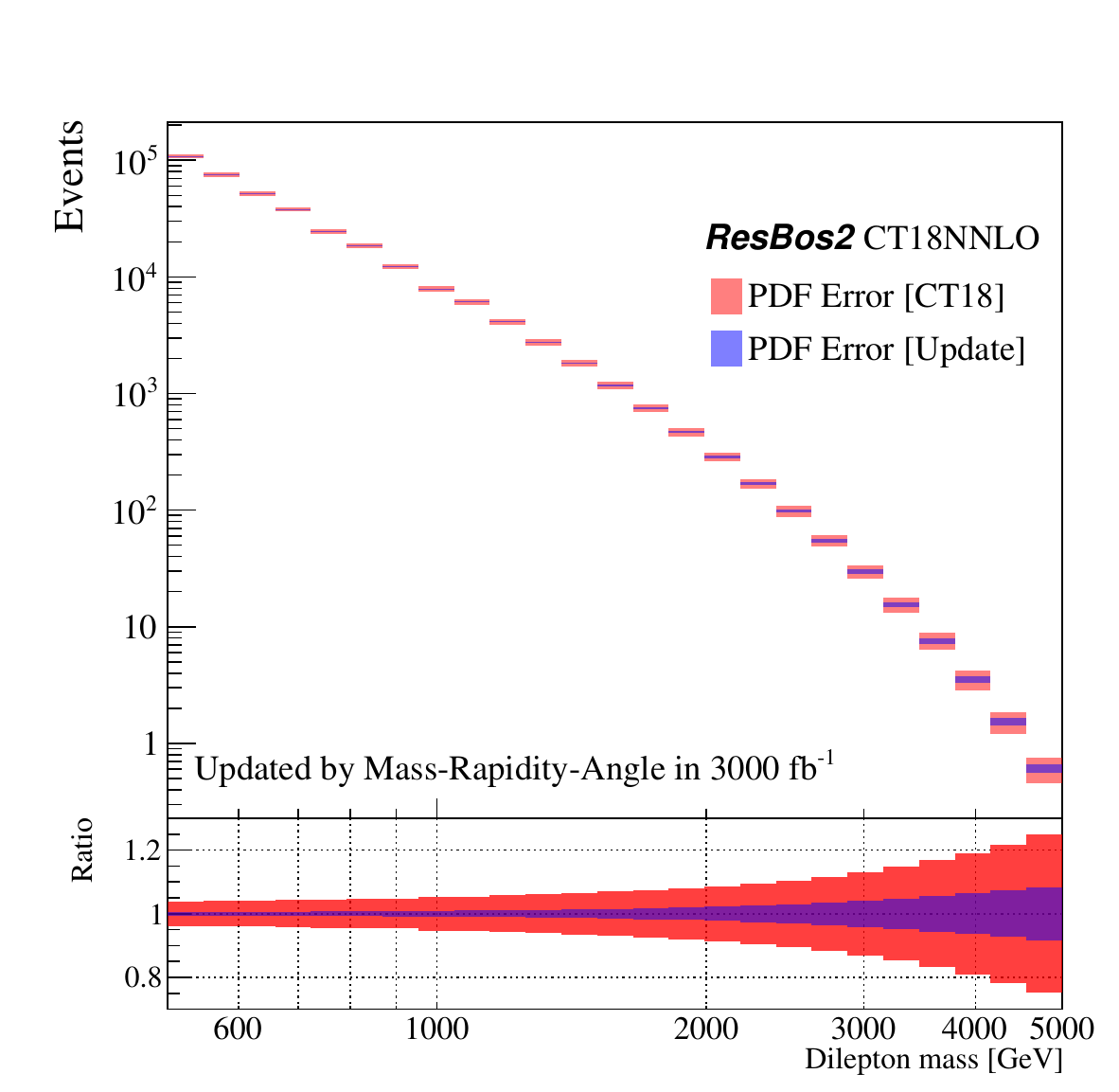}
  }
  \subfigure[]{
    \label{M_CC_3000}
    \includegraphics[width=0.48\textwidth]{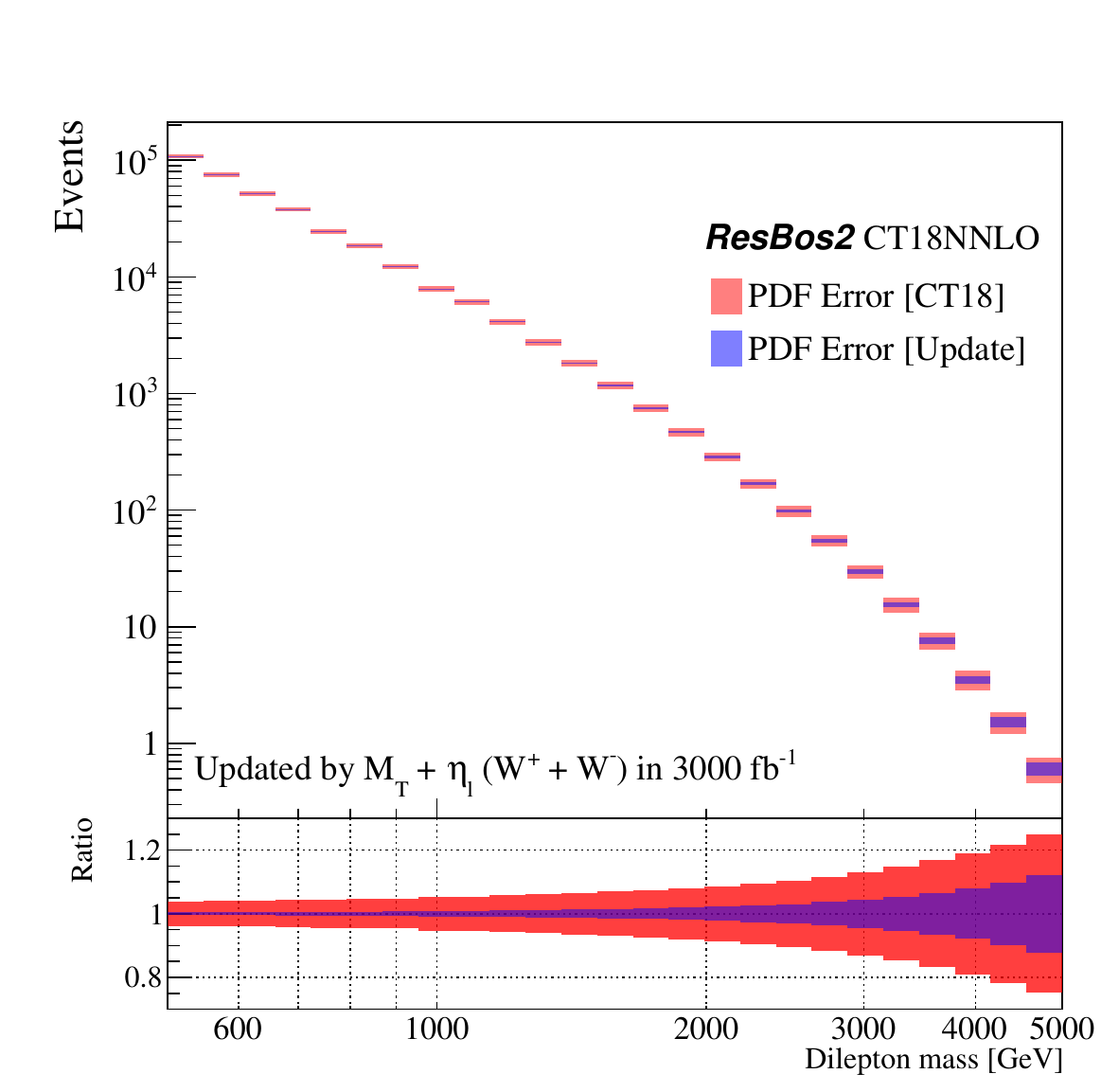}
  }
  \caption{$\dfrac{d\sigma}{dm_{\ell^+\ell^-}}$ and SM uncertainties appropriate for a BSM NC search using (a) NCDY templates assuming 3000~\fb ~and (b) CCDY templates assuming 3000~\fb.}
  \label{WHighMassMT_updated-by-WMT-Eta-Upto500.pdf}
\end{figure}

The reason for this improvement can be seen in Figure~\ref{partonsNC18}. Here effects of successively adding the input grids to \texttt{ePump} are explicitly shown. One can clearly see that incrementally adding more granular grids  from the 1-dimensional $\dfrac{d\sigma}{dm_{\ell\ell}}$, to the two dimensional $\dfrac{d^2\sigma}{dm_{\ell\ell}d|y_{\ell\ell}|}$, and finally to the complete three dimensional $\dfrac{d^3\sigma}{dm_{\ell\ell}d|y_{\ell\ell}|d\cos\theta^*}$ leads to the aforementioned reduction in uncertainty especially in the $x$ range of interest for both the valence and sea distributions. Finally, Fig.~\ref{M_NC_3000} shows the resulting reduction in PDF uncertainty using the \texttt{CT18}
PDFs as the baseline.

\section{Neutral and Charged Current DY}

In addition to updating the solely NCDY results to the more modern \texttt{CT18} fits, we expand this approach to include CCDY data toward three results. 
\begin{enumerate}
\item  First, we use CCDY data to create templates for PDF sets and fit for new distributions with \texttt{ePump}. The resulting boutique PDFs improve knowledge of up and adds knowledge for down quark parton distributions. Our NCDY templates are three dimensional in mass, rapidity and CS angle and  our CCDY templates are two dimensional in rapidity and transverse mass.
\item Next, we combine our strategy for creating three dimensional templates from NCDY final states and the two dimensional templates CCDY final states and use \texttt{ePump} to create new DY PDF sets. We apply them to quantify improvements in background uncertainty in regions of BSM interest  for NCDY and CCDY final states and report three results for each combination: graphical results of the reduction in $u$, $\bar{u}$, $d$, and $\bar{d}$ parton densities, graphical results for $\dfrac{d\sigma}{dm_{\ell^+\ell^-}}$ or $\dfrac{d\sigma}{dm_T}$, and tabular results for their contribution to error reduction in each of four searches backgrounds.
\item Finally, for completeness, in Appendix~\ref{appendix} we break out all of the charge state combinations for templates from the six combinations of NC [0] alone, CC [$+$, $-$, ($+\text{ \& } -)]$ alone, and NC [0] plus CC [$+$, $-$, ($+\text{ \& } -)]$ together as inputs. 
\end{enumerate}
For all results we assume that the High Luminosity LHC (HL-LHC) accumulates 3000~fb$^{-1}$ of integrated luminosity per experiment and so the inputs available for fitting amount to 3000~fb$^{-1}$ in all.

\subsection{Addition of CC to NC Inputs: Result \#1}
The CCDY project is different in a variety of ways from the NCDY approach. First, the CCDY cross-section is larger, so there are more events for a given integrated luminosity. Second, there are two charge states that in fact add different information to PDF precision: the easily signed $W^-$ final states probe the $d$ and $\bar{u}$ partons, while the $W^+$  final states probe the $u$ and $\bar{d}$ partons. So the latter are adding directly to the up parton distribution precision while the former adds complementary information to the less-potent NCDY production from down partons.

The NCDY strategy led us to create templates below a threshold where only SM physics is expected to be involved which we chose to be $\dfrac{d\sigma}{dm_{\ell^+\ell^-}}< 1$~TeV. In parallel, we chose $\dfrac{d\sigma}{dm_T}<500$~GeV to be a ``safe'' region to create SM templates for \texttt{ePump} in the CCDY project. This choice is supported by the observation that at leading order (LO), a single $W^\pm$ boson with a large $m_T$ is produced in association with a QCD jet, where the transverse momentum of the jet is approximately equal to that of the $W^\pm$ boson ($p_T^W$). Therefore, the hard scale of the event is roughly $m_T + p_T^W$.

\begin{figure}[!th]
  \centering
  \subfigure[]{
    \label{LABEL1}
    \includegraphics[width=0.48\textwidth]{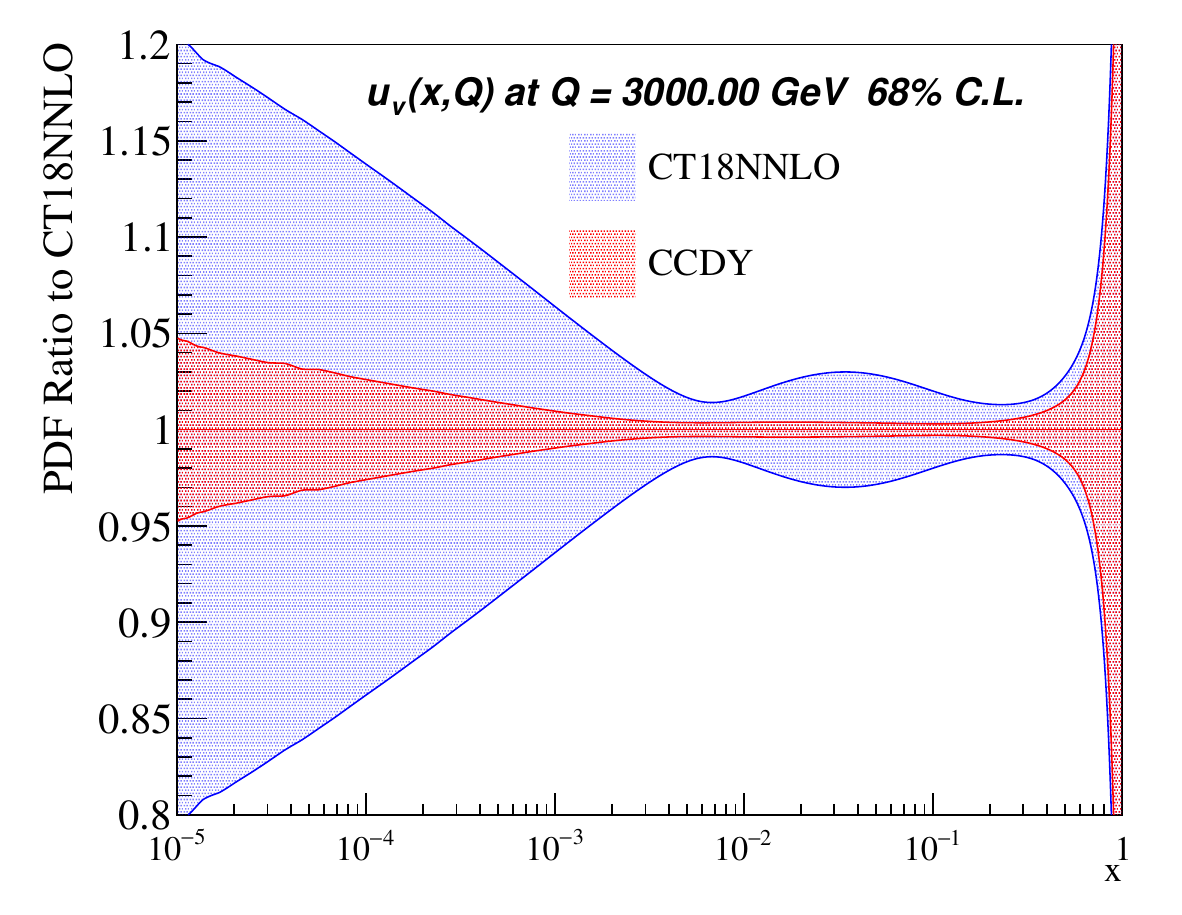}
  }
  \subfigure[]{
    \label{LABEL2}
    \includegraphics[width=0.48\textwidth]{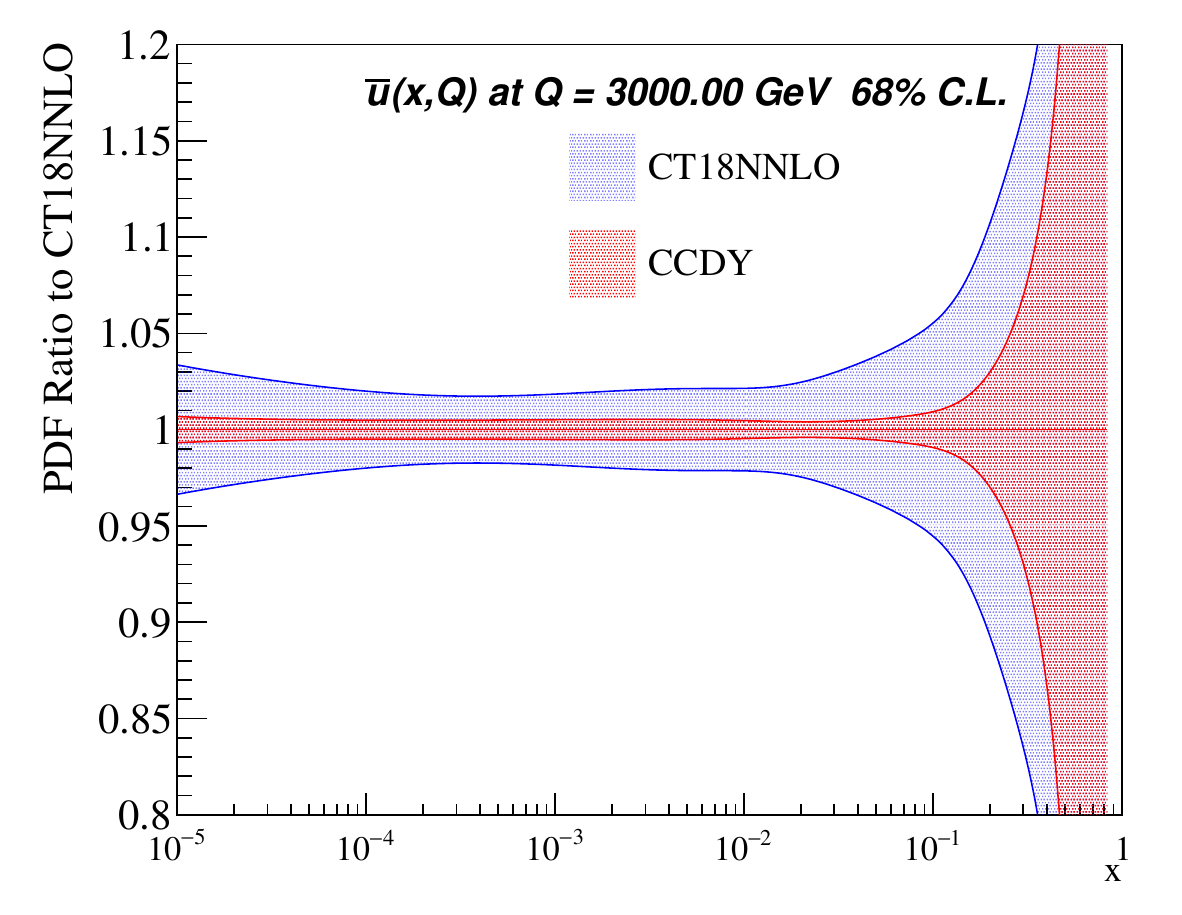}
  }
    \subfigure[]{
    \label{LABEL1}
    \includegraphics[width=0.48\textwidth]{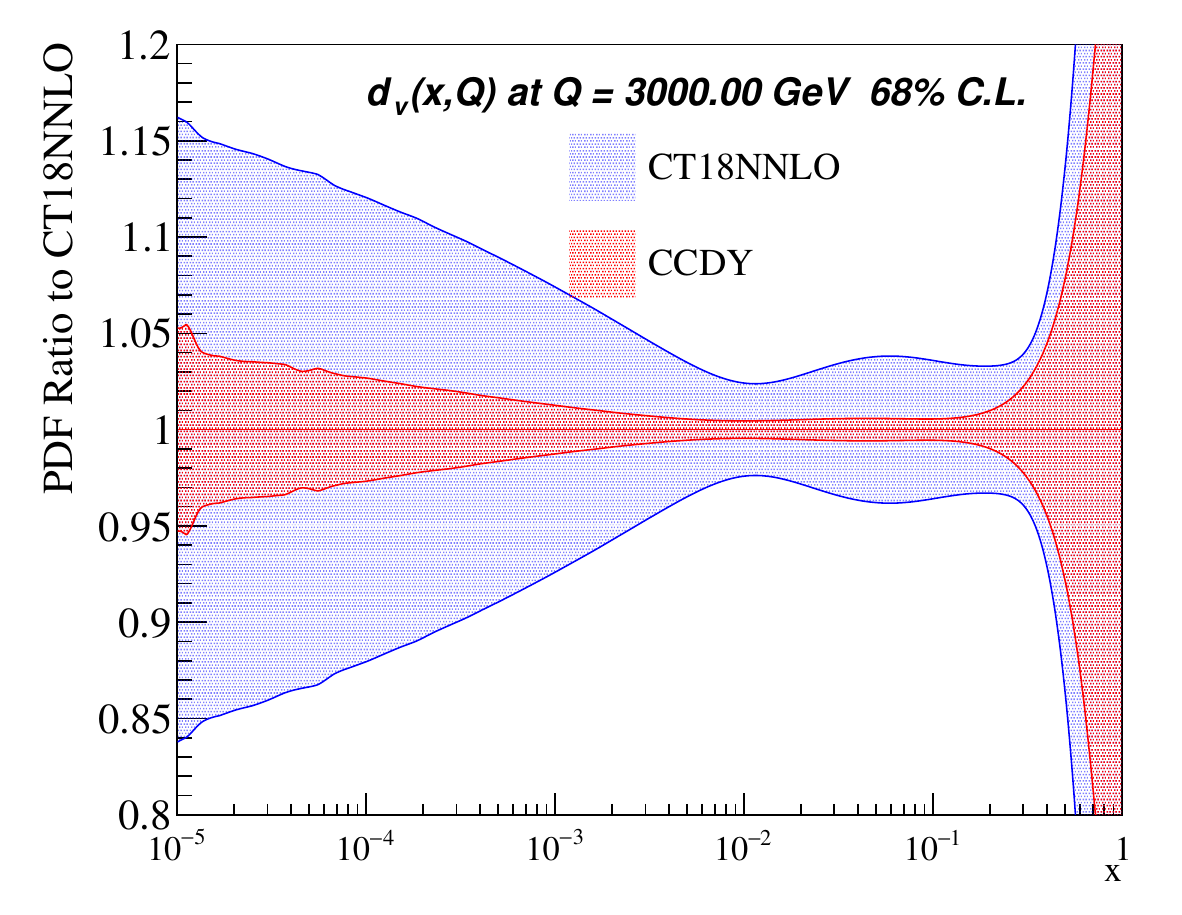}
  }
  \subfigure[]{
    \label{LABEL2}
    \includegraphics[width=0.48\textwidth]{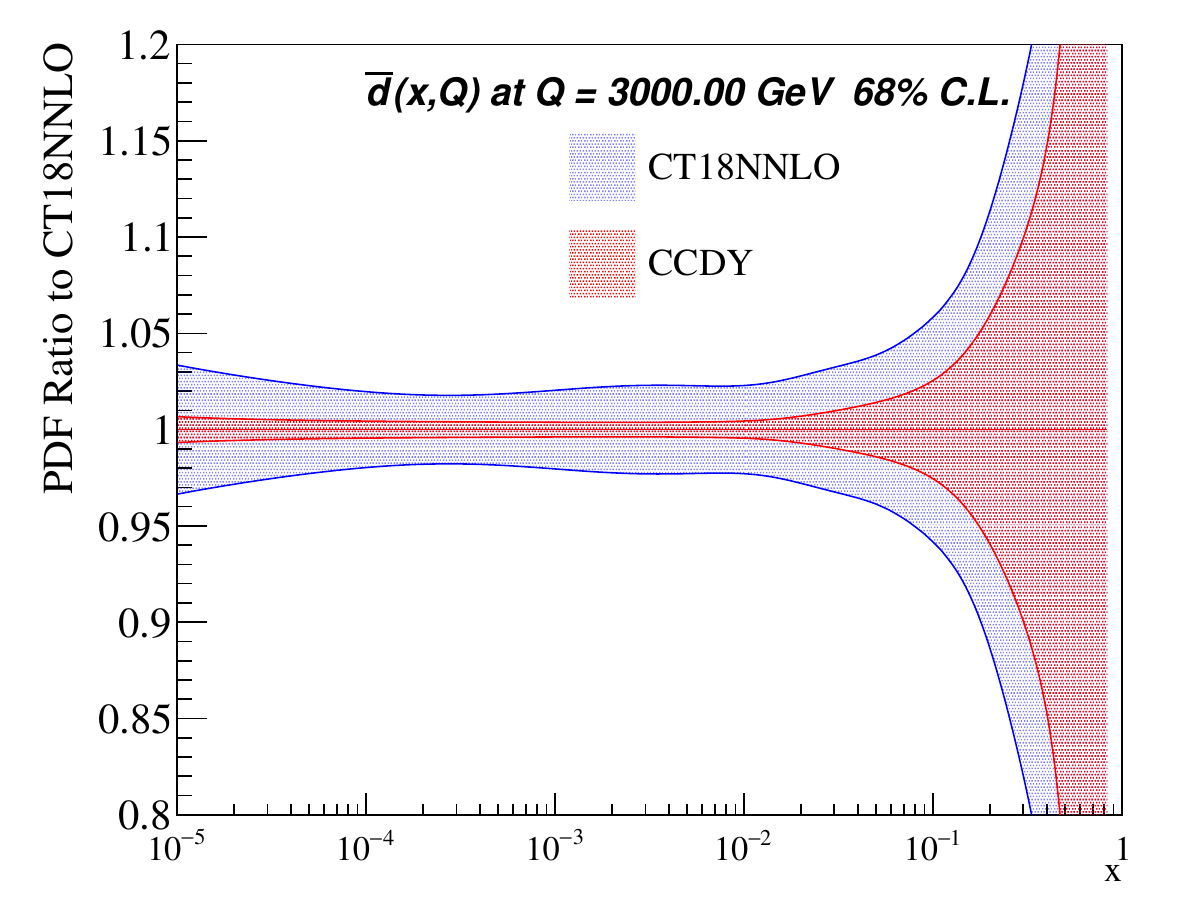}
  }
  \caption{Relevant parton distribution improvements for the CCDY project, relative to \texttt{CT18}.}
  \label{partonsCC18}
\end{figure}

Figure~\ref{partonsCC18} show the results of adding CCDY inputs to the CT18 PDFs using \texttt{ePump} for the up and down valence and sea quarks. Signficant reduction in the uncertainty is evident in especially the $\bar{u}$ and $\bar{d}$ distributions. Note that one would expect that fits using $W^-$ ($W^+$) would enhance knowledge of the up (down) partons and that's exactly what we see. 

To mimic experimental acceptance, lepton kinematics are required to satisfy $p_T^\ell>30$~GeV and $|\eta_\ell|<2.47$. Lepton plus neutrino events are further required to satisfy $E_T^{\rm miss}>30$~GeV and $m_T>50$~GeV. The bins used in two dimensional distribution include 20 bins for $60<m_T<100$~GeV, 5 bins for $100<m_T<500$~GeV, and 20 bins for $0.0 < |\eta_\ell| < 2.4$.

\begin{figure}[!th]
  \centering
  \subfigure[]{
    \label{LABEL1}
    \includegraphics[width=0.48\textwidth]{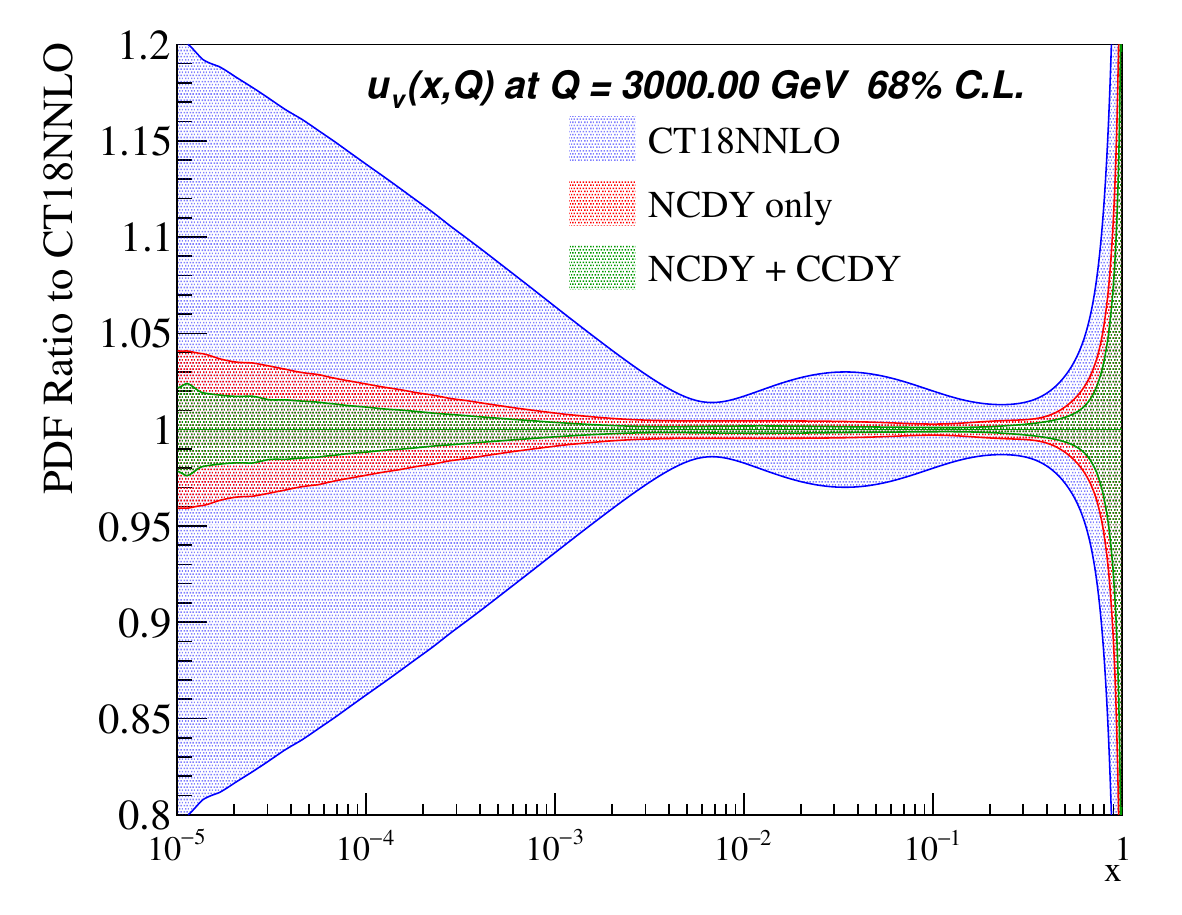}
  }
  \subfigure[]{
    \label{LABEL2}
    \includegraphics[width=0.48\textwidth]{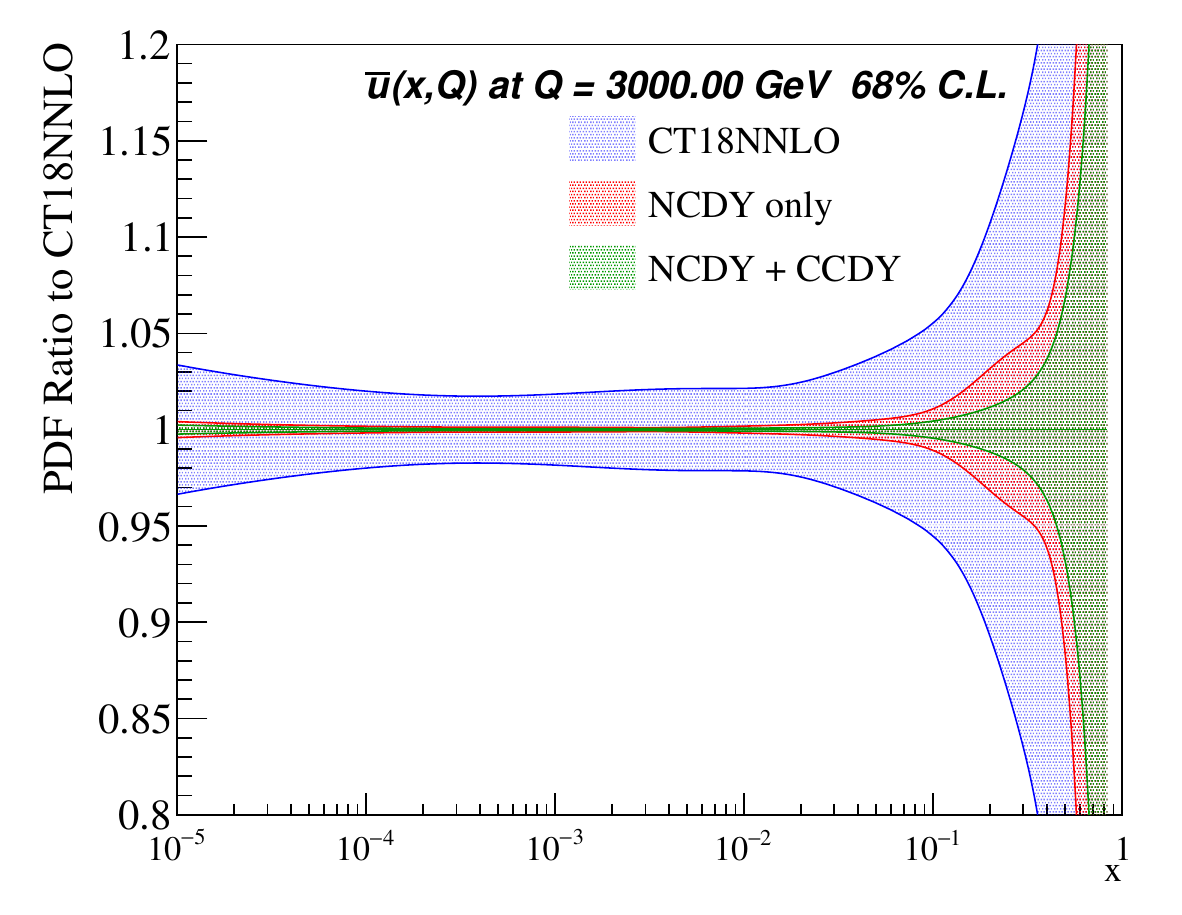}
  }
    \subfigure[]{
    \label{LABEL1}
    \includegraphics[width=0.48\textwidth]{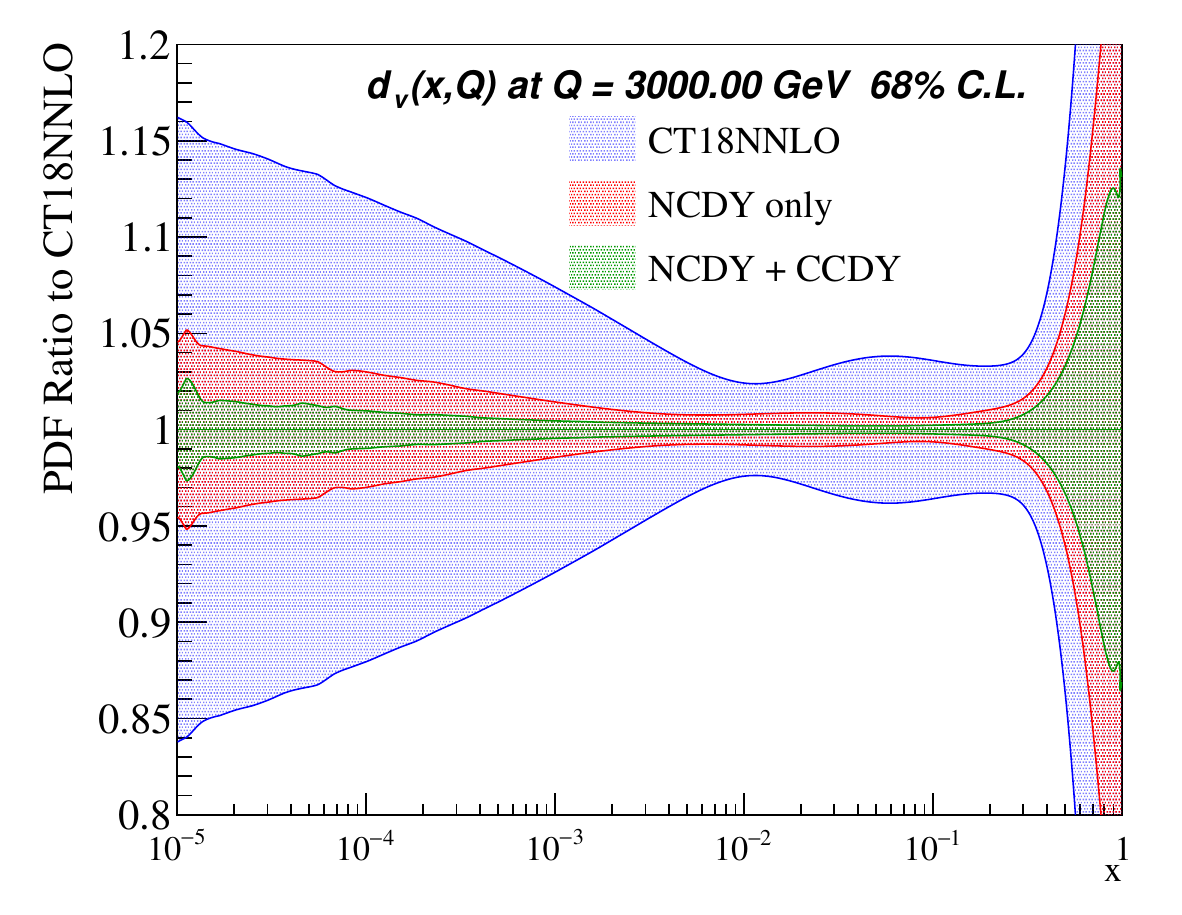}
  }
  \subfigure[]{
    \label{LABEL2}
    \includegraphics[width=0.48\textwidth]{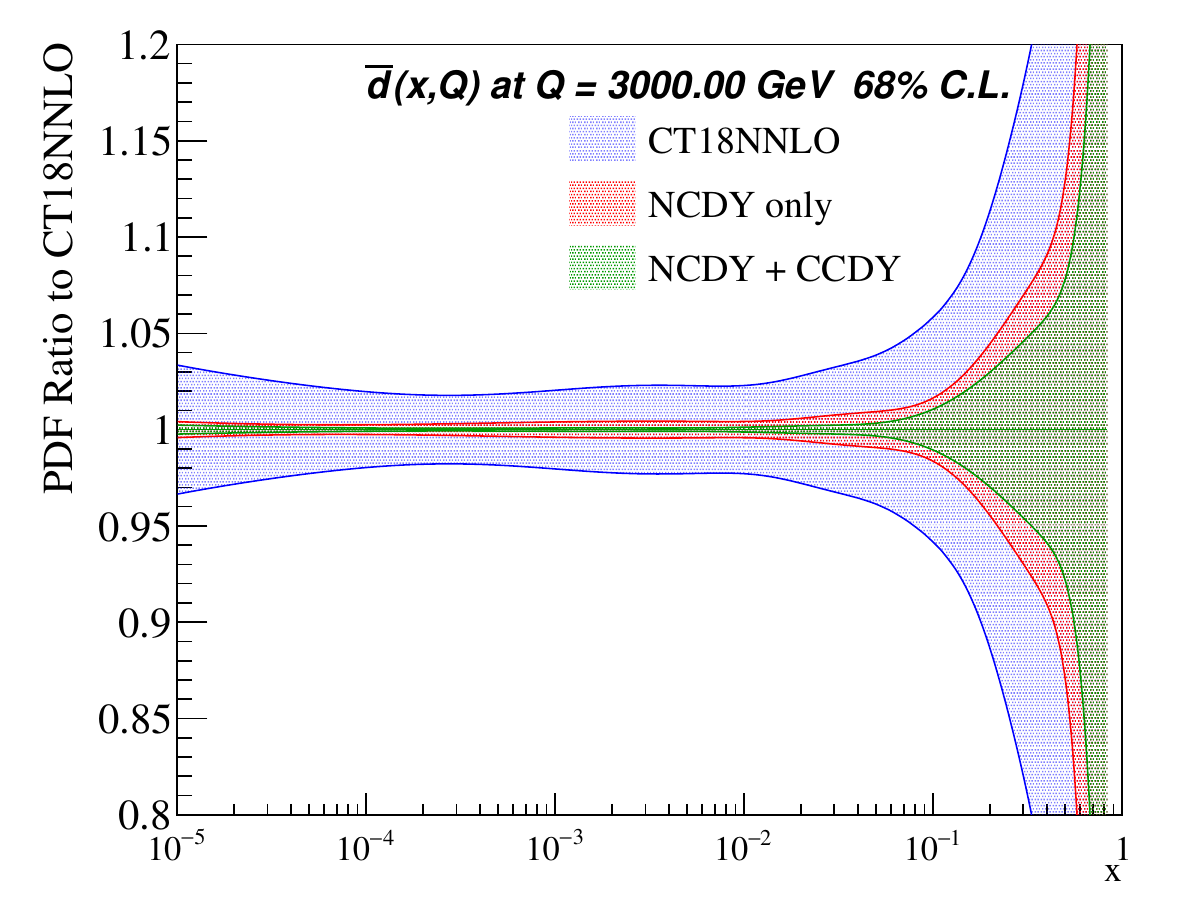}
  }
  \caption{The improvement, relative to \texttt{CT18}, for $u_V$, $\bar{u}$, $d_V$, and $\bar{d}$ for with the addition of NCDY results, and the addition of both NCDY and CCDY, for 6000~\fb.}
  \label{partonsNCCC18}
\end{figure}
\begin{figure}[!th]
  \centering
  \subfigure[]{
    \label{LABEL1}
    \includegraphics[width=0.48\textwidth]{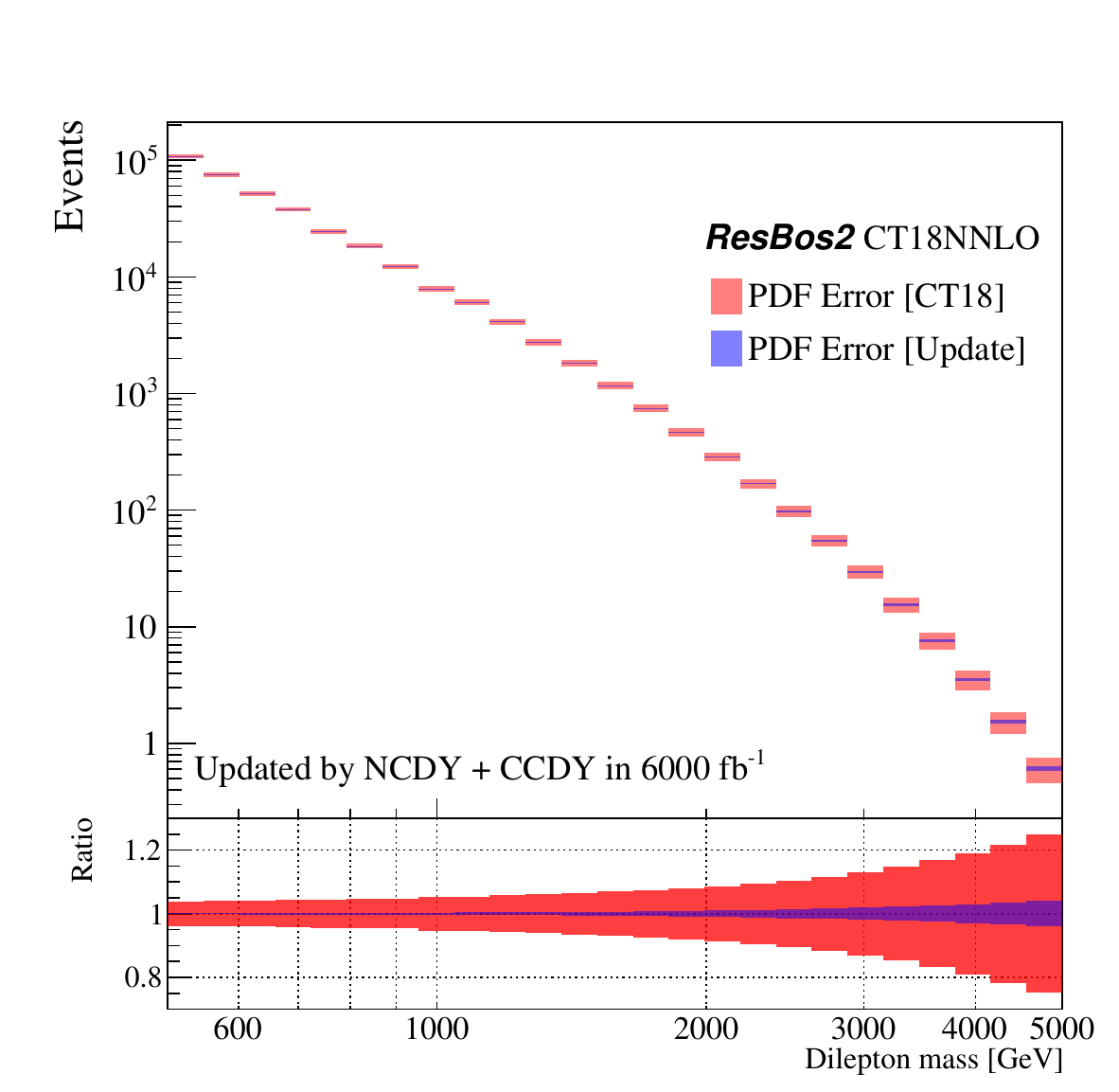}
  }
  \subfigure[]{
    \label{LABEL2}
    \includegraphics[width=0.48\textwidth]{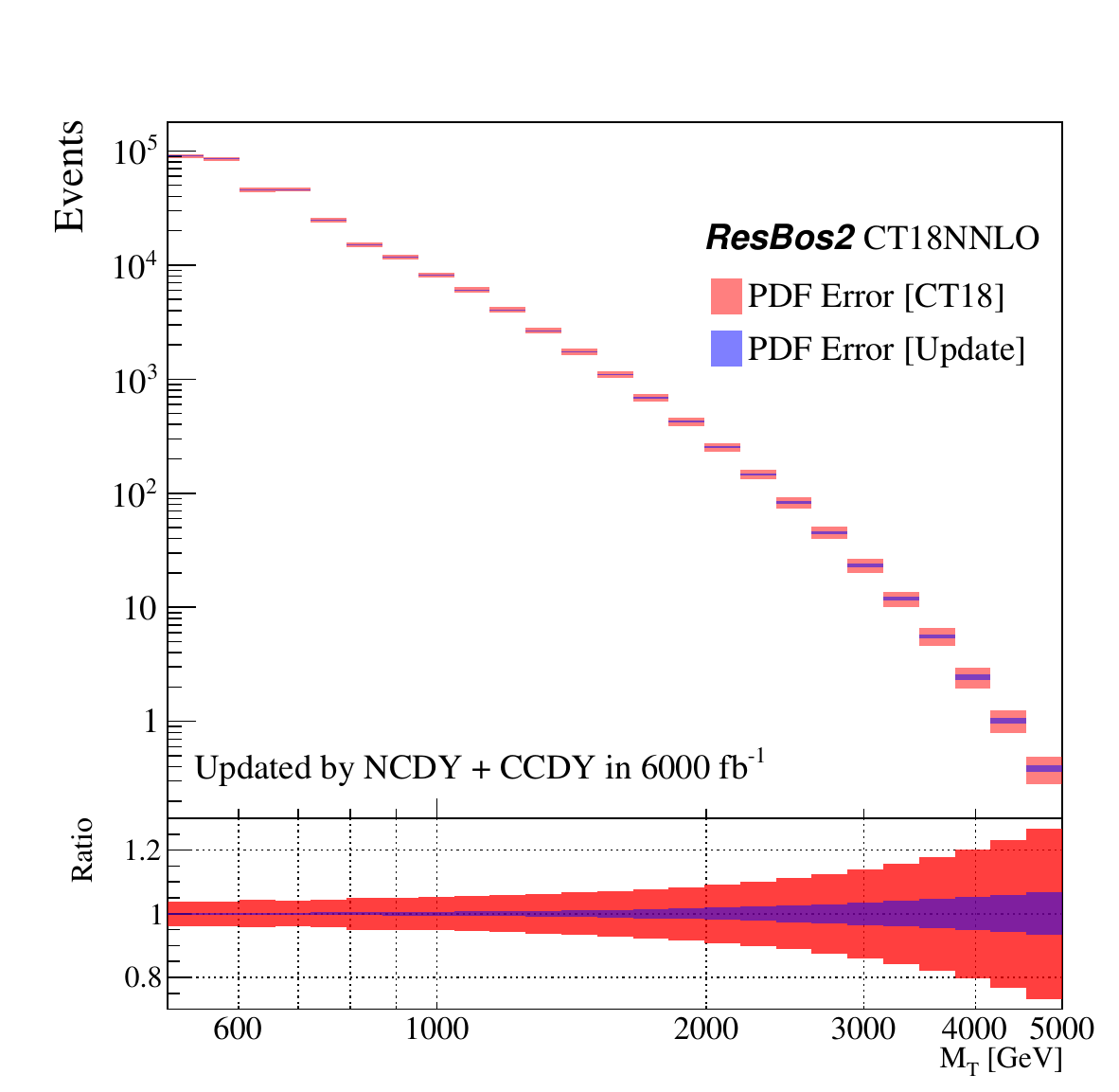}
  }
  \caption{Using both NCDY and CCDY templates assuming 6000~\fb. (a) $\dfrac{d\sigma}{dm_{\ell^+\ell^-}}$ and SM uncertainties appropriate for a BSM NC search and (b) $\dfrac{d\sigma}{dm_T}$ and SM uncertainties appropriate for a BSM CC search.}
  \label{massesCCNC3000}
\end{figure}

\subsection{Neutral and Charged Current DY Sources: Result \#2}
We now show the effects of combining the use of the new NCDY plus CCDY PDF sets and their effects on an overall PDF contribution to the background of high mass searches in $\dfrac{d\sigma}{dm_{\ell^+\ell^-}}$ and $\dfrac{d\sigma}{dm_T}$. These calculations form the major result from this paper. The improvement in background uncertainty for $\dfrac{d\sigma}{dm_{\ell^+\ell^-}}$ and $\dfrac{d\sigma}{dm_T}$ are shown in Figure~\ref{massesCCNC3000}. Finally, these results are summarized in Table~\ref{tab:PDFUnc_Xsec3D} which can be read as follows: Column one indicates masses ($m_{\ell^+\ell^-}$ or $m_{T}$)  and columns two through five report the inclusive improvement in uncertainty from the bare CT18 result ($\delta^{PDF}_{pre}$)  in percent and the new uncertainties from this analysis, ($\delta^{PDF}_{post}$). Column two corresponds to Figure~\ref{massesCCNC3000}~(a) and column three corresponds to Figure~\ref{massesCCNC3000}~(b). Columns four and five break out the individual charge states for $\dfrac{d\sigma(W^\pm)}{dm_T}$ separately. So for example, the uncertainty for a NCDY search background in $m_{\ell^+\ell^-}> 3$~TeV was 15.4\% using CT18 alone and using the strategy outlined here, improves to 2.3\%.  

\begin{figure}[H]
  \centering
  \subfigure[]{
    \label{L2Sensitivity_uv_NCCC}
    \includegraphics[width=0.55\textwidth]{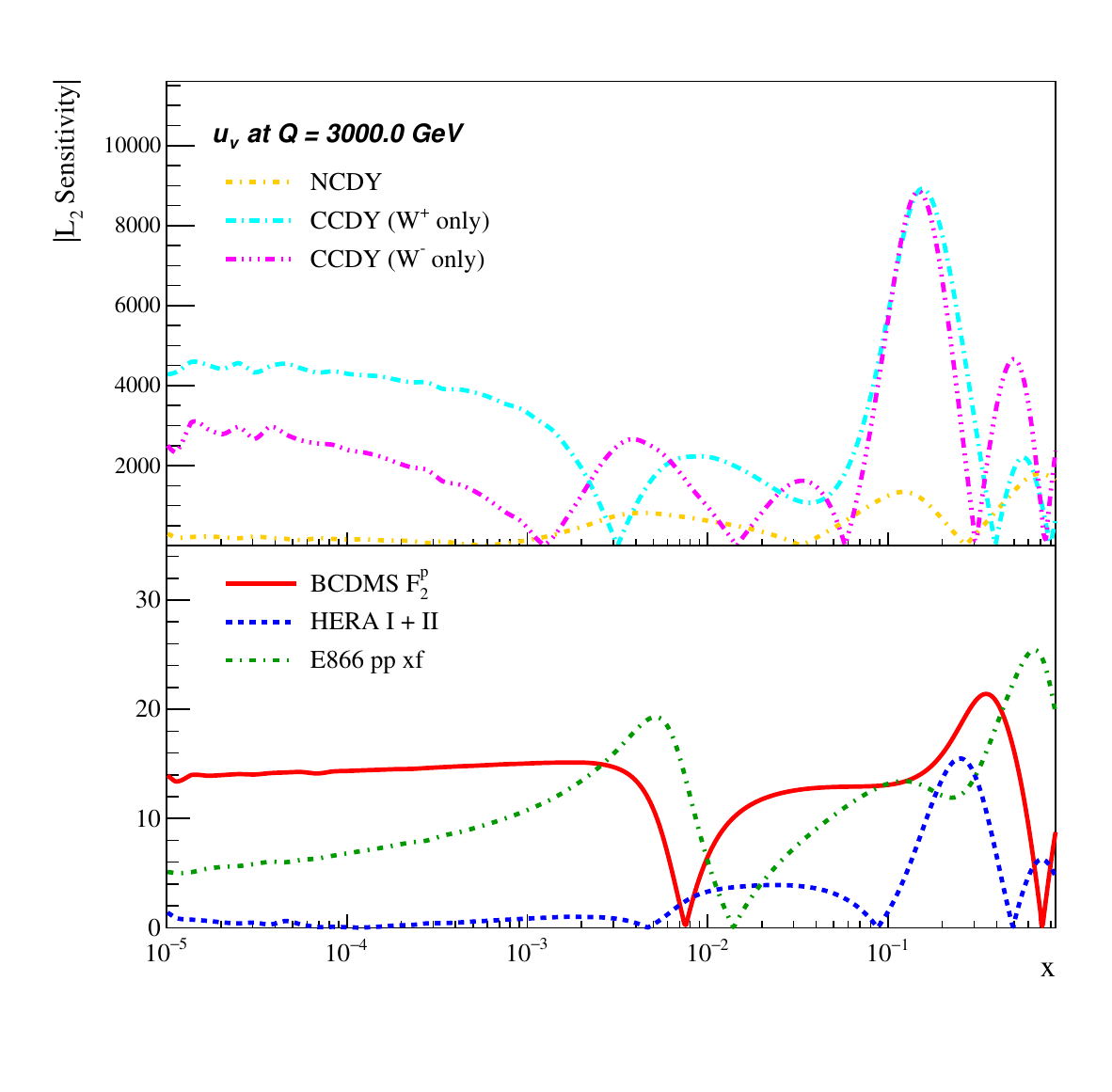}
  }
  \subfigure[]{
    \label{L2Sensitivity_ubar_NCCC}
    \includegraphics[width=0.55\textwidth]{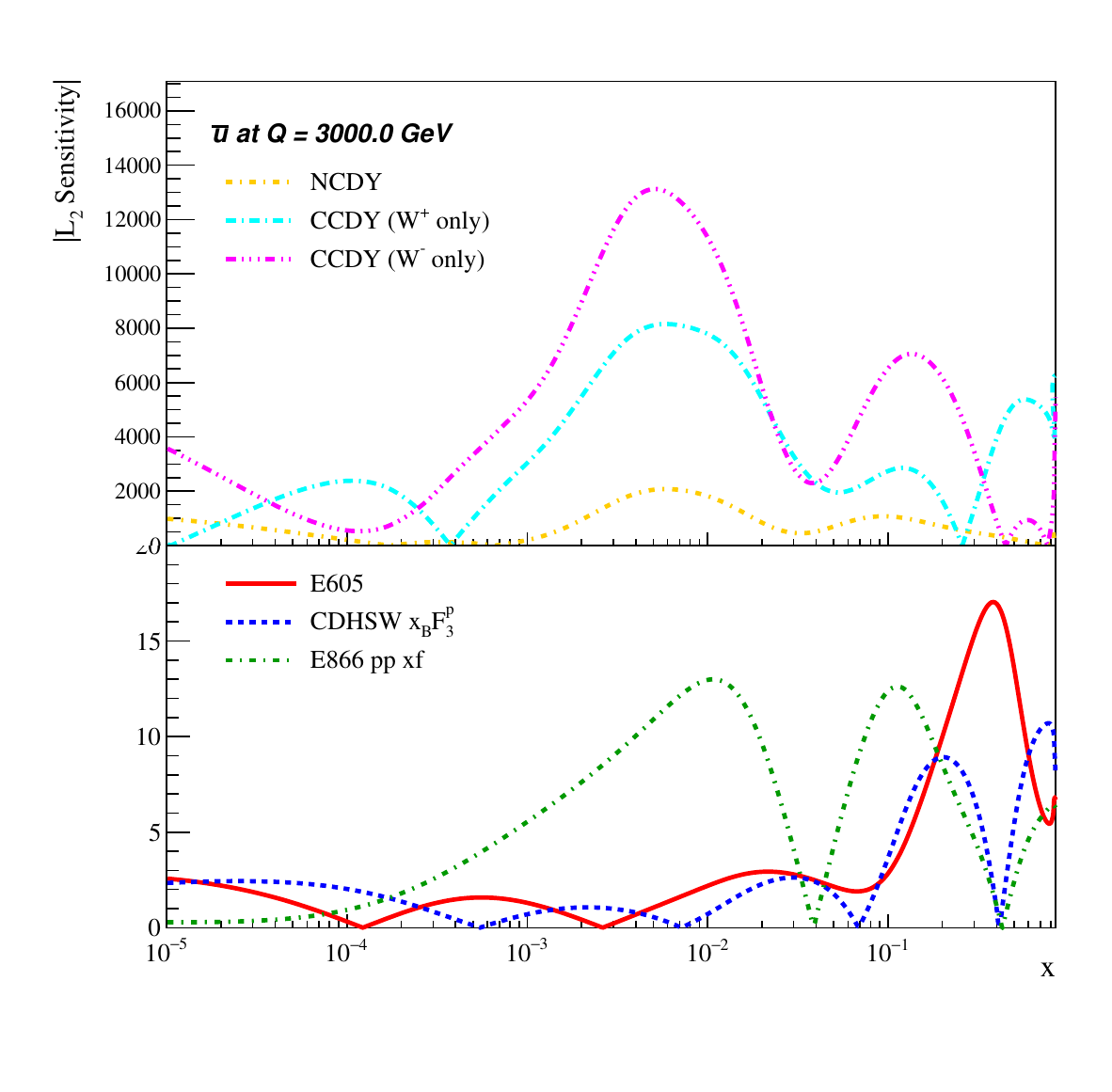}
  }
  \caption{Revisiting the sensitivity $|L_2|$ in light of the strategy adopted in this paper. (a) corresponds to $u_v$ and (b) to $\bar{u}$, for $Q=3$~TeV. In the top graph of each are the sensitivities due to the NCDY and CCDY contributions to a PDF fit resulting from our strategy, while the bottom graph repeats $|L_2|$ for the three most dominant experiments from Figures~\ref{L2Sensitivity_uv} and \ref{L2Sensitivity_ubar}, prior to this paper.}
  \label{fig:L2_newdata}
\end{figure}

To date, the only experiments with influence in the $x$ region of importance for BSM DY production are very low $Q$ and decades old. For the first time, our strategy provides a path for high scale LHC data to contribute.  We can see this by revisiting the  $L_2$ significance contributions to PDF fitting presented in Figures~\ref{L2Sensitivity_uv} and \ref{L2Sensitivity_ubar} which can now be amended to show the inclusion of  our boutique PDFs. This is shown in Figs.~\ref{L2Sensitivity_uv_NCCC} and \ref{L2Sensitivity_ubar_NCCC}. Notice that the scales are 1000 times different. This confirms that our strategy uniquely targets the kinematical region appropriate to backgrounds in high mass which are relevant to the search of BSM spin 1 bosons, extending  beyond the coverage of the older data inputs used in \texttt{CT18} global fitting.

\begin{table*}[th]
    \centering
    \begin{tabular}{c|cc|cc|cc|cc}
    \hline
    \hline
     & \multicolumn{2}{|c}{$m_{\ell\ell}$ (NCDY) [TeV]} 
     & \multicolumn{2}{|c}{$M_T$ (CCDY) [TeV]} 
     & \multicolumn{2}{|c}{$M_T$ ($W^+$) [TeV]} 
     & \multicolumn{2}{|c}{$M_T$ ($W^-$) [TeV]} \\
     $m_{\ell\ell}(M_T)$ 
     & $\delta^{\text{PDF}}_{\text{pre}}$ [\%] & $\delta^{\text{PDF}}_{\text{post}}$ [\%] 
     & $\delta^{\text{PDF}}_{\text{pre}}$ [\%] & $\delta^{\text{PDF}}_{\text{post}}$ [\%] 
     & $\delta^{\text{PDF}}_{\text{pre}}$ [\%] & $\delta^{\text{PDF}}_{\text{post}}$ [\%] 
     & $\delta^{\text{PDF}}_{\text{pre}}$ [\%] & $\delta^{\text{PDF}}_{\text{post}}$ [\%] \\
    \hline
        $>1$ & 5.9 & 0.5 & 5.9 & 0.9 & 6.2 & 1.0 & 7.9 & 0.9 \\
        $>2$ & 9.8 & 1.3 & 10.4 & 2.4 & 11.3 & 2.8 & 13.4 & 2.0 \\
        $>3$ & 15.4 & 2.3 & 16.4 & 4.2 & 18.0 & 4.7 & 20.7 & 3.4 \\
        $>4$ & 22.2 & 3.5 & 23.7 & 6.0 & 25.7 & 6.6 & 29.5 & 4.9 \\
        $>5$ & 30.8 & 5.1 & 33.0 & 8.0 & 35.0 & 8.7 & 40.1 & 6.8 \\
    \hline
    \hline
    \end{tabular}
    \caption{For $6000~\text{fb}^{-1}$ of data from two experiments at the HL-LHC, the PDF uncertainty contributions to backgrounds for $\frac{d\sigma}{dm_{\ell^+\ell^-}}$ and $\frac{d\sigma}{dm_T}$ (first two column groups) and $\frac{d\sigma}{dm_T}$ for $W^+$ and $W^-$ only in the last two column groups. $\delta^{\text{PDF}}_{\text{pre}}$ refers to the \textsc{cteq} results alone, while $\delta^{\text{PDF}}_{\text{post}}$ is the uncertainty after including both the NCDY and CCDY templates in the overall fit. The mass ranges are inclusive, so $>3$ means the uncertainty integrated above $3$~TeV.}
    \label{tab:PDFUnc_Xsec3D}
\end{table*}

\subsection{Potential for Run 3 Improvements}

The above results were calculated for HL-LHC data with integrated luminosities of $2\times 3000\text{ fb}^{-1}$ assuming contributions from both ATLAS and CMS NCDY and CCDY data for the \texttt{ePump} templates. But with Run 3 nearly complete, both experiments are planning publication of partial Run 3 results and so we have also calculated 600~fb$^{-1}$ luminosities appropriate from both ATLAS and CMS data as inputs.

The improvements from  \texttt{CT18}   by adding NCDY and CCDY templates using our strategy for partial Run 3 data are shown in Table~\ref{tab:PDFUnc_Xsec3D_600}. We judge improvements of a factor of 3 for NCDY and 2 for CCDY from data currently taken as an important suggestion that we need not wait for HL-LHC running to improve the SM backgrounds in regions of interest for BSM DY searches.

\begin{table*}[ht]
    \centering
    \begin{tabular}{c|cc|cc|cc|cc}
    \hline
    \hline
     & \multicolumn{2}{|c}{$m_{\ell\ell}$ (NCDY) [TeV]} 
     & \multicolumn{2}{|c}{$M_T$ (CCDY) [TeV]} 
     & \multicolumn{2}{|c}{$M_T$ ($W^+$) [TeV]} 
     & \multicolumn{2}{|c}{$M_T$ ($W^-$) [TeV]} \\
     $m_{\ell\ell}(M_T)$ 
     & $\delta^{\text{PDF}}_{\text{pre}}$ [\%] & $\delta^{\text{PDF}}_{\text{post}}$ [\%] 
     & $\delta^{\text{PDF}}_{\text{pre}}$ [\%] & $\delta^{\text{PDF}}_{\text{post}}$ [\%] 
     & $\delta^{\text{PDF}}_{\text{pre}}$ [\%] & $\delta^{\text{PDF}}_{\text{post}}$ [\%] 
     & $\delta^{\text{PDF}}_{\text{pre}}$ [\%] & $\delta^{\text{PDF}}_{\text{post}}$ [\%] \\
    \hline
        $>1$ & 5.9 & 1.1 & 5.9 & 1.8 & 6.2 & 2.1 & 7.9 & 2.0 \\
        $>2$ & 9.8 & 2.8 & 10.4 & 4.8 & 11.3 & 5.7 & 13.4 & 4.4 \\
        $>3$ & 15.4 & 4.9 & 16.4 & 8.5 & 18.0 & 9.8 & 20.7 & 7.3 \\
        $>4$ & 22.2 & 7.7 & 23.7 & 12.5 & 25.7 & 14.0 & 29.5 & 10.8 \\
        $>5$ & 30.8 & 11.5 & 33.0 & 17.2 & 35.0 & 18.9 & 40.1 & 15.1 \\
    \hline
    \hline
    \end{tabular}
    \caption{From data accumulated during partial Run 3 at the LHC, the PDF uncertainty contributions to backgrounds for $\frac{d\sigma}{dm_{\ell^+\ell^-}}$ and $\frac{d\sigma}{dm_T}$ (first two columns) and $\frac{d\sigma}{dm_T}$ for $W^+$ and $W^-$ only in the last two columns. $\delta^{\text{PDF}}_{\text{pre}}$ refers to the \textsc{cteq} results alone while $\delta^{\text{PDF}}_{\text{post}}$ is the uncertainty after including both the NCDY and CCDY templates in the overall fit. The mass ranges are inclusive, so $>3$ means the uncertainty above $3$~TeV.}
    \label{tab:PDFUnc_Xsec3D_600}
\end{table*}

\section{Conclusions}
In this work we lay out a strategy to create PDF sets uniquely characterized to attack the largest SM background PDF uncertainties in high-mass BSM IVB searches in both  neutral current (``$Z'\,$'' searches) and charged current (``$W'\,$'') final states. We show that for HL-LHC running, these uncertainties can be reduced by a factor of 7 for $m_{\ell \ell}$ above 4~TeV and a factor of 4 for $m_{\ell \nu}$ above 5~TeV. In addition even the partial LHC Run 3 data can yield substantial improvements. We urge that the strategy we outlined be pursued by the PDF fitting groups and that the LHC experiments provide them with three dimensional differential cross section grids for neutral current measurements and two dimensional grids for charged current measurements.

This is the third in a series of four papers on PDF uncertainties for high mass DY physics. The fourth paper will focus on possible improvements for ``non-resonant'' searches using an Effective Field Theory strategy for evidence of BSM states at very high masses, inaccessible for direct observation. 

\newpage \newpage

\appendix

\section{The Complete Suite of Results} \label{appendix}

In the body of the paper, we combined both signs of $W$ bosons and presented the best fit scenario. For completeness, in Appendix~\ref{app_pdfs} we show individual $u$, $\bar{u}$, $d$, and $\bar{d}$ PDF improvements due to our fitting strategy broken out by the individual charge states of the $W^\pm$ boson. In Appendix~\ref{app_tables} we present the resulting improvements in $\delta^{PDF}$ versus $m_{\ell\ell}$ from these individual PDF results due to the PDFs  in Appendix~\ref{app_pdfs}.

\subsection{Up and Down PDFs Separated by $W$ Boson Charge} \label{app_pdfs}

\begin{figure}[H]
  \centering
  \subfigure[]{
    \label{LABEL1}
    \includegraphics[width=0.48\textwidth]{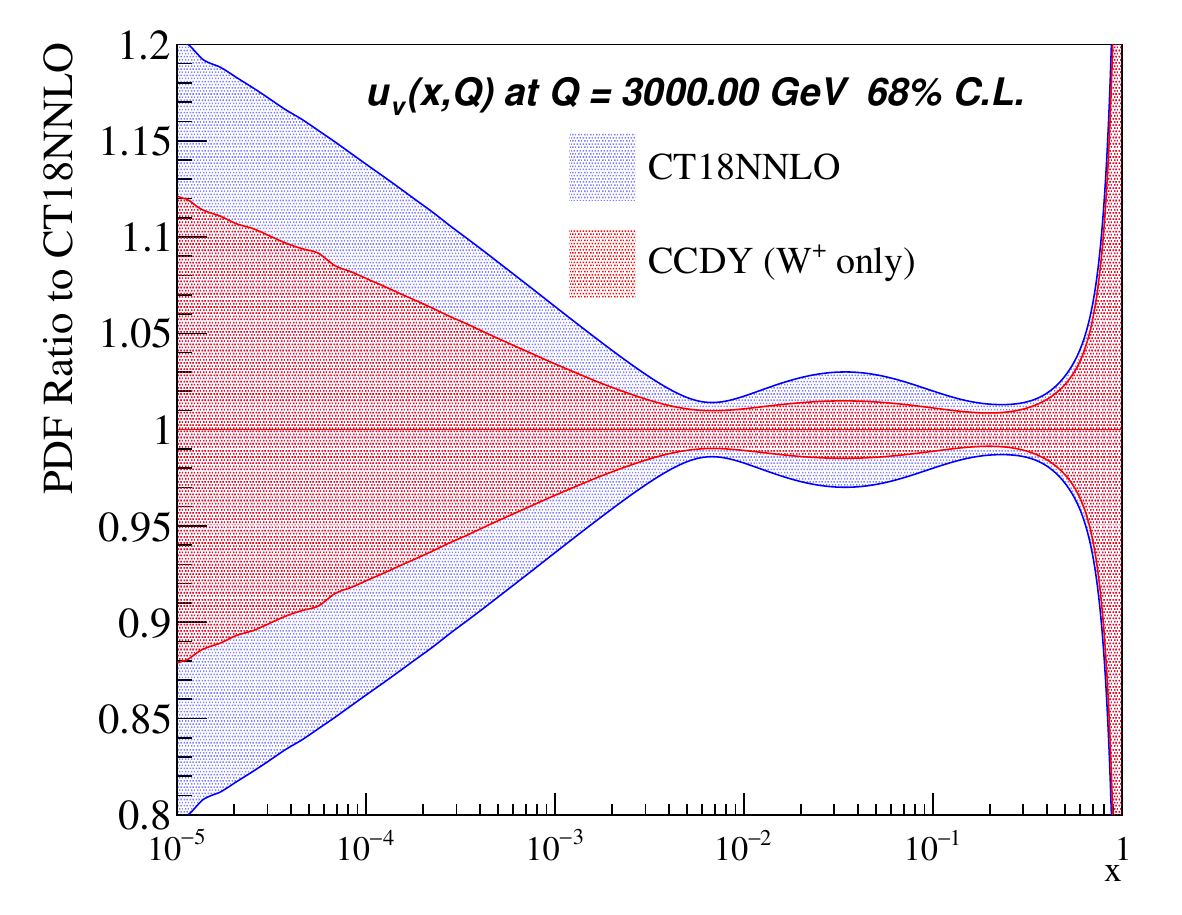}
  }
  \subfigure[]{
    \label{LABEL2}
    \includegraphics[width=0.48\textwidth]{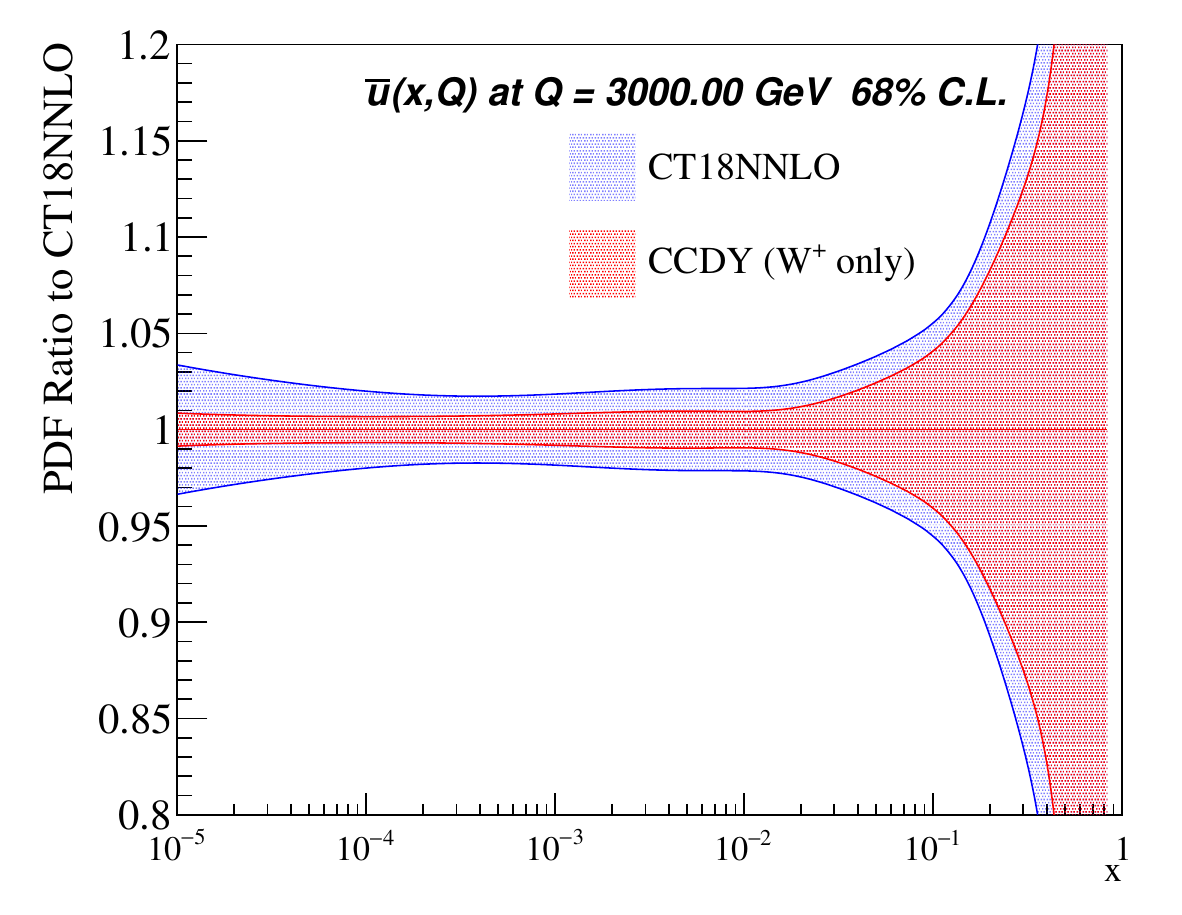}
  }
    \subfigure[]{
    \label{LABEL1}
    \includegraphics[width=0.48\textwidth]{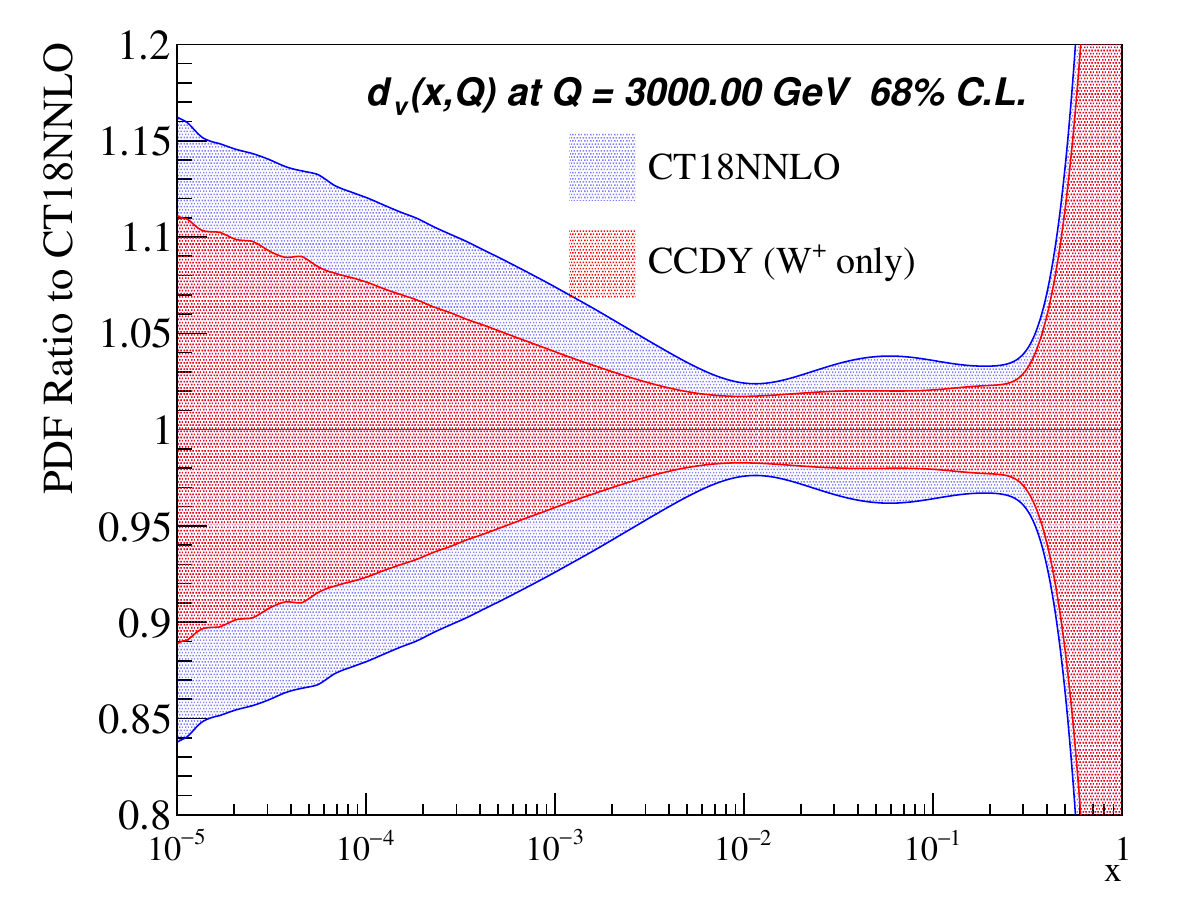}
  }
  \subfigure[]{
    \label{LABEL2}
    \includegraphics[width=0.48\textwidth]{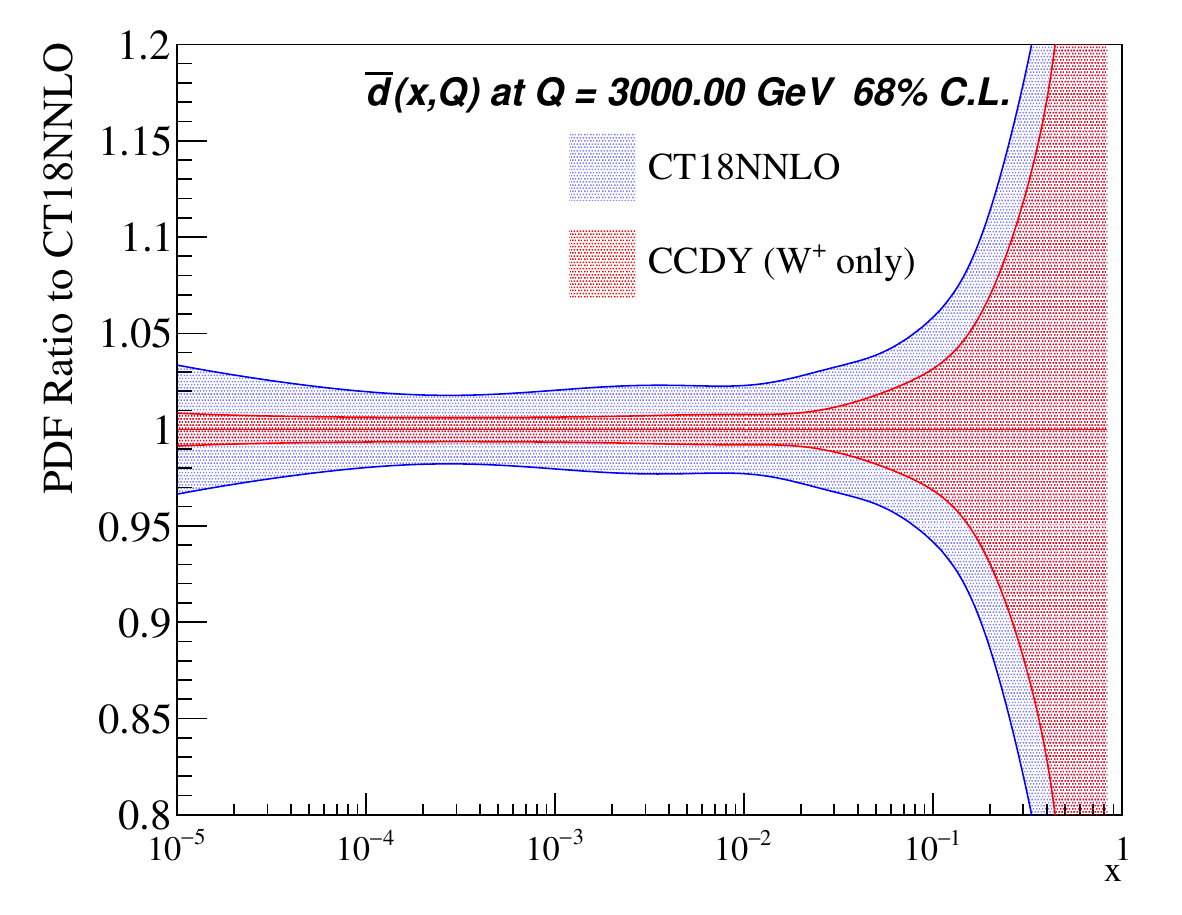}
  }
  \caption{The improvement, relative to  \texttt{CT18}, for $u_V$, $\bar{u}$, $d_V$,  with the  addition of $W^+$, for 6000~\fb.}
  \label{partonsWPlus18}
\end{figure}

\begin{figure}[H]
  \centering
  \subfigure[]{
    \label{LABEL1}
    \includegraphics[width=0.48\textwidth]{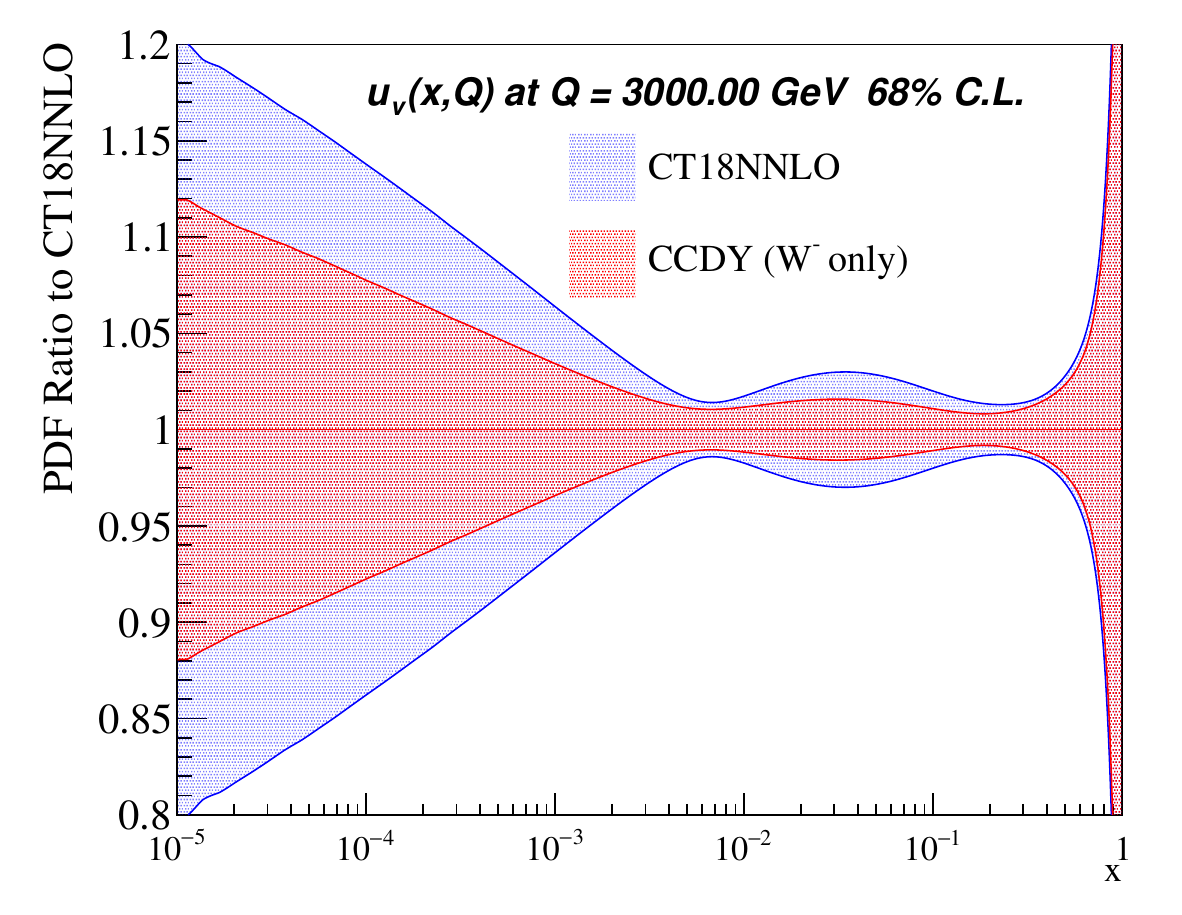}
  }
  \subfigure[]{
    \label{LABEL2}
    \includegraphics[width=0.48\textwidth]{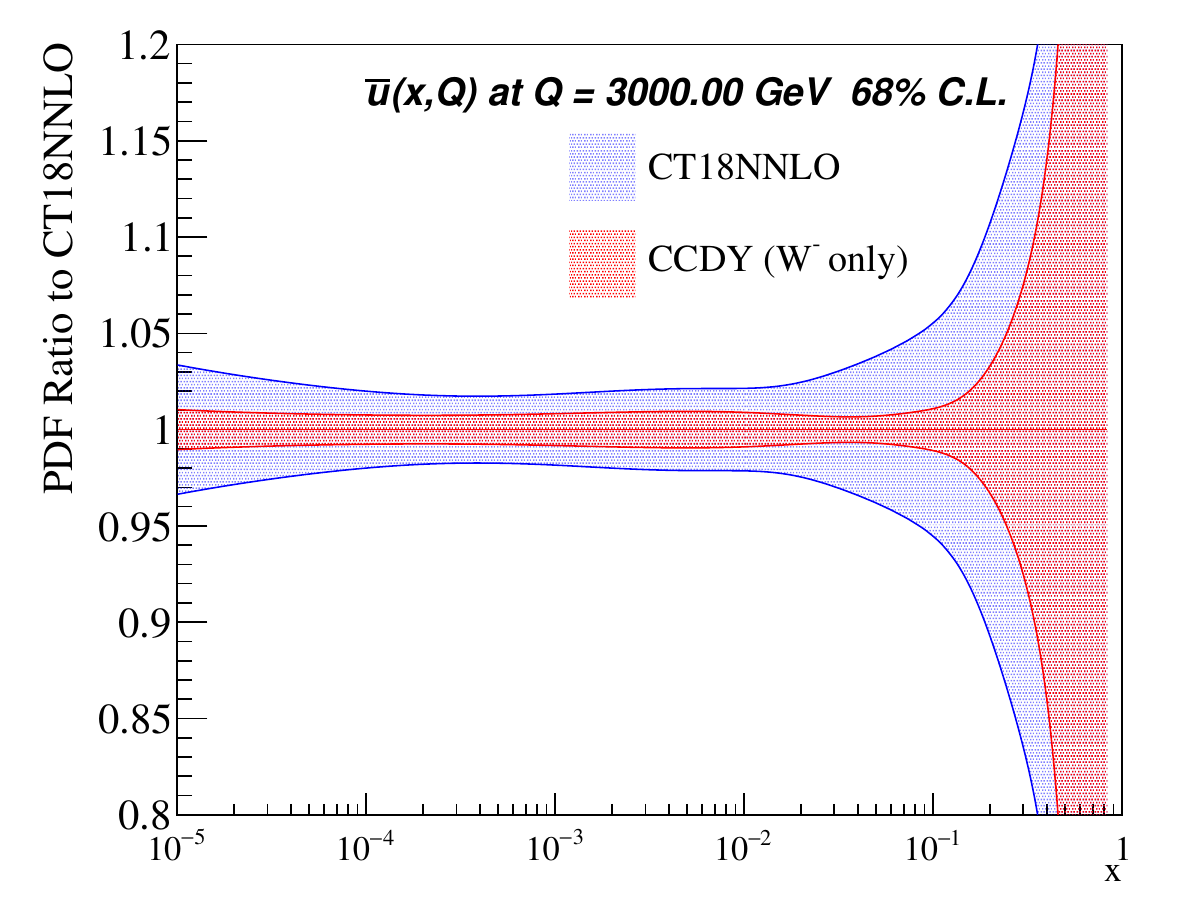}
  }
    \subfigure[]{
    \label{LABEL1}
    \includegraphics[width=0.48\textwidth]{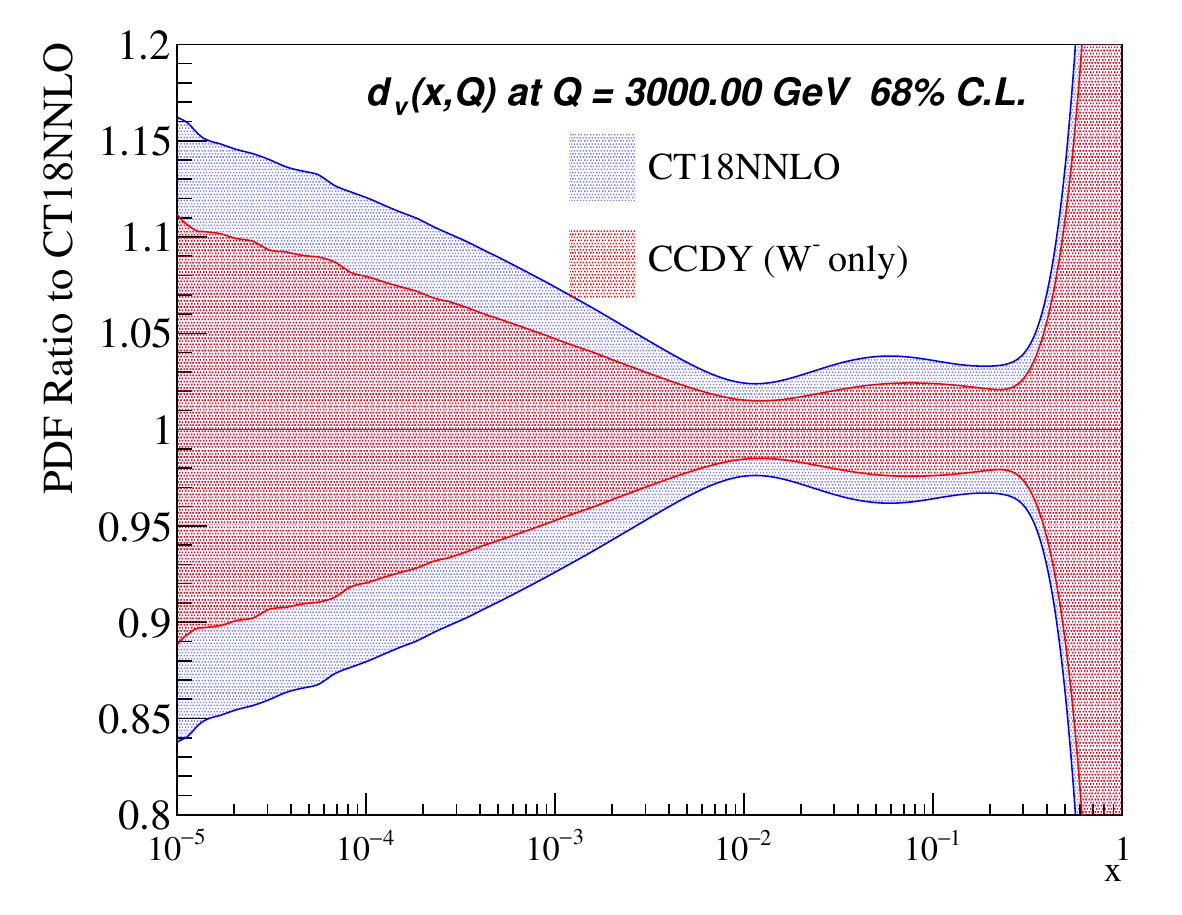}
  }
  \subfigure[]{
    \label{LABEL2}
    \includegraphics[width=0.48\textwidth]{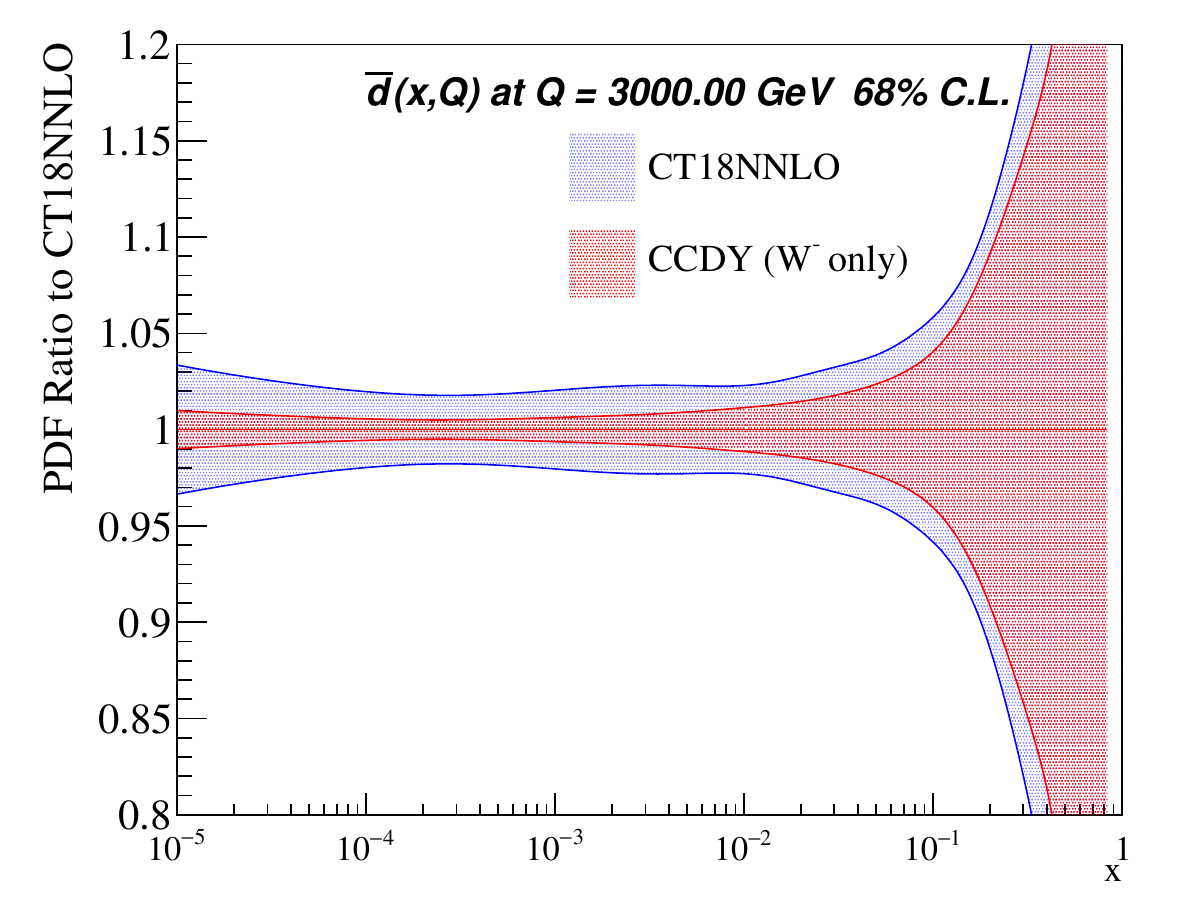}
  }
  \caption{The improvement, relative to \texttt{CT18}, for $u_V$, $\bar{u}$, $d_V$, with the addition of $W^-$, for 6000~\fb.}
  \label{partonsWMinus18}
\end{figure}

\begin{figure}[H]
  \centering
  \subfigure[]{
    \label{LABEL1}
    \includegraphics[width=0.48\textwidth]{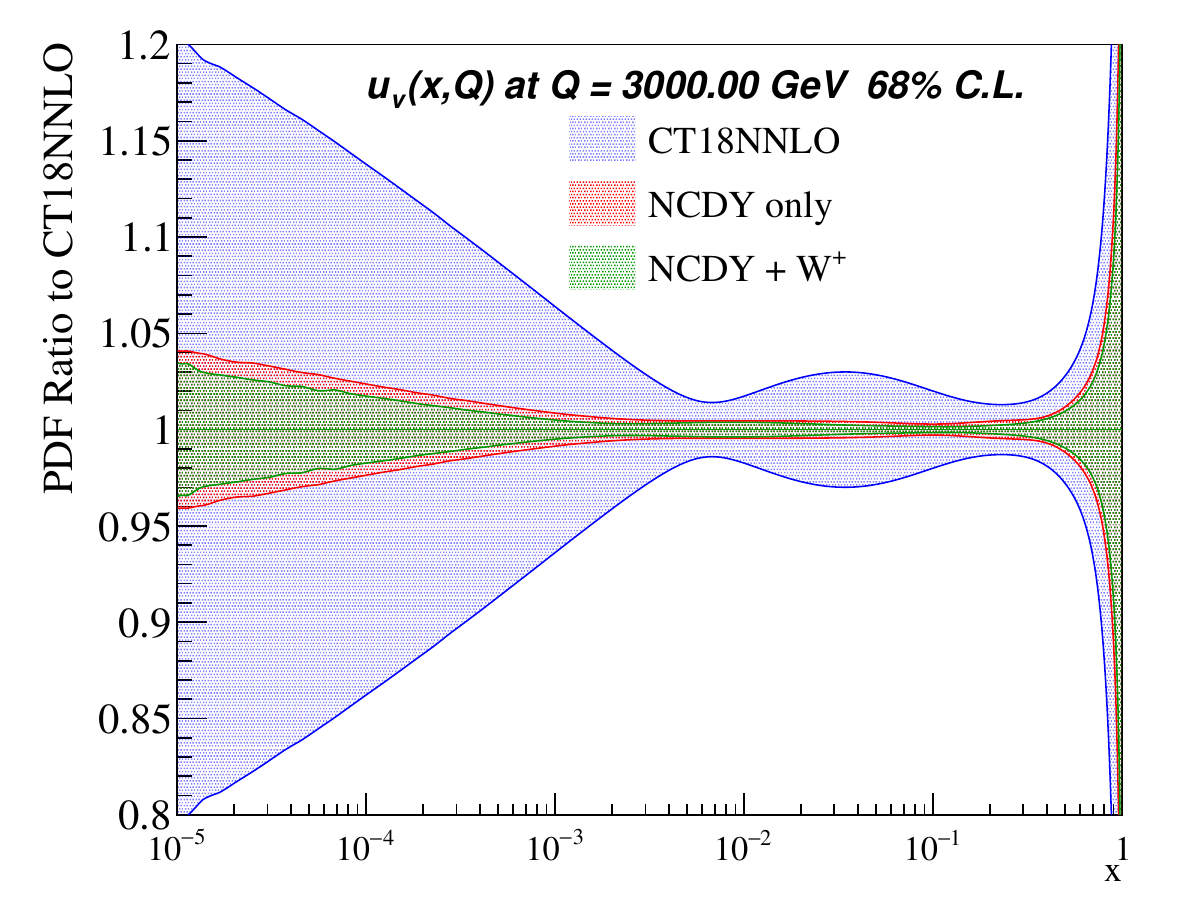}
  }
  \subfigure[]{
    \label{LABEL2}
    \includegraphics[width=0.48\textwidth]{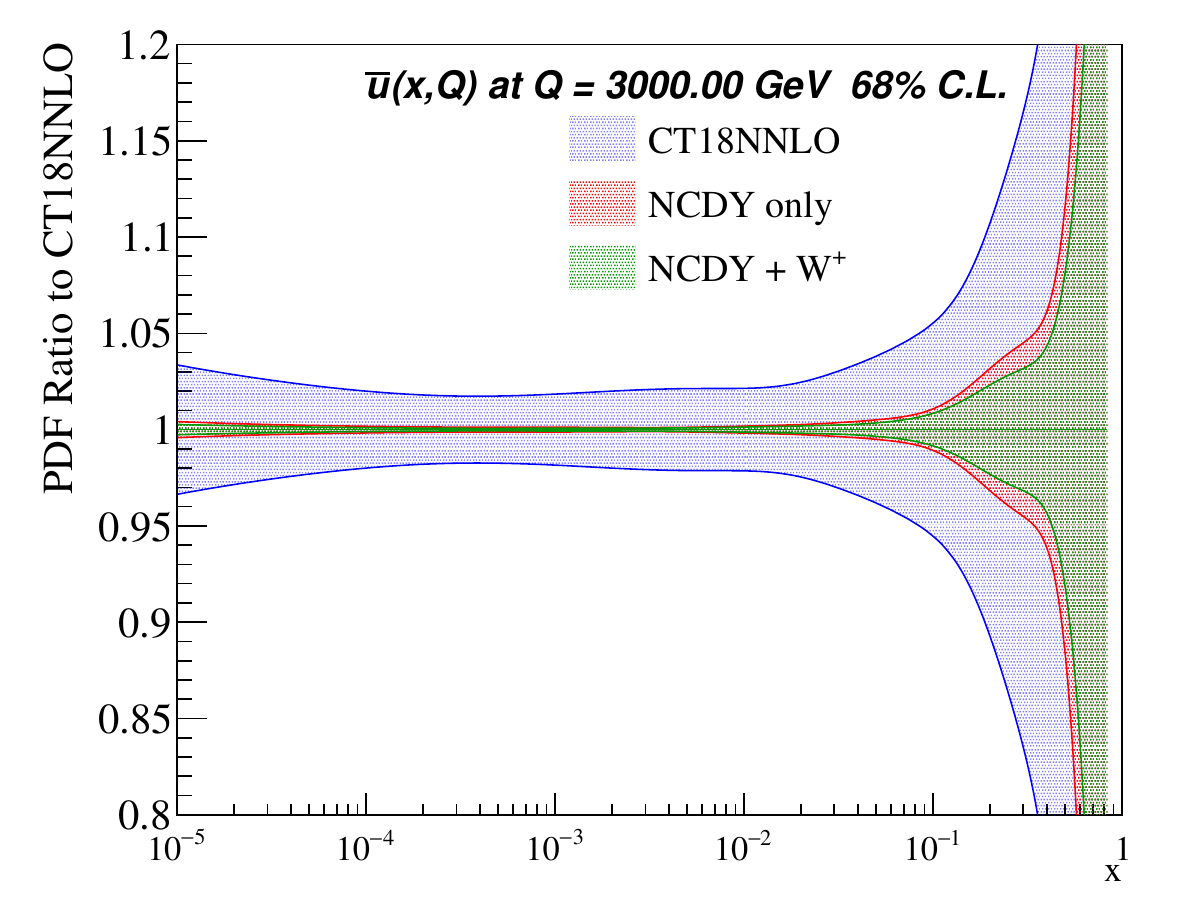}
  }
    \subfigure[]{
    \label{LABEL1}
    \includegraphics[width=0.48\textwidth]{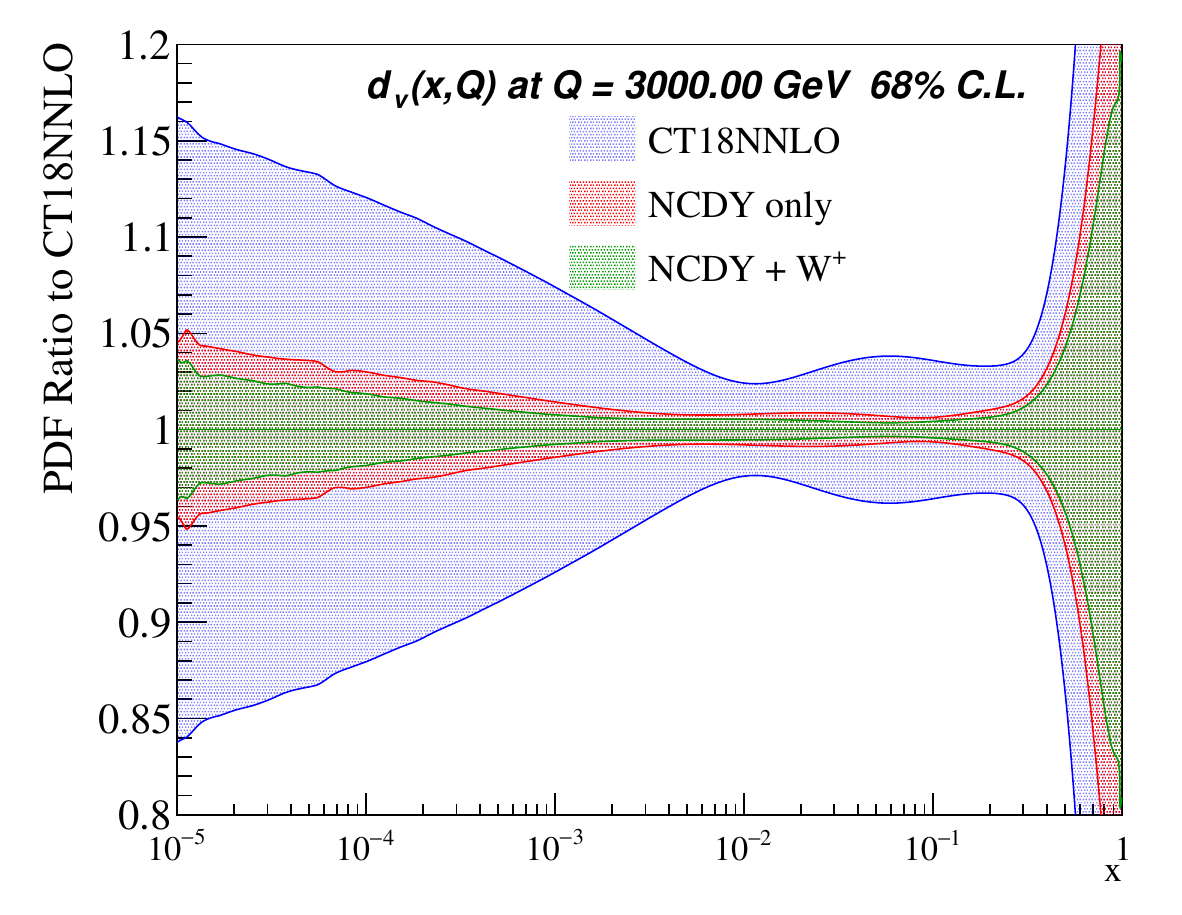}
  }
  \subfigure[]{
    \label{LABEL2}
    \includegraphics[width=0.48\textwidth]{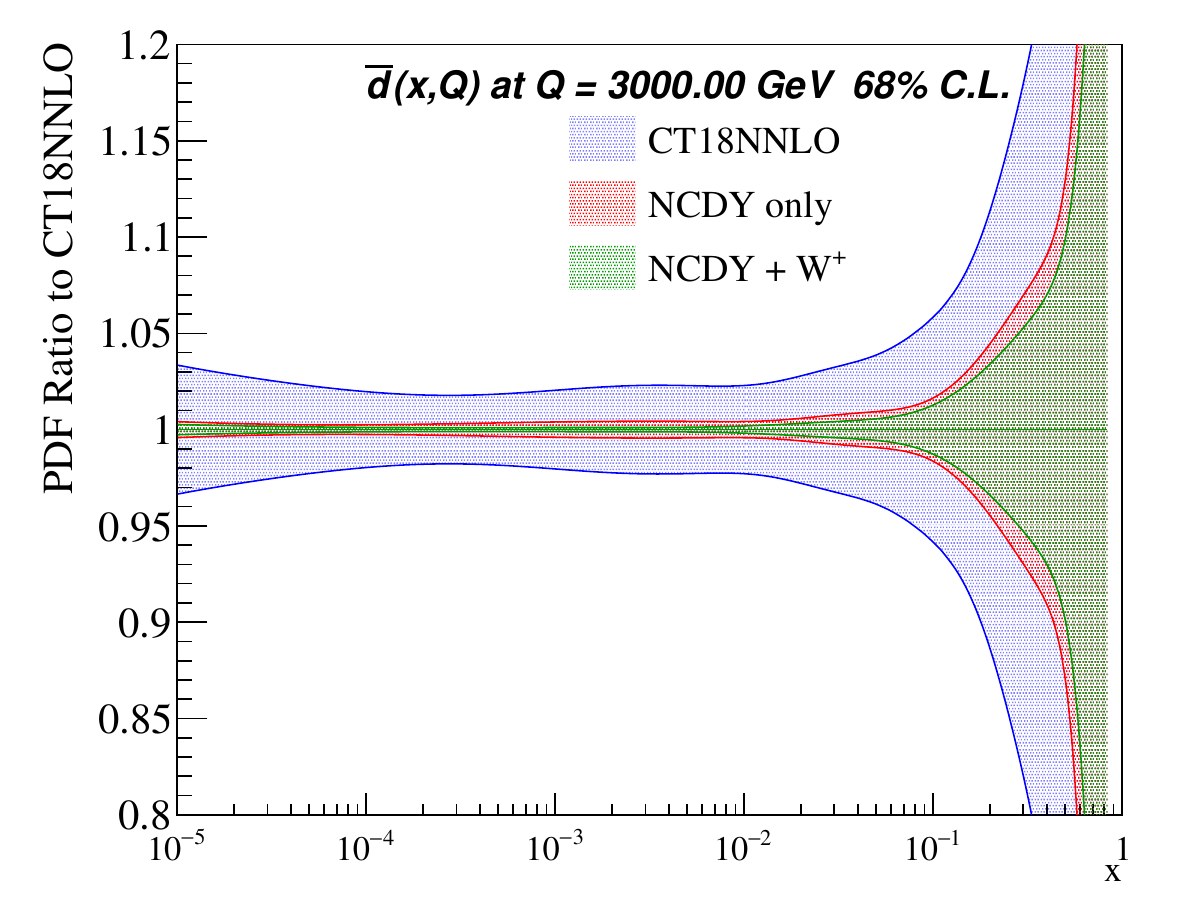}
  }
  \caption{The improvement. relative to \texttt{CT18}, for $u_V$, $\bar{u}$, $d_V$, with the addition of NCDY results, and the addition of both NCDY and $W^+$, for 6000~\fb.}
  \label{partonsNCWPlus18}
\end{figure}

\begin{figure}[H]
  \centering
  \subfigure[]{
    \label{LABEL1}
    \includegraphics[width=0.48\textwidth]{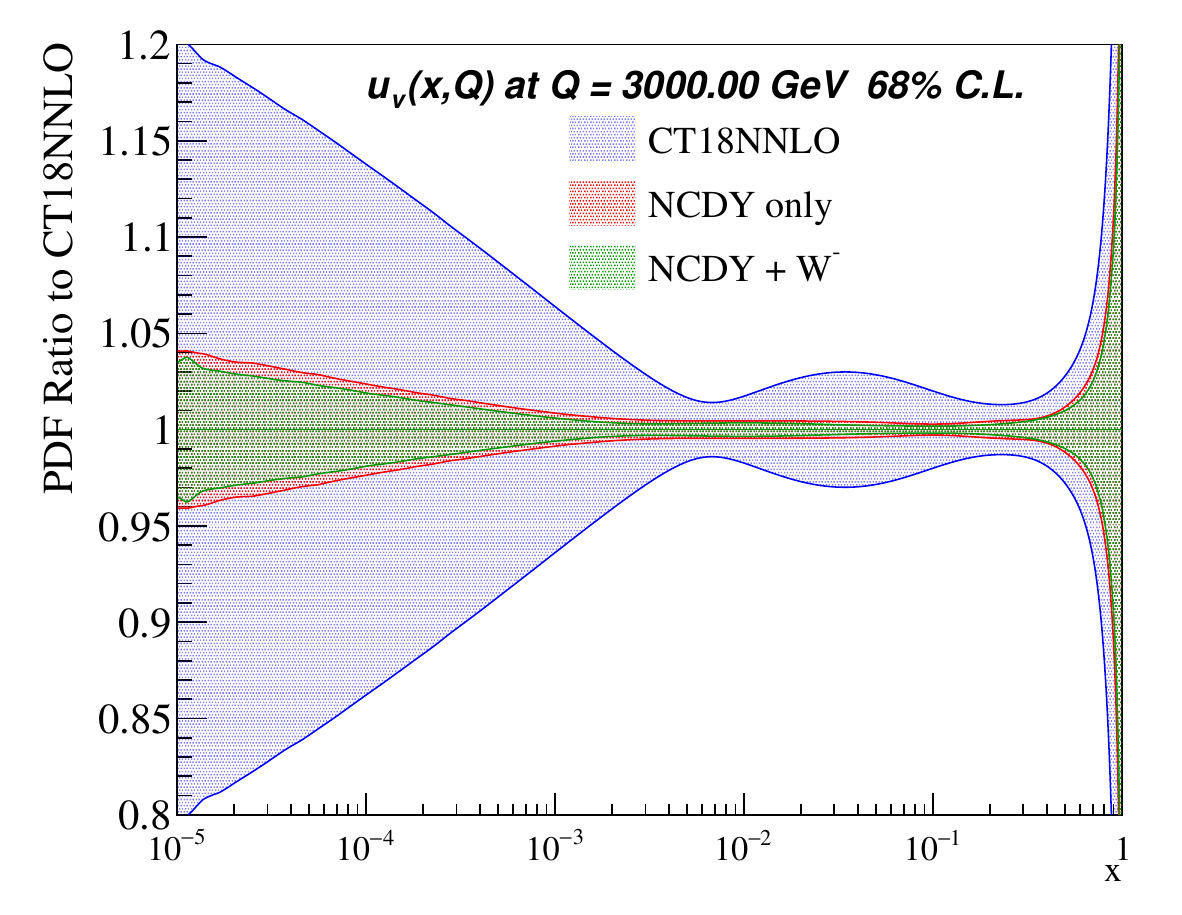}
  }
  \subfigure[]{
    \label{LABEL2}
    \includegraphics[width=0.48\textwidth]{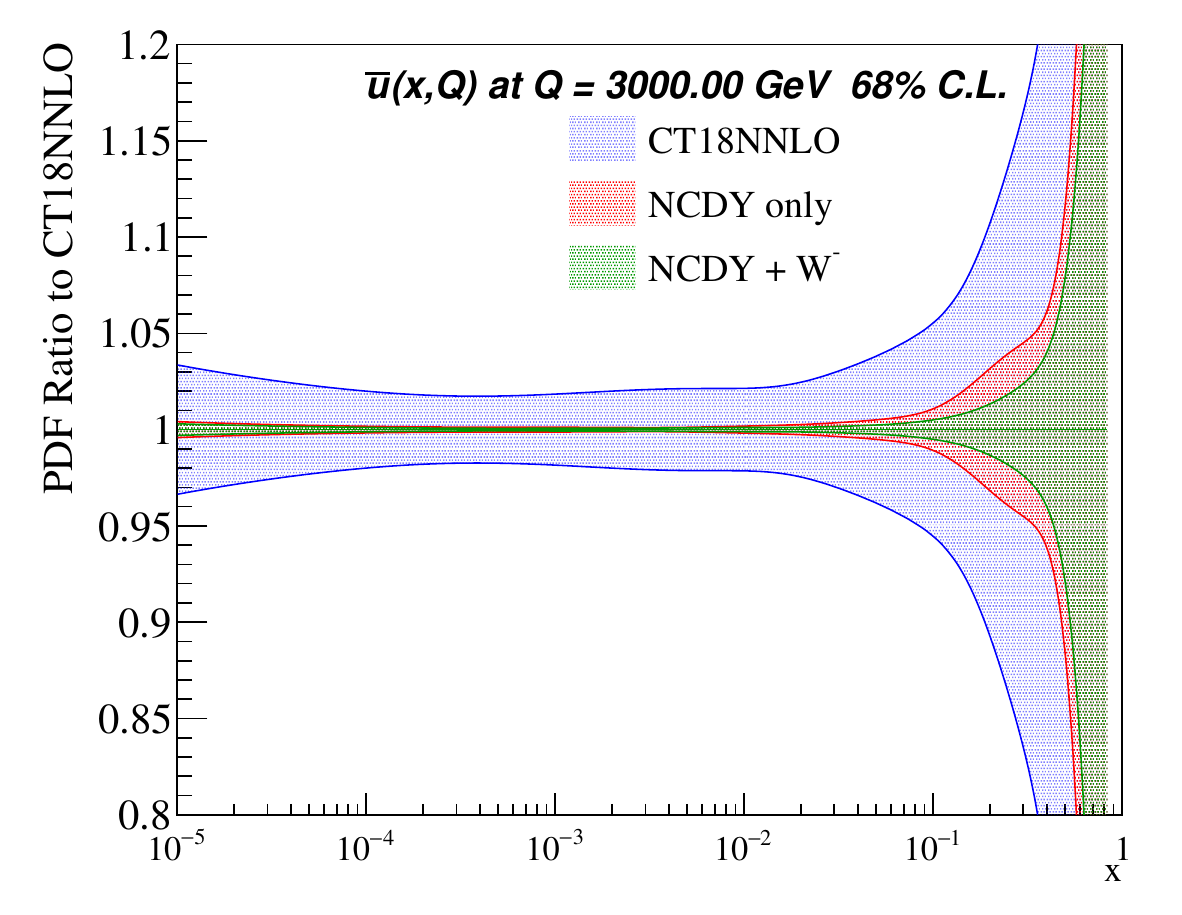}
  }
    \subfigure[]{
    \label{LABEL1}
    \includegraphics[width=0.48\textwidth]{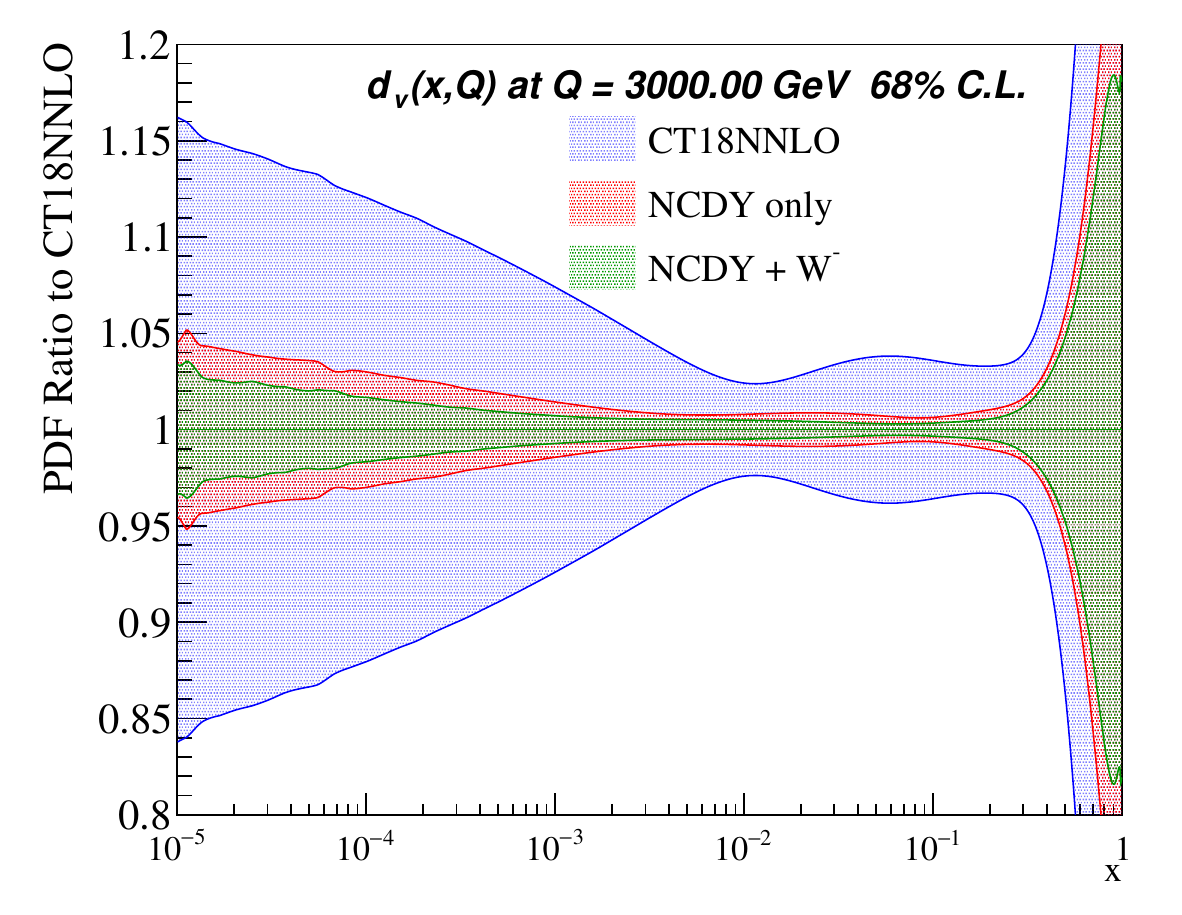}
  }
  \subfigure[]{
    \label{LABEL2}
    \includegraphics[width=0.48\textwidth]{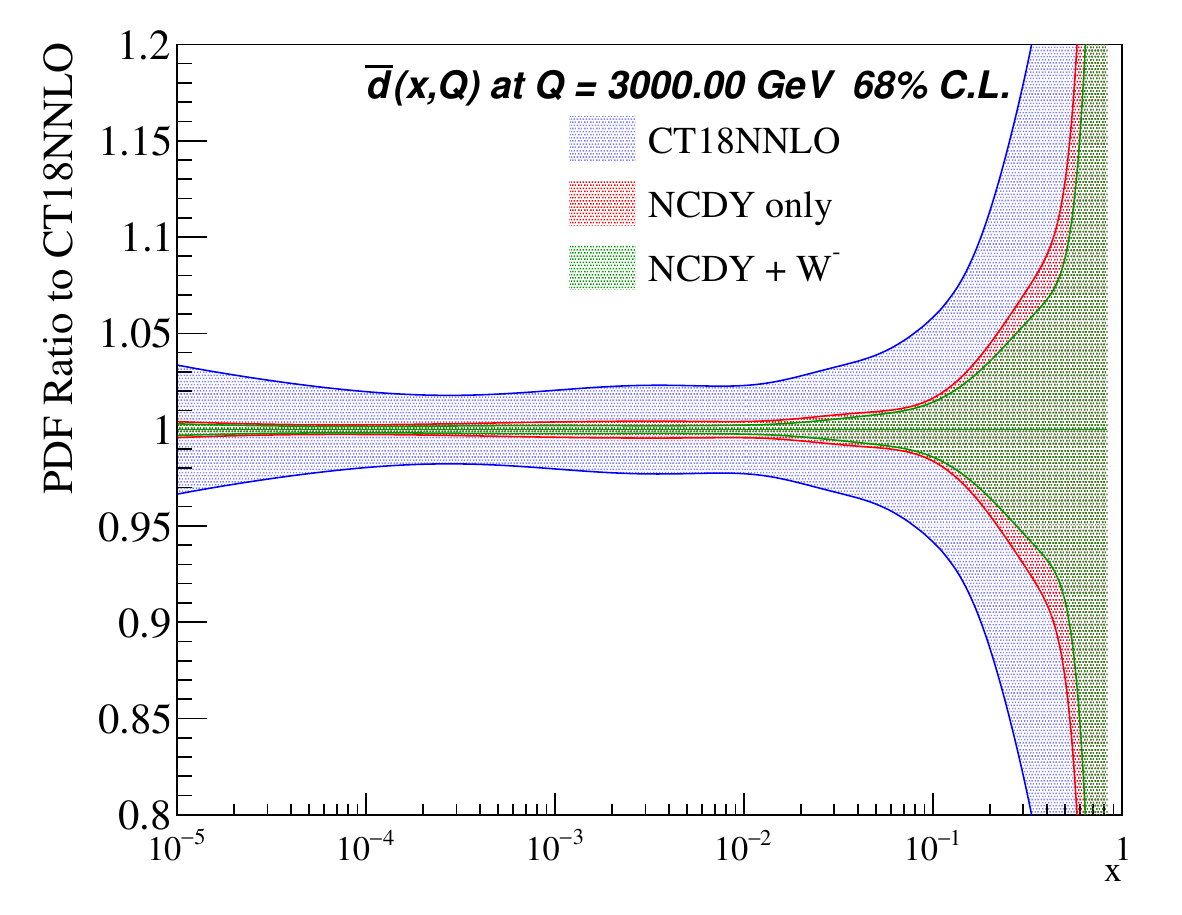}
  }
  \caption{The improvement, relative to  \texttt{CT18}, for $u_V$, $\bar{u}$, $d_V$, and $\bar{d}$,  with the addition of NCDY results, and the addition of both NCDY and $W^-$, for 6000~\fb.}
  \label{partonsNCWMinus18}
\end{figure}

\subsection{Improvements in $\delta^{PDF}$ versus $m_{\ell\ell}$, separated by $W$ Boson Charge} \label{app_tables}

\begin{table}[H]
    \centering
    \begin{tabular}{c|cc|cc|cc|cc}
    \hline
    \hline
     & \multicolumn{2}{|c}{$m_{\ell\ell}$ (NCDY) [TeV]} & \multicolumn{2}{|c}{$M_T$ (CCDY) [TeV]} & \multicolumn{2}{|c}{$M_T$ ($W^+$) [TeV]} & \multicolumn{2}{|c}{$M_T$ ($W^-$) [TeV]} \\
     $m_{\ell\ell}(M_T)$ & $\delta^{PDF}_{pre}$ [\%] & $\delta^{PDF}_{post}$ [\%] & $\delta^{PDF}_{pre}$ [\%] & $\delta^{PDF}_{post}$ [\%] & $\delta^{PDF}_{pre}$ [\%] & $\delta^{PDF}_{post}$ [\%] & $\delta^{PDF}_{pre}$ [\%] & $\delta^{PDF}_{post}$ [\%] \\
     \hline
        $>1$ & 5.9 & 3.3 & 5.9 & 2.9 & 6.2 & 2.8 & 7.9 & 4.8 \\
        $>2$ & 9.8 & 7.0 & 10.4 & 6.4 & 11.3 & 6.5 & 13.4 & 9.7 \\
        $>3$ & 15.4 & 11.8 & 16.4 & 10.9 & 18.0 & 11.1 & 20.7 & 15.8 \\
        $>4$ & 22.2 & 17.1 & 23.7 & 16.5 & 25.7 & 16.6 & 29.5 & 22.8 \\
        $>5$ & 30.8 & 23.2 & 33.0 & 23.4 & 35.0 & 23.5 & 40.1 & 30.8 \\
    \hline
    \hline
    \end{tabular}
    \caption{Similar as Table~\ref{tab:PDFUnc_Xsec3D}, but the PDF is updated by $W^+$ only, for 6000 fb$^{-1}$.}
    \label{tab:PDFUnc_WPlus}
\end{table}

\begin{table}[H]
    \centering
    \begin{tabular}{c|cc|cc|cc|cc}
    \hline
    \hline
     & \multicolumn{2}{|c}{$m_{\ell\ell}$ (NCDY) [TeV]} & \multicolumn{2}{|c}{$M_T$ (CCDY) [TeV]} & \multicolumn{2}{|c}{$M_T$ ($W^+$) [TeV]} & \multicolumn{2}{|c}{$M_T$ ($W^-$) [TeV]} \\
     $m_{\ell\ell}(M_T)$ & $\delta^{PDF}_{pre}$ [\%] & $\delta^{PDF}_{post}$ [\%] & $\delta^{PDF}_{pre}$ [\%] & $\delta^{PDF}_{post}$ [\%] & $\delta^{PDF}_{pre}$ [\%] & $\delta^{PDF}_{post}$ [\%] & $\delta^{PDF}_{pre}$ [\%] & $\delta^{PDF}_{post}$ [\%] \\
     \hline
        $>1$ & 5.9 & 1.5 & 5.9 & 2.9 & 6.2 & 3.8 & 7.9 & 2.3 \\
        $>2$ & 9.8 & 3.3 & 10.4 & 7.2 & 11.3 & 8.7 & 13.4 & 5.6 \\
        $>3$ & 15.4 & 6.5 & 16.4 & 12.5 & 18.0 & 14.3 & 20.7 & 10.7 \\
        $>4$ & 22.2 & 11.2 & 23.7 & 18.3 & 25.7 & 20.1 & 29.5 & 17.7 \\
        $>5$ & 30.8 & 17.6 & 33.0 & 25.0 & 35.0 & 26.6 & 40.1 & 26.1 \\
    \hline
    \hline
    \end{tabular}
    \caption{Similar as Table~\ref{tab:PDFUnc_Xsec3D}, but the PDF is updated by $W^-$ only, for 6000 fb$^{-1}$.}
    \label{tab:PDFUnc_WMinus}
\end{table}

\begin{table}[H]
    \centering
    \begin{tabular}{c|cc|cc|cc|cc}
    \hline
    \hline
     & \multicolumn{2}{|c}{$m_{\ell\ell}$ (NCDY) [TeV]} & \multicolumn{2}{|c}{$M_T$ (CCDY) [TeV]} & \multicolumn{2}{|c}{$M_T$ ($W^+$) [TeV]} & \multicolumn{2}{|c}{$M_T$ ($W^-$) [TeV]} \\
     $m_{\ell\ell}(M_T)$ & $\delta^{PDF}_{pre}$ [\%] & $\delta^{PDF}_{post}$ [\%] & $\delta^{PDF}_{pre}$ [\%] & $\delta^{PDF}_{post}$ [\%] & $\delta^{PDF}_{pre}$ [\%] & $\delta^{PDF}_{post}$ [\%] & $\delta^{PDF}_{pre}$ [\%] & $\delta^{PDF}_{post}$ [\%] \\
     \hline
        $>1$ & 5.9 & 0.8 & 5.9 & 1.1 & 6.2 & 1.3 & 7.9 & 1.4 \\
        $>2$ & 9.8 & 1.9 & 10.4 & 2.8 & 11.3 & 3.2 & 13.4 & 3.2 \\
        $>3$ & 15.4 & 3.3 & 16.4 & 4.7 & 18.0 & 5.4 & 20.7 & 4.9 \\
        $>4$ & 22.2 & 4.7 & 23.7 & 6.8 & 25.7 & 7.7 & 29.5 & 6.5 \\
        $>5$ & 30.8 & 6.3 & 33.0 & 9.2 & 35.0 & 10.3 & 40.1 & 8.3 \\
    \hline
    \hline
    \end{tabular}
    \caption{Similar as Table~\ref{tab:PDFUnc_Xsec3D}, but the PDF is updated by NCDY and $W^+$ templates, for 6000 fb$^{-1}$.}
    \label{tab:PDFUnc_WPlus_NCDY}
\end{table}

\begin{table}[H]
    \centering
    \begin{tabular}{c|cc|cc|cc|cc}
    \hline
    \hline
     & \multicolumn{2}{|c}{$m_{\ell\ell}$ (NCDY) [TeV]} & \multicolumn{2}{|c}{$M_T$ (CCDY) [TeV]} & \multicolumn{2}{|c}{$M_T$ ($W^+$) [TeV]} & \multicolumn{2}{|c}{$M_T$ ($W^-$) [TeV]} \\
     $m_{\ell\ell}(M_T)$ & $\delta^{PDF}_{pre}$ [\%] & $\delta^{PDF}_{post}$ [\%] & $\delta^{PDF}_{pre}$ [\%] & $\delta^{PDF}_{post}$ [\%] & $\delta^{PDF}_{pre}$ [\%] & $\delta^{PDF}_{post}$ [\%] & $\delta^{PDF}_{pre}$ [\%] & $\delta^{PDF}_{post}$ [\%] \\
     \hline
        $>1$ & 5.9 & 0.6 & 5.9 & 1.1 & 6.2 & 1.5 & 7.9 & 1.1 \\
        $>2$ & 9.8 & 1.5 & 10.4 & 2.9 & 11.3 & 3.4 & 13.4 & 2.6 \\
        $>3$ & 15.4 & 2.6 & 16.4 & 4.9 & 18.0 & 5.7 & 20.7 & 4.5 \\
        $>4$ & 22.2 & 3.9 & 23.7 & 7.0 & 25.7 & 7.8 & 29.5 & 6.7 \\
        $>5$ & 30.8 & 5.5 & 33.0 & 9.2 & 35.0 & 10.2 & 40.1 & 9.2 \\
    \hline
    \hline
    \end{tabular}
    \caption{Similar as Table~\ref{tab:PDFUnc_Xsec3D}, but the PDF is updated by NCDY and $W^-$ templates, for 6000 fb$^{-1}$.}
    \label{tab:PDFUnc_WMinus_NCDY}
\end{table}

\begin{acknowledgments}
	We thank our CTEQ-TEA colleagues for support and discussions.
	This work was supported by the U.S. National Science Foundation under Grants No. PHY-2309950 and PHY-2310291.
	C.-P. Yuan is also grateful for the support from the Wu-Ki Tung Professorship in Particle Physics.
\end{acknowledgments}
\bibliography{ZWprime_1_bib} 



%
\end{document}